# On the Local Correctness of $\ell^1$-minimization for Dictionary Learning

Quan Geng[*,◇], Huan Wang[†,◇], and John Wright[◇]

[*]Department of Electrical Engineering, University of Illinois at Urbana-Champaign
[†]Department of Computer Science, Yale University
[◇]Visual Computing Group, Microsoft Research Asia

October 30, 2018

**Abstract**

The idea that many important classes of signals can be well-represented by linear combinations of a small set of atoms selected from a given dictionary has had dramatic impact on the theory and practice of signal processing. For practical problems in which an appropriate sparsifying dictionary is not known ahead of time, a very popular and successful heuristic is to search for a dictionary that minimizes an appropriate sparsity surrogate over a given set of sample data. While this idea is appealing, the behavior of these algorithms is largely a mystery; although there is a body of empirical evidence suggesting they do learn very effective representations, there is little theory to guarantee when they will behave correctly, or when the learned dictionary can be expected to generalize. In this paper, we take a step towards such a theory. We show that under mild hypotheses, the dictionary learning problem is locally well-posed: the desired solution is indeed a local minimum of the $\ell^1$ norm. Namely, if $\boldsymbol{A} \in \mathbb{R}^{m\times n}$ is an incoherent (and possibly overcomplete) dictionary, and the coefficients $\boldsymbol{X} \in \mathbb{R}^{n\times p}$ follow a random sparse model, then with high probability $(\boldsymbol{A}, \boldsymbol{X})$ is a local minimum of the $\ell^1$ norm over the manifold of factorizations $(\boldsymbol{A}', \boldsymbol{X}')$ satisfying $\boldsymbol{A}'\boldsymbol{X}' = \boldsymbol{Y}$, provided the number of samples $p = \Omega(n^3 k)$. For overcomplete $\boldsymbol{A}$, this is the first result showing that the dictionary learning problem is locally solvable. Our analysis draws on tools developed for the problem of completing a low-rank matrix from a small subset of its entries, which allow us to overcome a number of technical obstacles; in particular, the absence of the restricted isometry property.[1]

## 1 Introduction

To a great extent, progress in signal processing over the past four decades has been driven by the quest for ever more effective signal representations. The development of increasingly powerful, relevant representations for natural images, from Fourier and DCT bases [ANR74] to Wavelets [MG84], Curvelets [CDDY06] and beyond, has significantly enriched our understanding of the structure of images, and has also spurred the development of influential practical coding standards [Wal91]. Because of this, hand design of signal representations has been a dominant paradigm in signal processing and applied mathematics. Indeed, it is difficult to overstate the intellectual and practical impact of this quest.

However, there are voices of dissent. One competing train of thought, dating at least back to the advent of the Karhunen-Loève transform in the 1970's, suggests that rather than meticulously designing an appropriate representation for each class of signals we encounter, it may be possible to simply learn an appropriate representation from large sets of sample data. This idea has several appeals: Given the recent proliferation of new and exotic types of data (images, videos, web and bioinformatic data, ect.), it may not be possible to invest the intellectual effort required to develop optimal representations for each new class of signal we encounter. At the same time, data are becoming increasingly high-dimensional, a fact which stretches the limitations of our human intuition, potentially limiting our ability to develop effective data representations. It may be possible for an automatic procedure to discover useful structure in the data that is not readily apparent to us.

[1]This work was partially performed while Q. Geng and H. Wang were interns in the Visual Computing Group at Microsoft Research Asia. The authors would like to thank Dan Spielman of Yale University and Yi Ma of MSRA for helpful discussions.

Spurred by this promise, researchers have invested a great amount of effort in developing algorithms that can automatically derive good representations for sample data. In particular, much recent effort has been focused on *sparse linear representations*. A signal $\boldsymbol{y} \in \mathbb{R}^m$ is said to have a sparse representation in terms of a given dictionary of basis signals $\boldsymbol{A} = [\boldsymbol{A}_1, \ldots, \boldsymbol{A}_n] \in \mathbb{R}^{m \times n}$ if $\boldsymbol{y} \approx \boldsymbol{A}\boldsymbol{x}$, where $\boldsymbol{x} \in \mathbb{R}^n$ is a coefficient vector with only a few nonzero entries ($k = \|\boldsymbol{x}\|_0 \ll n$). This notion of sparsity has emerged in the past 10 years as a dominant idea in signal processing [BDE09]. This is due both to the ubiquity of sparsity (or near-sparsity) in practical problems, as well as a line of fundamental theoretical results [DE03, Fuc04, CT05, ZY06, MY09] that assert that if $\boldsymbol{y}$ is known to be sparse in a known basis $\boldsymbol{A}$ satisfying certain technical conditions, the sparse coefficients $\boldsymbol{x}_0$ can be very accurately estimated (sometimes perfectly so!) by solving an $\ell^1$ minimization problem:

$$\text{minimize} \quad \|\boldsymbol{x}\|_1 \quad \text{subject to} \quad \boldsymbol{y} = \boldsymbol{A}\boldsymbol{x}. \tag{1.1}$$

These theoretical results allows us to deploy tools from sparse signal representation with great confidence: if the signal $\boldsymbol{y}$ has a sparse representation, then efficient algorithms are guaranteed to recover it.

When facing a new class of signals, however, it is not clear how to begin: what basis $\boldsymbol{A}$ might allow typical signals $\boldsymbol{y}$ to be sparsely represented? A popular heuristic is to search for a basis $\boldsymbol{A}$ that allows a given set of examples $\boldsymbol{Y} = [\boldsymbol{y}_1, \ldots, \boldsymbol{y}_p] \in \mathbb{R}^{m \times p}$ to be represented as compactly as possible. That is, we attempt to solve the following model problem, often referred to as "dictionary learning":

> Given samples $\boldsymbol{Y} = [\boldsymbol{y}_1, \ldots, \boldsymbol{y}_p] \in \mathbb{R}^{m \times p}$ all of which can be sparsely represented in terms of some unknown dictionary $\boldsymbol{A}$ ($\boldsymbol{Y} = \boldsymbol{A}\boldsymbol{X}$, for some $\boldsymbol{X}$ with sparse columns), recover $\boldsymbol{A}$.

A number of algorithms have been proposed for this problem [OF96, EAHH99, KDMR+03, AEB06, MBPS10] (see the survey [RBE10] for a more thorough review). Exploiting sparsity in learned dictionaries has led to practical success in a number of important problems in signal acquisition and processing [EA06, BE08, MBP+08, RS08, YWHM10]. On the other hand, relatively little theory is available to explain when and why dictionary learning algorithms succeed. There is also little in the way of guidelines to tell practitioners when the learned dictionary is expected to generalize beyond the given sample set $\boldsymbol{Y}$. This stands in contrast to the situation with hand-designed dictionaries, which often come with proofs of optimality for important classes of signals.

In this paper, we take a step towards closing this gap. We study a model optimization approach to dictionary learning:

$$\text{minimize } \|\boldsymbol{X}\|_1 \quad \text{subject to} \quad \boldsymbol{Y} = \boldsymbol{A}\boldsymbol{X}, \ \|\boldsymbol{A}_i\|_2 = 1 \ \forall\, i. \tag{1.2}$$

Here, $\|\cdot\|_1$ denotes the sum of magnitudes, $\|\boldsymbol{X}\|_1 = \sum_{ij} |X_{ij}|$. This optimization problem was first studied by Gribonval and Schnass [GS10], as a natural abstraction of popular dictionary learning algorithms (we will dicsuss the results of [GS10] in more detail in Section 2.1). Notice that while the objective function in (1.2) is convex, the constraint is not. Hence, in general it may seem that all we can hope for is a local optimum. This is a common feature of dictionary learning algorithms. Indeed, it is a classical observation in source separation that if we take a permutation matrix $\boldsymbol{\Pi} \in \mathbb{R}^{n \times n}$ and a diagonal matrix of signs $\boldsymbol{\Sigma}$, then whenever $(\boldsymbol{A}, \boldsymbol{X})$ solves the above problem, so does $(\boldsymbol{A}\boldsymbol{\Pi}\boldsymbol{\Sigma}, \boldsymbol{\Sigma}\boldsymbol{\Pi}^* \boldsymbol{X})$. This "sign-permutation ambiguity" implies corresponding to every local minimum of (1.2), there is a class of $2^n n!$ equivalent solutions. Moreover, a-priori there is nothing to prevent the existence of exponentially large classes of local minima. This might lead one to a dispiriting conclusion: "the problem (1.2) is impossible to solve in general; moreover, nothing rigorous can be said about its solution."

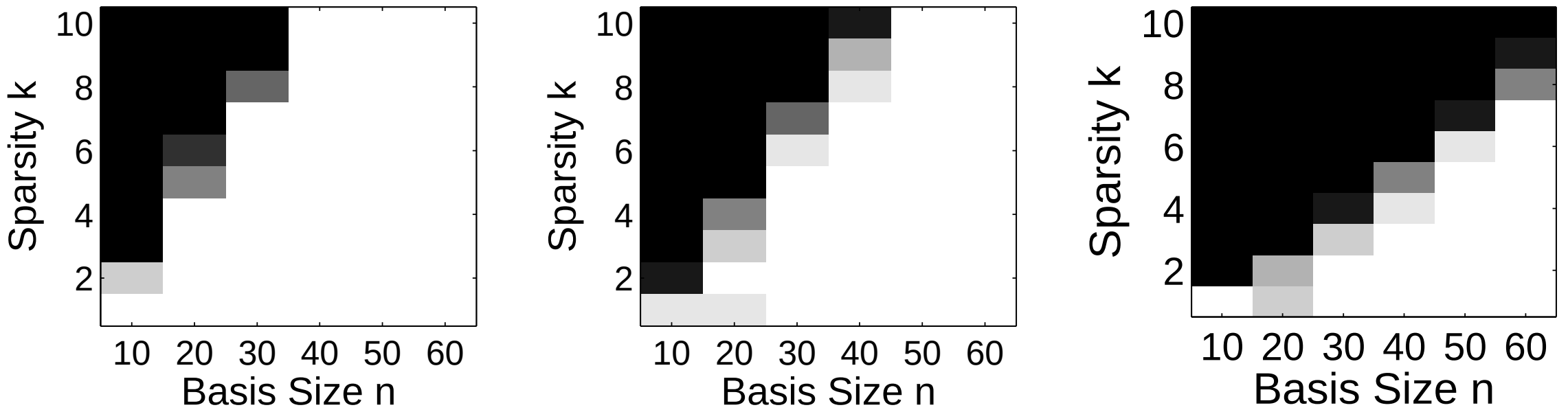

Figure 1: **Phase transitions in dictionary recovery?** We test whether locally minimizing the $\ell^1$ norm correctly recovers the dictionary $\boldsymbol{A} \in \mathbb{R}^{m\times n}$ and sparse coefficients $\boldsymbol{X} \in \mathbb{R}^{n\times p}$, for varying sparsity levels $k$ and problem size $n$. Left: $m = n$. Middle: $m = .8 \times n$. Right: $m = .6 \times n$. Here, $p = 5n\log(n)$. Trials are judged successful if the relative error $\|\hat{\boldsymbol{A}} - \boldsymbol{A}\|_F / \|\boldsymbol{A}\|_F$ in the recovered $\hat{\boldsymbol{A}}$ is smaller than $10^{-5}$. We average over 10 trials; white corresponds to success in all trials, black to failure in all trials. The problems are solved using an algorithm outlined in [GWW11].

Part of the goal of this paper is to dispel such pessimism. Figure 1 shows why there might be reason for hope. In it, we solve various synthetic instances of the problem (1.2), with varying problem size and sparsity level. The figure plots fractions of correct recoveries, for various aspect ratios $m/n$ of $\boldsymbol{A} \in \mathbb{R}^{m\times n}$. We observe a very intriguing phenomenon:

> Empirically, optimization algorithms for dictionary learning succeed when the the problem is well-structured ($\boldsymbol{X}$ is sufficiently sparse), and fail otherwise. Moreover, in simulated examples, the transition between these two modes of operation is fairly sharp.

This suggests that, similar to the results for $\ell^1$-minimization discussed above, there are important classes of dictionary learning problems that can be solved exactly by efficient (polynomial time) algorithms.

Fully understanding this phenomenon is a long-term goal. Although local optimization approaches to dictionary learning have repeatedly demonstrated good empirical behavior, the aforementioned difficulties of non-convexity and sign-permutation ambiguity raise significant technical obstacles to developing a theory of their correctness. Nevertheless, a step in this direction was taken by Gribonval and Schnass [GS10], who showed that if $\boldsymbol{A}$ is square ($m = n$), then for certain random coefficient models, the desired solution is indeed a local minimum of the $\ell^1$-norm with high probability. In this paper, we show that this is true for a wider range of matrices, including *overcomplete dictionaries* $\boldsymbol{A}$ with more columns than rows. We prove:

> If the matrix $\boldsymbol{A}$ is appropriately *incoherent* and the coefficients $\boldsymbol{X}$ are drawn from a random sparsity model, then after seeing polynomially many samples (say, $\Omega(n^3)$), with high probability the desired solution is indeed locally recoverable.

For non-square matrices, this is the first result suggesting that correct recovery is possible by $\ell^1$-minimization, even locally. Establishing it seems to demand a different set of technical tools and ideas from [GS10]. We will see that understanding the local properties of (1.2) essentially requires us to study a certain equality-constrained $\ell^1$ norm minimization problem, which arises by linearizing the nonlinear constraint in (1.2) at the desired solution $(\boldsymbol{A}_\star, \boldsymbol{X}_\star)$. While $\ell^1$ norm minimization has been widely studied, and its correctness for recovering sparse representations in known bases (i.e., problem (1.1)) is increasingly well-understood, the particular $\ell^1$ minimization problem encountered in dictionary learning raises new challenges. In particular, we will see that the linear constraints in this problem do not satisfy the Restricted Isometry Property (RIP) [CT05], a fact which significantly complicates their analysis. We are instead inspired by an analogy to the problem of completing a low-rank matrix from an observation consisting of a small subset of its entries [CR08], another problem in which the RIP (and analogues) fails. In particular, our analysis is inspired by the golfing scheme of David Gross [Gro09], which has proved useful for a variety of problems where the RIP is absent [CLMW09, CP10]. We also make heavy use of the convenient and powerful operator Chernoff bounds of Joel Tropp [Tro10], whose work builds on an approach introduced by Ahlswede and Winter [AW02].

### 1.1 Organization

This paper is organized as follows. In Section 2, we describe in greater detail the model studied here, and formally state our main result, and discuss its implications. In particular, in Section 2.1 we discuss its relationship to existing results. The remainder of the paper comprises a proof of this result. Section 3 develops optimality conditions, phrased in terms of the existence of a certain dual certificate. In Section 4, we construct this dual certificate. The success of the construction relies on a certain balancedness property of the linearized subproblem at the optimum; we formally state and prove this property in Section 5.

### 1.2 Notation

For matrices, $\boldsymbol{X}^*$ will denote the transpose of $\boldsymbol{X}$. $\|\boldsymbol{X}\|$ will denote the $\ell^2$ operator norm. $\|\boldsymbol{X}\|_F = \sqrt{\operatorname{tr}[\boldsymbol{X}^*\boldsymbol{X}]}$ will denote the Frobenius norm. By slight abuse of notation, $\|\boldsymbol{X}\|_1$ and $\|\boldsymbol{X}\|_\infty$ will denote the $\ell^1$ and $\ell^\infty$ norms of the matrix, viewed as a large vector:

$$\|\boldsymbol{X}\|_1 = \sum_{ij} |X_{ij}|, \qquad \|\boldsymbol{X}\|_\infty = \max_{ij} |X_{ij}|. \tag{1.3}$$

For vectors $\boldsymbol{x}$, the notation $\|\boldsymbol{x}\|$ will mean the $\ell^2$ norm $\sqrt{\boldsymbol{x}^*\boldsymbol{x}}$. $\|\boldsymbol{x}\|_1$ and $\|\boldsymbol{x}\|_\infty$ will denote the usual $\ell^1$ and $\ell^\infty$ norms, respectively. $[n]$ denotes the first $n$ positive integers, $\{1, \dots, n\}$. The symbols $\boldsymbol{e}_1, \dots, \boldsymbol{e}_d$ will denote the $d$ standard basis vectors for $\mathbb{R}^d$; their dimension will be clear from context. Throughout, the symbols $C_1, C_2, \dots,\ c_1, c_2, \dots$ refer to numerical constants. When used in different sections, they need not refer to the same constant. For a linear subspace $V \subset \mathbb{R}^d$, we will let $\boldsymbol{P}_V \in \mathbb{R}^{d \times d}$ denote the projection matrix onto $V$. For a linear subspace $V$ contained in a more general linear space (say, $V \subset \mathbb{R}^{d \times d'}$), we will let $\mathcal{P}_V$ denote the projection operator onto this space. We will slightly abuse notation, and define, for $I \subseteq [d]$, $\boldsymbol{P}_I$ to be the projection matrix onto the subspace of vectors supported on $I$; similarly, for $\Omega \subseteq [d] \times [d']$, $\mathcal{P}_\Omega : \mathbb{R}^{d \times d'} \to \mathbb{R}^{d \times d'}$ will denote the projection operator onto $\Omega$, which retains the entries indexed by $\Omega$, and sets the rest to zero. As usual, $\boldsymbol{A} \otimes \boldsymbol{B}$ denotes the Kronecker product between matrices $\boldsymbol{A}$ and $\boldsymbol{B}$. For $\boldsymbol{B} \in \mathbb{R}^{a \times b}$, $\operatorname{vec}[\boldsymbol{B}] \in \mathbb{R}^{ab}$ is defined by stacking $\boldsymbol{B}$ as a vector, columnwise.

## 2 Main Result

As described in the previous section, this paper is dedicated to better understanding the good behavior of $\ell^1$ minimization for dictionary learning. In particular, we would like to assert that under natural, easily-satisfied conditions, the desired solution can be recovered, at least locally. Of course, whether this is true will depend strongly on the properties of the dictionary $\boldsymbol{A}$ to be recovered, as well as the sparse coefficients $\boldsymbol{X}$ that generate our observation $\boldsymbol{Y} = \boldsymbol{A}\boldsymbol{X}$. In this paper, we restrict our attention to dictionaries $\boldsymbol{A}$ whose columns have unit $\ell^2$ norm. We will adopt the simple assumption that the columns of $\boldsymbol{A}$ are well-spread in the observation space $\mathbb{R}^m$, i.e., the *mutual coherence* [DE03]

$$\mu(\boldsymbol{A}) = \max_{i \neq j} |\langle \boldsymbol{A}_i, \boldsymbol{A}_j \rangle| \tag{2.1}$$

is small. Classical results [DE03, GN03, Fuc04] show that if $\boldsymbol{A}$ is a (known) dictionary, then $\ell^1$ minimization recovers any sparse representation with up to $1/2\mu(\boldsymbol{A})$ nonzeros:

$$\|\boldsymbol{x}_0\|_0 < \tfrac{1}{2}(1 + 1/\mu(\boldsymbol{A})) \quad \Longrightarrow \quad \boldsymbol{x}_0 = \arg\min \|\boldsymbol{x}\|_1 \quad \text{subject to} \quad \boldsymbol{A}\boldsymbol{x} = \boldsymbol{A}\boldsymbol{x}_0. \tag{2.2}$$

This result, while pessimistic compared to typical-case behavior [CP09], is powerful because its assumptions on $\boldsymbol{A}$ are reasonable; it does not seem particularly onerous to assume that $\mu(\boldsymbol{A})$ will be small for learned dictionaries.[2]

The next question is how to model the sparse coefficients $\boldsymbol{X}$. In analogy to results in sparse representation, we would like to assert that dictionary learning algorithms function correctly when their assumptions are met, i.e., when the coefficients $\boldsymbol{X}$ are sufficiently sparse. However, it is also clear that by itself sparsity of $\boldsymbol{X}$ is not sufficient for $(\boldsymbol{A}, \boldsymbol{X})$ to be a local minimum. As a very simple example, imagine that there is some $i$ for which all of the $X_{ij}$ are zero. In this paper, we assume that the sparsity pattern of $\boldsymbol{X}$ is random, and that the values of the nonzero entries are Gaussian.

More precisely, we assume that each of the columns $\boldsymbol{x}_1, \dots, \boldsymbol{x}_p$ of $\boldsymbol{X} \in \mathbb{R}^{n \times p}$ is generated iid by first choosing $k$ out of its $n$ entries uniformly at random to be nonzero, and letting the magnitude of these nonzero entries be independent Gaussians with zero mean and common standard deviation $\sigma$. The choice of a Gaussian model is one of mathematical convenience; the results in this paper are easily generalized to wider classes of symmetric distributions. However, the assumptions of zero mean and common variance are more essential to our analysis. We can state the above model more formally as follows. We assume that the observations $\boldsymbol{Y} = [\boldsymbol{y}_1, \dots, \boldsymbol{y}_p] \in \mathbb{R}^{m \times p}$ are generated iid, $\boldsymbol{y}_j = \boldsymbol{A}\boldsymbol{x}_j$, where $\boldsymbol{x}_j \in \mathbb{R}^n$ satisfies a Gaussian-random-sparsity model:

$$\Omega_j \;\sim\; \mathrm{uni}\binom{[n]}{k} \tag{2.3}$$

and

$$\boldsymbol{x}_j = \boldsymbol{P}_{\Omega_j} \boldsymbol{v}_j, \tag{2.4}$$

where

$$v_{ij} \;\sim_{iid}\; \mathcal{N}\left(0, \sigma^2\right), \qquad \sigma = \sqrt{n/kp}. \tag{2.5}$$

That is, $\boldsymbol{X} = \mathcal{P}_\Omega[\boldsymbol{V}]$, where $\Omega = \{(i,j) \mid j \in [p], i \in \Omega_j\}$ is the overall support set. The advantage to writing $\boldsymbol{X}$ in this manner is that it makes independence of $\Omega$ and $\boldsymbol{V}$ clear. The scaling on $v_{ij}$ plays no essential role in our proof – the normalization in (2.5) is simply notationally convenient because it implies that the spectral norm, $\|\boldsymbol{X}\|$, is approximately one when $p$ is large.

In dictionary learning, we do not observe $\boldsymbol{A}$ or $\boldsymbol{X}$, but rather their product $\boldsymbol{Y} = \boldsymbol{A}\boldsymbol{X} \in \mathbb{R}^{m \times p}$. Corresponding to this observation $\boldsymbol{Y}$, there exists a manifold of possible factorizations

$$\mathcal{M} = \{(\boldsymbol{A}, \boldsymbol{X}) \mid \boldsymbol{A}\boldsymbol{X} = \boldsymbol{Y},\ \|\boldsymbol{A}_i\|_2 = 1\ \forall\, i\} \subset \mathbb{R}^{m \times n} \times \mathbb{R}^{n \times p}. \tag{2.6}$$

In this notation, our model approach (1.2) can be can be viewed as a nonsmooth optimization over this smooth submanifold:

$$\text{minimize } f(\boldsymbol{x}) \quad \text{subject to} \quad \boldsymbol{x} \in \mathcal{M}. \tag{2.7}$$

Our main result states that if $\boldsymbol{x} = (\boldsymbol{A}, \boldsymbol{X})$ satisfies the above assumptions, then provided the number of samples is large enough, with high probability $\boldsymbol{x}$ will be a local minimum of $f$. More precisely:

[2]Conversely, there is less a-priori reason to believe that dictionaries encountered from sample data will satisfy more powerful assumptions such as the RIP. The absence of RIP in $\boldsymbol{A}$ should not, however, be confused with the absence of RIP in the local analysis of dictionary learning, which as we will see arises not from the properties of $\boldsymbol{A}$ per se, but rather from the structure of the tangent space to the constraint manifold.

**Theorem 2.1.** *There exist numerical constants $C_1, C_2, C_3 > 0$ such that the following occurs. If $\boldsymbol{x} = (\boldsymbol{A}, \boldsymbol{X})$ satisfy the probability model* (2.3)-(2.5) *with*

$$k \;\le\; \min\{C_1/\mu(\boldsymbol{A}), C_2 n\}, \tag{2.8}$$

*Then $\boldsymbol{x}$ is a local minimum of the $\ell^1$ norm over $\mathcal{M}$, with probability at least*

$$1 - C_3\|\boldsymbol{A}\|^2 n^{3/2} k^{1/2} p^{-1/2} (\log p). \tag{2.9}$$

This result implies that from polynomially many samples (say $p = \omega(n^3 k)$), the dictionary learning problem becomes locally well-posed, i.e., the desired solution becomes a local minimum of the $\ell^1$-norm. One can see that sparsity demanded by Theorem 2.1 mimics that of (2.2). Indeed, this result implies that under essentially the same conditions as the classical bound for sparse recovery (2.2), one can (locally) recover all of the sparse coefficients $\boldsymbol{X}$, as well as the sparsifying basis $\boldsymbol{A}$.

## 2.1 Comparison to existing results

As mentioned in the introduction, the theory of dictionary learning is only beginning to develop. The most direct point of comparison for our result is the very nice paper of Gribonval and Schnass [GS10] (henceforth "G-S"). That work proposed to study the optimization (1.2), and developed conditions for a given solution $\boldsymbol{x} = (\boldsymbol{A}, \boldsymbol{X})$ to be a local minimum. These conditions essentially demand that $\boldsymbol{x}$ be optimal over the tangent space to the constraint manifold at $\boldsymbol{x}$. While we do not directly use the optimality conditions of G-S, the duality condition that we base our approach on is essentially equivalent. However, the subsequent analysis uses a completely different set of tools and approaches.

Aside from developing optimality conditions, the major contribution of [GS10] is a probabilistic analysis of the case when $\boldsymbol{A} \in \mathbb{R}^{n\times n}$ is square and the coefficients $\boldsymbol{X}$ are iid Bernoulli-Gaussian, i.e., each $X_{ij}$ is nonzero with probability $\rho$, and the nonzero entries are conditionally Gaussian. Using arguments from geometry and concentration of measure, G-S show in this situation $(\boldsymbol{A}, \boldsymbol{X})$ is a local optimum with high probability provided $p = \Omega(n \log n/\rho)$.

Our Theorem 2.1 is more general, since it encompasses cases where $\boldsymbol{A}$ is nonsquare (i.e., an overcomplete dictionary). However, the number of samples stipulated by our bound is larger. Indeed, if we take $k = O(1)$, and set $\rho = k/n$ for purposes of comparison, then for square matrices, G-S's result guarantees correct recovery from $n^2 \log n$ samples. Our result requires at least $n^3$ samples, but applies to general matrices. It is possible that the gap between the two orders of growth might be further closed with a more refined analysis of the construction proposed in this paper.

## 2.2 Discussion

While we find these results quite encouraging, there is still much to do. In fact, there remains a wealth of fascinating open problems just involving the linearized subproblem. One natural question is whether the assumption of hard sparsity in $\boldsymbol{X}$ can be relaxed to a Bernoulli-Gaussian model, with similar probability of each coefficient being nonzero; i.e., $\rho \approx k/n$. In this case, care will need to be taken because a small number of columns of $\boldsymbol{X}$ may be so dense as to not be optimal. However, we see no essential obstacle to extending the approach used here to deal with this case. Another, more difficult question, is what will happen if the number of nonzero entries dramatically exceeds $C_1/\mu(\boldsymbol{A})$. In this case, again, many of the individual columns of $\boldsymbol{X}$ may be suboptimal, but it is still likely that the basis $\boldsymbol{A}$ is a local minimum. We believe that the golfing scheme of Section 4 will again provide a relevant tool. However, more work will need to be done to ensure that the balancedness condition in Theorem 5.1 still holds. Even more interesting from an application perspective would be to show that noise does not significantly affect the local optimality of the desired solution $\boldsymbol{x}_\star$. The framework of Negahban and collaborators may be relevant here [NRWY09].

# 3 Local Properties and the Linearized Subproblem

As we saw in the previous section, our main result concerns the local optimality of the desired solution $\boldsymbol{x} = (\boldsymbol{A}, \boldsymbol{X})$ over the smooth submanifold $\mathcal{M} \subset \mathbb{R}^{m\times n} \times \mathbb{R}^{n\times p}$. A key role in this result will be played by the tangent space $T_{\boldsymbol{x}}\mathcal{M}$ to $\mathcal{M}$ at $\boldsymbol{x}$, which can be identified[3] with the space of all perturbations $(\boldsymbol{\Delta}_A, \boldsymbol{\Delta}_X) \in \mathbb{R}^{m\times n} \times \mathbb{R}^{n\times p}$ satisfying

$$\boldsymbol{\Delta}_A \boldsymbol{X} + \boldsymbol{A}\boldsymbol{\Delta}_X = \boldsymbol{0}, \qquad \langle \boldsymbol{A}_i, \boldsymbol{\Delta}_{Ai} \rangle = 0 \; \forall\, i. \tag{3.1}$$

The first equation comes from differentiating the bilinear constraint $\boldsymbol{Y} = \boldsymbol{A}\boldsymbol{X}$, while the second comes from differentiating the constraint $\|\boldsymbol{A}_i\|^2 = 1$.

Intuitively, we might hope to study the local properties of $f$ by studying how it behaves on the tangent space at $\boldsymbol{x}$. Replacing $\mathcal{M}$ with its linearization about $\boldsymbol{x}$ yields the following optimization problem:

$$\text{minimize } f(\boldsymbol{x} + \boldsymbol{\delta}) \quad \text{subject to} \quad \boldsymbol{\delta} \in T_{\boldsymbol{x}}\mathcal{M}. \tag{3.2}$$

Using the above characterization of $T_{\boldsymbol{x}}\mathcal{M}$, this can be written a bit more concretely as

$$\text{minimize } \|\boldsymbol{X} + \boldsymbol{\Delta}_X\|_1 \quad \text{subject to} \quad \boldsymbol{\Delta}_A \boldsymbol{X} + \boldsymbol{A}\boldsymbol{\Delta}_X = \boldsymbol{0},\ \langle \boldsymbol{A}_i, \boldsymbol{\Delta}_{Ai} \rangle = 0,\ \forall\, i. \tag{3.3}$$

This linearized subproblem is convex. In particular, it is easy to see that under an appropriate change of variables, it is equivalent to an equality constrained $\ell^1$ minimization problem,

$$\text{minimize } \|\boldsymbol{z}\|_1 \quad \text{subject to} \quad \boldsymbol{B}\boldsymbol{z} = \boldsymbol{B}\boldsymbol{z}_0. \tag{3.4}$$

This should give us reason for optimism: as alluded to in the introduction, a great deal of effort has gone into developing technical tools for understanding the solutions to $\ell^1$-minimization problems. The following lemma tells us that in order to determine if $\boldsymbol{x}$ is a local minimum, it is enough to ask whether $\boldsymbol{\delta} = \boldsymbol{0}$ is the unique optimal solution to the linearized subproblem (3.2):

**Lemma 3.1.** *Suppose that $\boldsymbol{x} \in \mathcal{M}$ is such that $\boldsymbol{\delta} = \boldsymbol{0}$ is the unique optimal solution to* (3.2). *Then $\boldsymbol{x}$ is a local minimum of the function $f(\cdot)$ over $\mathcal{M}$. Conversely, if $\boldsymbol{x}$ is a local minimum, then $\boldsymbol{\delta} = \boldsymbol{0}$ is an optimal solution to* (3.2).

*Proof.* Please see Appendix D. □

We will prove our main result, Theorem 2.1 by showing that under the stated conditions the zero perturbation $(\boldsymbol{\Delta}_A, \boldsymbol{\Delta}_X) = (\boldsymbol{0}, \boldsymbol{0})$ is indeed the unique optimal solution to (3.3). To do so, we need to study an equality constrained $\ell^1$-minimization problem of the same form as (3.4). In the absence of specific assumptions on the distribution of $\boldsymbol{B}$ (such as Gaussianity [DT09]), the dominant tool for doing this is the Restricted Isometry Property (RIP), which holds with order $k$ and constant $0 \le \delta < 1$ if

$$(1-\delta)\|\boldsymbol{z}\|^2 \;\le\; \|\boldsymbol{B}\boldsymbol{z}\|^2 \;\le\; (1+\delta)\|\boldsymbol{z}\|^2 \quad \forall\, \boldsymbol{z} \text{ such that } \|\boldsymbol{z}\|_0 \le k. \tag{3.5}$$

When the RIP holds (with appropriate $k, \delta$), the $\ell^1$-minimization (3.4) recovers any sufficiently sparse $\boldsymbol{z}_0$, and noise-aware versions perform stably [Can08]. Thus, if we could show that the equality constraints in (3.3) satisfy an appropriate RIP variant, we would be done.

Unfortunately, this is not the case: *the RIP fails for our problem of interest.* We sketch why this is true. At a high level, the RIP states that the operator $\boldsymbol{B}$ respects the geometry of all sparse vectors; in particular, there are no sparse vectors near the nullspace of $\boldsymbol{B}$. In our case, $\boldsymbol{B}$ is specified by the equality constraints in (3.3). Take any permutation matrix $\boldsymbol{\Pi} \in \mathbb{R}^{n\times n}$ with no fixed point, and set

$$\boldsymbol{\Delta}_A = -\boldsymbol{A}\boldsymbol{\Pi}, \quad \boldsymbol{\Delta}_X = \boldsymbol{\Pi}\boldsymbol{X}. \tag{3.6}$$

[3]In this paper, our space of optimization $\mathcal{M}$ is most naturally specified as a submanifold of $\mathbb{R}^D$. We will commit sins of notation such as identifying the tangent space $T_{\boldsymbol{x}}\mathcal{M}$ with a particular vector subspace of $\mathbb{R}^D$, and occasionally writing $\boldsymbol{x} + \boldsymbol{\delta} \in \mathbb{R}^D$, where $\boldsymbol{x} \in \mathcal{M}$ and $\boldsymbol{\delta} \in T_{\boldsymbol{x}}\mathcal{M}$. Given the relatively small role played by the *intrinsic* Riemannian structure of $\mathcal{M}$ in the paper, we believe these simplifications are justified.

Then, it is easy to see that $\boldsymbol{\Delta}_A\boldsymbol{X} + \boldsymbol{A}\boldsymbol{\Delta}_X = \mathbf{0}$. Moreover, for each $i$,

$$\langle \boldsymbol{A}_i, \boldsymbol{\Delta}_{Ai} \rangle = -\langle \boldsymbol{A}_i, \boldsymbol{A}_{\pi(i)} \rangle \approx 0, \tag{3.7}$$

which follows because $\pi(i) \neq i$ and $\boldsymbol{A}$ has incoherent columns. Thus, we have constructed a perturbation $(\boldsymbol{\Delta}_A, \boldsymbol{\Delta}_X)$ that lies very near the nullspace of $\boldsymbol{B}$, and such that $\boldsymbol{\Delta}_X$ has exactly the same sparsity as the desired solution $\boldsymbol{X}$. In fact, not only does the RIP not hold, but structured variants (for example, restricting to matrices $\boldsymbol{\Delta}_X$ with *sparse columns*, rather than general sparse matrices) also fail. We make this intuitive argument more precise in Section A of the Appendix.

This leaves us in a situation with less in common with compressed sensing, and much more in common with the difficult problem of *matrix completion* [CR08]. In that problem, we are shown a small set of entries $Q_{i_1,j_1}, \ldots, Q_{i_p,j_p}$ of an unknown low-rank matrix $\boldsymbol{Q}$. The goal is to fill in the missing values. There, the natural analogue of the RIP also fails, since the sampling operator completely misses some low-rank matrices (for example, those consisting of a single nonzero entry that is not in the observed set). This fact significantly complicates analysis [CT09]. Motivated by applications in quantum information theory, recent papers by Gross and collaborators have introduced a number of technical tools that significantly ease the analysis of matrix completion [Gro09], allowing them to derive near-optimal recovery guarantees in a clear and simple manner. Moreover, the ideas of [Gro09] appear to be useful in a variety of settings beyond matrix completion [CLMW09, CP10]. In this paper, we use similar proof techniques to analyze the linearized subproblem (3.3). While the details necessarily differ quite a bit from Gross's work, our inspiration is very much the success of these tools in other non-RIP settings.

To describe this scheme in more detail, however, it is easiest to start at the very beginning. We wish to establish that $(\mathbf{0}, \mathbf{0})$ is optimal for (3.3). To do so, we recall the KKT conditions for this problem, which imply that $(\mathbf{0}, \mathbf{0})$ is optimal if and only if there exist two dual variables, a matrix $\boldsymbol{\Lambda} \in \mathbb{R}^{m\times p}$ (corresponding to the constraint $\boldsymbol{\Delta}_A\boldsymbol{X} + \boldsymbol{A}\boldsymbol{\Delta}_X = \mathbf{0}$) and a diagonal matrix $\boldsymbol{\Gamma} \in \mathbb{R}^{n\times n}$ (corresponding to the constraint $\langle \boldsymbol{A}_i, \boldsymbol{\Delta}_{Ai} \rangle = 0$) satisfying

$$\boldsymbol{A}^*\boldsymbol{\Lambda} \in \partial\|\cdot\|_1(\boldsymbol{X}) \tag{3.8}$$

$$\boldsymbol{\Lambda}\boldsymbol{X}^* = \boldsymbol{A}\boldsymbol{\Gamma}. \tag{3.9}$$

The interested reader can easily derive these conditions; we will provide a rigorous proof of a more useful variant below in Lemma 3.2. The first constraint simply asserts that each column $\boldsymbol{x}_j$ of $\boldsymbol{X}$ is the minimum $\ell^1$ norm solution to $\boldsymbol{A}\boldsymbol{x} = \boldsymbol{y}_j$. Indeed, writing $\Omega = \mathrm{support}(\boldsymbol{X})$ and $\boldsymbol{\Sigma} = \mathrm{sign}(\boldsymbol{X})$, we recall that

$$\partial\|\cdot\|_1(\boldsymbol{X}) = \{\boldsymbol{\Sigma} + \boldsymbol{W} \mid \mathcal{P}_\Omega[\boldsymbol{W}] = \mathbf{0},\ \|\boldsymbol{W}\|_\infty \leq 1\}. \tag{3.10}$$

Then, (3.8) holds if and only if $\exists\, \boldsymbol{w}_1, \ldots, \boldsymbol{w}_p \in \mathbb{R}^m$ such that

$$\boldsymbol{A}^*\boldsymbol{\lambda}_j = \boldsymbol{\Sigma}_j + \boldsymbol{w}_j, \quad \boldsymbol{P}_{\Omega_j}\boldsymbol{w}_j = 0,\ \|\boldsymbol{w}_j\|_\infty \leq 1. \tag{3.11}$$

This constraint is quite familiar from $\ell^1$-minimization: duality, and in particular the construction of dual certificates $\boldsymbol{\lambda}_j$ plays a crucial role in a number of works on the correctness of $\ell^1$ minimization [Fuc04, CT05, WM10].

On the other hand, the second constraint (3.9) is less familiar. It essentially asserts that locally we cannot improve our situation by changing the basis $\boldsymbol{A}$. Notice that it demands that each column of $\boldsymbol{\Lambda}\boldsymbol{X}^*$ is proportional to the corresponding column of $\boldsymbol{A}$; we find it convenient to introduce an operator $\Phi : \mathbb{R}^{m\times n} \to \mathbb{R}^{m\times n}$ that projects each column onto the orthogonal complement of the corresponding column of $\boldsymbol{A}$:

$$\Phi[\boldsymbol{M}] = [(\boldsymbol{I} - \boldsymbol{A}_1\boldsymbol{A}_1^*)\boldsymbol{M}_1 \mid \cdots \mid (\boldsymbol{I} - \boldsymbol{A}_n\boldsymbol{A}_n^*)\boldsymbol{M}_n], \tag{3.12}$$

giving an equivalent constraint $\Phi[\boldsymbol{\Lambda X}^*] = \mathbf{0}$. This constraint still places demands on all of the dual vectors $\boldsymbol{\lambda}_j$ simultaneously, making it potentially more difficult to satisfy than (3.8).

In the following lemma, we trade off between the two constraints, showing that if we tighten our demands on (3.8), we can correspondingly loosen the demand on (3.9):

**Lemma 3.2.** *Let $\boldsymbol{A}$ be a matrix with no $k$-sparse vectors in its nullspace. Suppose that there exists $\alpha > 0$ such that for all pairs $(\boldsymbol{\Delta}_A, \boldsymbol{\Delta}_X)$ satisfying* (3.1),

$$\|\mathcal{P}_{\Omega^c}\boldsymbol{\Delta}_X\|_F \;\geq\; \alpha\|\boldsymbol{\Delta}_A\|_F. \tag{3.13}$$

*Then if there exists $\boldsymbol{\Lambda} \in \mathbb{R}^{m\times p}$ such that*

$$\mathcal{P}_\Omega[\boldsymbol{A}^*\boldsymbol{\Lambda}] = \boldsymbol{\Sigma}, \quad \|\mathcal{P}_{\Omega^c}[\boldsymbol{A}^*\boldsymbol{\Lambda}]\|_\infty \leq 1/2, \tag{3.14}$$

*and*

$$\|\Phi[\boldsymbol{\Lambda X}^*]\|_F < \frac{\alpha}{2}, \tag{3.15}$$

*we conclude that $(\boldsymbol{\Delta}_A, \boldsymbol{\Delta}_X) = (\mathbf{0}, \mathbf{0})$ is the unique optimal solution to* (3.3).

*Proof.* Consider any feasible $(\boldsymbol{\Delta}_A, \boldsymbol{\Delta}_X)$. Choose $\boldsymbol{H} \in \partial\|\cdot\|_1(\boldsymbol{X})$ such that $\langle \boldsymbol{H}, \mathcal{P}_{\Omega^c}\boldsymbol{\Delta}_X\rangle = \|\mathcal{P}_{\Omega^c}\boldsymbol{\Delta}_X\|_1$, and notice that $\mathcal{P}_\Omega \boldsymbol{H} = \boldsymbol{\Sigma}$. Then

$$\|\boldsymbol{X} + \boldsymbol{\Delta}_X\|_1 \geq \|\boldsymbol{X}\|_1 + \langle \boldsymbol{H}, \boldsymbol{\Delta}_X\rangle. \tag{3.16}$$

Notice that since $(\boldsymbol{\Delta}_A, \boldsymbol{\Delta}_X)$ is feasible,

$$\begin{aligned} \langle \boldsymbol{\Delta}_A, \boldsymbol{\Lambda X}^*\rangle + \langle \boldsymbol{\Delta}_X, \boldsymbol{A}^*\boldsymbol{\Lambda}\rangle &= \langle \boldsymbol{\Delta}_A \boldsymbol{X}, \boldsymbol{\Lambda}\rangle + \langle \boldsymbol{A}\boldsymbol{\Delta}_X, \boldsymbol{\Lambda}\rangle \\ &= \langle \boldsymbol{\Delta}_A\boldsymbol{X} + \boldsymbol{A}\boldsymbol{\Delta}_X, \boldsymbol{\Lambda}\rangle = \langle \mathbf{0}, \boldsymbol{\Lambda}\rangle = 0. \end{aligned}$$

Hence,

$$\begin{aligned} \|\boldsymbol{X} + \boldsymbol{\Delta}_X\|_1 &\geq \|\boldsymbol{X}\|_1 + \langle \boldsymbol{H}, \boldsymbol{\Delta}_X\rangle - \langle \boldsymbol{A}^*\boldsymbol{\Lambda}, \boldsymbol{\Delta}_X\rangle - \langle \boldsymbol{\Lambda X}^*, \boldsymbol{\Delta}_A\rangle & (3.17)\\ &= \|\boldsymbol{X}\|_1 + \langle \boldsymbol{H} - \boldsymbol{A}^*\boldsymbol{\Lambda}, \boldsymbol{\Delta}_X\rangle - \langle \boldsymbol{\Lambda X}^*, \boldsymbol{\Delta}_A\rangle & (3.18)\\ &= \|\boldsymbol{X}\|_1 + \langle \mathcal{P}_\Omega[\boldsymbol{H} - \boldsymbol{A}^*\boldsymbol{\Lambda}], \mathcal{P}_\Omega\boldsymbol{\Delta}_X\rangle + \langle \mathcal{P}_{\Omega^c}[\boldsymbol{H} - \boldsymbol{A}^*\boldsymbol{\Lambda}], \mathcal{P}_{\Omega^c}\boldsymbol{\Delta}_X\rangle - \langle \Phi[\boldsymbol{\Lambda X}^*], \boldsymbol{\Delta}_A\rangle & (3.19)\\ &= \|\boldsymbol{X}\|_1 + \langle \mathcal{P}_{\Omega^c}[\boldsymbol{H} - \boldsymbol{A}^*\boldsymbol{\Lambda}], \mathcal{P}_{\Omega^c}\boldsymbol{\Delta}_X\rangle - \langle \Phi[\boldsymbol{\Lambda X}^*], \boldsymbol{\Delta}_A\rangle & (3.20)\\ &\geq \|\boldsymbol{X}\|_1 + \|\mathcal{P}_{\Omega^c}\boldsymbol{\Delta}_X\|_1/2 - \|\boldsymbol{\Delta}_A\|_F\|\Phi[\boldsymbol{\Lambda X}^*]\|_F & (3.21)\\ &\geq \|\boldsymbol{X}\|_1 + \left(\tfrac{1}{2} - \alpha^{-1}\|\Phi[\boldsymbol{\Lambda X}^*]\|_F\right)\|\mathcal{P}_{\Omega^c}\boldsymbol{\Delta}_X\|_1. & (3.22) \end{aligned}$$

In (3.19), we have used the fact that since $\boldsymbol{\Delta}_A$ is feasible, each column of $\boldsymbol{\Delta}_A$ is orthogonal to the corresponding column of $\boldsymbol{A}$, and so $\Phi[\boldsymbol{\Delta}_A] = \boldsymbol{\Delta}_A$. Furthermore, it is easily verified that $\Phi$ is self-adjoint, and so $\langle \boldsymbol{\Lambda X}^*, \Phi[\boldsymbol{\Delta}_A]\rangle = \langle \Phi[\boldsymbol{\Lambda X}^*], \boldsymbol{\Delta}_A\rangle$. In (3.20), we have used that since $\boldsymbol{H} \in \partial\|\cdot\|_1$, $\mathcal{P}_\Omega\boldsymbol{H} = \boldsymbol{\Sigma} = \boldsymbol{P}_\Omega[\boldsymbol{A}^*\boldsymbol{\Lambda}]$.

The right hand side of (3.22) is strictly greater than $\|\boldsymbol{X}\|_1$ provided that (i) $\|\Phi[\boldsymbol{\Lambda X}^*]\|_F < \alpha/2$ and (ii) $\mathcal{P}_{\Omega^c}\boldsymbol{\Delta}_X \neq \mathbf{0}$. The assumptions (3.13) and our assumption on the nullspace of $\boldsymbol{A}$ imply (ii). □

The remainder of the argument will show that the hypotheses of this lemma indeed hold. In Section 4, we give a construction of a dual matrix $\boldsymbol{\Lambda}$ that always satisfies (3.14), and satisfies (3.15) with high probability, provided $p$ is large enough. This is the content of Theorem 4.1. In Section 5, we show that with high probability the balancedness property (3.13) indeed holds with nonzero $\alpha$ (in particular, we can take $\alpha = C/\|\boldsymbol{A}\|^2$). This is done in Theorem 5.1. Combining these two results with Lemma 3.2 completes the proof of Theorem 2.1. The proofs of these key lemmas make repeated use of bounds on singular values of submatrices of an incoherent matrix. For completeness, we assemble these required results in Appendix C.

# 4 Certification Process

In this section, we show how to construct the dual certificate demanded by Lemma 3.2. In particular, we prove the following result:

**Theorem 4.1.** *There exist numerical constants $C_1, C_2, C_3 > 0$ such that the following occurs. Whenever*

$$k \;\le\; \min\left\{ \frac{C_1}{\mu(\boldsymbol{A})}, C_2 n \right\}, \tag{4.1}$$

*then for any $\alpha > 0$, there exists $\boldsymbol{\Lambda} \in \mathbb{R}^{m \times p}$ simultaneously satisfying the following three properties:*

$$\begin{aligned} \mathcal{P}_{\Omega}[\boldsymbol{A}^* \boldsymbol{\Lambda}] &= \boldsymbol{\Sigma}, & (4.2)\\ \|\mathcal{P}_{\Omega^c}[\boldsymbol{A}^* \boldsymbol{\Lambda}]\|_{\infty} &\le 1/2, & (4.3)\\ \|\Phi\left[\boldsymbol{\Lambda} \boldsymbol{X}^*\right]\|_F &< \alpha/2, & (4.4) \end{aligned}$$

*with probability at least*

$$1 - C_3 \alpha^{-1} n^{3/2} k^{1/2} p^{-1/2} (\log p). \tag{4.5}$$

## 4.1 First Steps

The following lemma goes most of the way to establishing Theorem 4.1:

**Lemma 4.2.** *Fix any $p > 0$ and let $\boldsymbol{x}_1, \dots, \boldsymbol{x}_p$ be independent and identically distributed random vectors with $\boldsymbol{x}_j = \boldsymbol{P}_{\Omega_j} \boldsymbol{v}_j$, where the $\Omega_j \subset [n]$ are uniform random subsets of size $k$ and $\boldsymbol{v}_j \sim_{iid} \mathcal{N}(0, n/kp)$. Then there exists a positive integer $t_\star \in [\lfloor (p-1)/2 \rfloor, p]$ and a sequence of random vectors $\boldsymbol{\lambda}_1, \dots, \boldsymbol{\lambda}_{t_\star}$ depending only on $\boldsymbol{x}_1, \dots, \boldsymbol{x}_{t_\star}$ such that*

$$\begin{aligned} \boldsymbol{P}_{\Omega_j} \boldsymbol{A}^* \boldsymbol{\lambda}_j &= \operatorname{sign}(\boldsymbol{x}_j), \qquad j = 1, \dots, t_\star, & (4.6)\\ \|\boldsymbol{P}_{\Omega_j^c} \boldsymbol{A}^* \boldsymbol{\lambda}_j\|_\infty &\le 1/2, \qquad j = 1, \dots, t_\star & (4.7)\\ \mathbb{E}\Big[\Big\|\Phi[\sum_{j=1}^{t_\star} \boldsymbol{\lambda}_j \boldsymbol{x}_j^*]\Big\|_F\Big] &\le C n^{3/2} k^{1/2} p^{-1/2}, & (4.8) \end{aligned}$$

*where $C$ is numerical.*

Section 4.2 proves Lemma 4.2 by giving an explicit construction of the desired dual certificates. Before describing this construction in greater detail, we first show that Theorem 4.1 follows as an easy consequence of Lemma 4.2, by dividing the sample set into subsets and then applying Lemma 4.2 to each subset.

*Proof of Theorem 4.1.* Choose $t_1$ according to Lemma 4.2, and let $\boldsymbol{\lambda}_1, \dots, \boldsymbol{\lambda}_{t_1}$ be the corresponding (random) dual vectors indicated by Lemma 4.2. Then

$$\mathbb{E}\Big[\Big\|\Phi[\sum_{j=1}^{t_1} \boldsymbol{\lambda}_j \boldsymbol{x}_j^*]\Big\|_F\Big] \le C n^{3/2} k^{1/2} p^{-1/2}. \tag{4.9}$$

Moreover, unless $p < 3$, $p - t_1 \le 3p/4$. Notice that the iid random vectors

$$\left(\tfrac{p}{p-t_1}\right)^{1/2} \boldsymbol{x}_{t_1+1}, \;\dots,\; \left(\tfrac{p}{p-t_1}\right)^{1/2} \boldsymbol{x}_p \tag{4.10}$$

again satisfy the hypotheses of Lemma 4.2. Hence, there exists $\delta \in [\lfloor (p-t_1-1)/2 \rfloor, p-t_1]$ and corresponding certificates $\boldsymbol{\lambda}_{t_1+1}, \dots, \boldsymbol{\lambda}_{t_1+\delta}$, again satisfying (4.6)-(4.7), such that if we set $t_2 = t_1+\delta$, we have

$$\mathbb{E}\Big[\Big\| \Phi[\sum_{j=t_1+1}^{t_2} \boldsymbol{\lambda}_j \boldsymbol{x}_j^*] \Big\|_F\Big] \;\le\; \left(\frac{p-t_1}{p}\right)^{1/2} C n^{3/2} k^{1/2} (p-t_1)^{-1/2} \;=\; C n^{3/2} k^{1/2} p^{-1/2}. \tag{4.11}$$

This leaves at most $p - t_2 \le \max((3/4)^2 p, 2)$ vectors $\boldsymbol{x}_{t_2+1}, \dots, \boldsymbol{x}_p$ to be certified. Repeating this construction $O(\log p)$ times yields a sequence of dual certificates $\boldsymbol{\lambda}_1, \dots, \boldsymbol{\lambda}_p$ satisfying (4.6)-(4.7), with

$$\mathbb{E}\Big[\Big\| \Phi[\sum_{j=1}^{p} \boldsymbol{\lambda}_j \boldsymbol{x}_j^*] \Big\|_F\Big] \;\le\; C' (\log p) n^{3/2} k^{1/2} p^{-1/2}.$$

The desired probability estimate follows from the Markov inequality. □

## 4.2 The construction

In this section, we outline a process that constructs the random sequence of certificates $\boldsymbol{\lambda}_1, \dots, \boldsymbol{\lambda}_{t_\star}$ described in Lemma 4.2. We will describe a construction of $p$ certificates $\boldsymbol{\lambda}_1, \dots, \boldsymbol{\lambda}_p$, and then choose $t_\star \in [\lfloor (p-1)/2 \rfloor, p]$ according to our analysis of this construction. Recall that we have defined $\Omega_j = \mathrm{support}(\boldsymbol{x}_j) \subset [n]$. Below, we will use $\boldsymbol{\Theta}_j \in \mathbb{R}^{m \times m}$ to denote the orthoprojector onto the orthogonal complement of the range of $\boldsymbol{A}_{\Omega_j}$:

$$\boldsymbol{\Theta}_j = \boldsymbol{I} - \boldsymbol{A}_{\Omega_j} (\boldsymbol{A}_{\Omega_j}^* \boldsymbol{A}_{\Omega_j})^{-1} \boldsymbol{A}_{\Omega_j}^*. \tag{4.12}$$

We will let $\boldsymbol{Q}_j$ denote the residual at time $j$:

$$\boldsymbol{Q}_j \doteq \sum_{l=1}^{j} \Phi[\boldsymbol{\lambda}_l \boldsymbol{x}_l^*]. \tag{4.13}$$

As above, let $\boldsymbol{\sigma}_j = \mathrm{sign}(\boldsymbol{x}_j(\Omega_j)) \in \{\pm 1\}^k$. Set

$$\boldsymbol{\zeta}_j \;=\; \begin{cases} \frac{1}{4} \frac{\boldsymbol{\Theta}_j \boldsymbol{Q}_{j-1} \boldsymbol{x}_j}{\|\boldsymbol{\Theta}_j \boldsymbol{Q}_{j-1} \boldsymbol{x}_j\|} & \boldsymbol{\Theta}_j \boldsymbol{Q}_{j-1} \boldsymbol{x}_j \neq \boldsymbol{0} \\ \boldsymbol{0} & \text{else} \end{cases} \tag{4.14}$$

$$\boldsymbol{\lambda}_j^{LS} \;=\; \boldsymbol{A}_{\Omega_j} (\boldsymbol{A}_{\Omega_j}^* \boldsymbol{A}_{\Omega_j})^{-1} \boldsymbol{\sigma}_j, \tag{4.15}$$

$$\boldsymbol{\lambda}_j \;=\; \boldsymbol{\lambda}_j^{LS} - \boldsymbol{\zeta}_j, \tag{4.16}$$

While it appears complicated, the rationale for the above procedure is actually quite simple. At each step $j$, we construct a certificate $\boldsymbol{\lambda}_j \in \mathbb{R}^m$. We would like to make $\boldsymbol{Q}_j = \Phi[\sum_{l=1}^{j} \boldsymbol{\lambda}_j \boldsymbol{x}_j^*]$ as small as possible, while still respecting the constraints $\boldsymbol{A}_{\Omega_j}^* \boldsymbol{\lambda}_j = \boldsymbol{\sigma}_j$, $\|\boldsymbol{A}_{\Omega_j^c}^* \boldsymbol{\lambda}_j\|_\infty \le 1/2$. The first term, $\boldsymbol{\lambda}_j^{LS}$ serves to ensure that the certification constraints are met. Notice that since $\boldsymbol{A}_{\Omega_j}^* \boldsymbol{\Theta}_j = \boldsymbol{0}$,

$$\boldsymbol{A}_{\Omega_j}^* \boldsymbol{\lambda}_j = \boldsymbol{A}_{\Omega_j}^* (\boldsymbol{\lambda}_j^{LS} - \boldsymbol{\zeta}_j) = \boldsymbol{A}_{\Omega_j}^* \boldsymbol{\lambda}_j^{LS} = \boldsymbol{\sigma}_j. \tag{4.17}$$

Moreover, for each $i \notin \Omega_j$,

$$|\boldsymbol{A}_i^* \boldsymbol{\lambda}_j^{LS}| \;=\; |\boldsymbol{A}_i^* \boldsymbol{A}_{\Omega_j} (\boldsymbol{A}_{\Omega_j}^* \boldsymbol{A}_{\Omega_j})^{-1} \boldsymbol{\sigma}_j| \;\le\; \|\boldsymbol{A}_i^* \boldsymbol{A}_{\Omega_j}\|_2 \|(\boldsymbol{A}_{\Omega_j}^* \boldsymbol{A}_{\Omega_j})^{-1}\| \|\boldsymbol{\sigma}_j\|_2. \tag{4.18}$$

Since $\boldsymbol{\sigma}_j \in \{\pm 1\}^k$, $\|\boldsymbol{\sigma}_j\|_2 = \sqrt{k}$. Under the assumption $k\mu(\boldsymbol{A}) < 1/2$, a standard argument (repeated as (C.4) of Appendix C) shows that $\|(\boldsymbol{A}_{\Omega_j}^* \boldsymbol{A}_{\Omega_j})^{-1}\| \le 2$. Finally, since $\boldsymbol{A}_i^* \boldsymbol{A}_{\Omega_j}$ is a vector of length $k$ with entries bounded by $\mu(\boldsymbol{A})$, $\|\boldsymbol{A}_i^* \boldsymbol{A}_{\Omega_j}\|_2 \le \mu(\boldsymbol{A})\sqrt{k}$. Combining bounds, we have

$$\|\boldsymbol{A}_{\Omega_j^c}^* \boldsymbol{\lambda}_j^{LS}\|_\infty \;=\; \max_{i \notin \Omega_j} |\boldsymbol{A}_i^* \boldsymbol{\lambda}_j^{LS}| \;\le\; 2k\mu(\boldsymbol{A}). \tag{4.19}$$

Hence, further assuming $k\mu(\boldsymbol{A}) < 1/8$, we obtain $\|\boldsymbol{A}^*_{\Omega_j^c}\boldsymbol{\lambda}_j^{LS}\|_\infty \leq 1/4$ and

$$\left\|\boldsymbol{A}^*_{\Omega_j^c}\boldsymbol{\lambda}_j\right\|_\infty \quad\leq\quad \left\|\boldsymbol{A}^*_{\Omega_j^c}\boldsymbol{\lambda}_j^{LS}\right\|_\infty + \left\|\boldsymbol{A}^*_{\Omega_j^c}\boldsymbol{\zeta}_j\right\|_\infty \quad\leq\quad \frac{1}{4} + \max_i \|\boldsymbol{A}_i\|_2\|\boldsymbol{\zeta}_j\|_2 \quad\leq\quad \frac{1}{2}. \tag{4.20}$$

The choice of $\boldsymbol{\zeta}_j$ is designed to deflate the residual $\boldsymbol{Q}$ as much as possible.[4] As we will see in the proof of Theorem 4.1, this process does succeed in controlling the norm of the residual $\boldsymbol{Q}_j$.

## 4.3 Analysis

The next question is how to analyze the order of growth of $\|\boldsymbol{Q}_j\|_F$, as a function of the matrix $\boldsymbol{A}$ and the sparsity $k$. To be more formal, let $\mathcal{F}_j$ be the $\sigma$-algebra generated by $\Omega_1, \dots, \Omega_j$ and $v_1, \dots, v_j$;

$$\mathcal{F}_1 \subset \mathcal{F}_2 \subset \cdots \subset \mathcal{F}_p$$

is the natural filtration. We will occasionally use the notation $\mathbb{E}_{\Omega_j}[f(\Omega, \boldsymbol{V})]$ to denote the expectation over $\Omega_j$, with all other variables fixed. More precisely,

$$\mathbb{E}_{\Omega_j}[f(\Omega, \boldsymbol{V})] \;\doteq\; \mathbb{E}[f(\Omega, \boldsymbol{V}) \mid \sigma(\Omega_1, \dots, \Omega_{j-1}, \Omega_{j+1}, \dots, \Omega_j, \boldsymbol{V})]. \tag{4.21}$$

We will use a similar notation $\mathbb{E}_{\boldsymbol{v}_j}[f(\Omega, \boldsymbol{V})]$ for the expectation over $\boldsymbol{v}_j$ with all other variables fixed. With these notations in mind, we embark on the proof of Lemma 4.2.

*Proof of Lemma 4.2.* We begin by using $\boldsymbol{Q}_j = \boldsymbol{Q}_{j-1} + \Phi[\boldsymbol{\lambda}_j \boldsymbol{x}_j^*]$ to write

$$\mathbb{E}\left[\|\boldsymbol{Q}_j\|_F^2 \mid \mathcal{F}_{j-1}\right] \quad=\quad \|\boldsymbol{Q}_{j-1}\|_F^2 + 2\,\mathbb{E}\left[\langle \boldsymbol{Q}_{j-1}, \Phi[\boldsymbol{\lambda}_j\boldsymbol{x}_j^*]\rangle \mid \mathcal{F}_{j-1}\right] + \mathbb{E}\left[\|\Phi[\boldsymbol{\lambda}_j\boldsymbol{x}_j^*]\|_F^2 \mid \mathcal{F}_{j-1}\right] \tag{4.22}$$

We will show that there exists $\varepsilon(p) > 0$ and $\tau(p)$ such that

$$\mathbb{E}\left[\langle \boldsymbol{Q}_{j-1}, \Phi[\boldsymbol{\lambda}_j\boldsymbol{x}_j^*]\rangle \mid \mathcal{F}_{j-1}\right] \quad\leq\quad -\varepsilon(p) \times \|\boldsymbol{Q}_{j-1}\|_F, \tag{4.23}$$

$$\mathbb{E}\left[\|\Phi[\boldsymbol{\lambda}_j\boldsymbol{x}_j^*]\|_F^2 \mid \mathcal{F}_{j-1}\right] \quad\leq\quad \tau(p). \tag{4.24}$$

Plugging in to (4.22) and taking the expectation of both sides yields

$$\mathbb{E}[\|\boldsymbol{Q}_j\|_F^2] \quad\leq\quad \mathbb{E}[\|\boldsymbol{Q}_{j-1}\|_F^2] - 2\varepsilon(p)\mathbb{E}[\|\boldsymbol{Q}_{j-1}\|_F] + \tau(p). \tag{4.25}$$

Summing from $j = 1, \dots, p$ and using that $\boldsymbol{Q}_0 = \boldsymbol{0}$, we have

$$\mathbb{E}[\|\boldsymbol{Q}_p\|_F^2] \quad\leq\quad p\tau(p) - 2\varepsilon(p)\sum_{j=1}^{p-1}\mathbb{E}[\|\boldsymbol{Q}_j\|_F] \tag{4.26}$$

In paragraphs (i)-(iii) below, we show that the quantities $\varepsilon$ and $\tau$ satisfy the following bounds:

$$\varepsilon(p) \;\geq\; C_1\sqrt{k/np}, \quad\text{and}\quad \tau(p) \;\leq\; C_2 nk/p. \tag{4.27}$$

For now, taking these bounds as given, we observe that by (4.26)

$$\mathbb{E}[\|\boldsymbol{Q}_1\|_F] \;\leq\; (\mathbb{E}[\|\boldsymbol{Q}_1\|_F^2])^{1/2} \;\leq\; \sqrt{\tau(p)}, \tag{4.28}$$

[4]Indeed, $\boldsymbol{\zeta}_j$ is a scaled version of a solution to the optimization problem

$$\text{minimize } \|\boldsymbol{Q}_{j-1} + \boldsymbol{\zeta}\boldsymbol{x}_j^*\|_F \quad \text{subject to} \quad \boldsymbol{A}^*_{\Omega_j}\boldsymbol{\zeta} = \boldsymbol{0}.$$

and hence the claim of Lemma 4.2 is verified in the case $p = 1$. On the other hand, if $p > 1$, then using the fact that the left hand side of (4.26) is nonnegative, we have

$$\frac{1}{p-1}\sum_{j=1}^{p-1}\mathbb{E}[\|\boldsymbol{Q}_j\|_F] \quad \leq \quad \frac{p}{p-1}\frac{\tau(p)}{2\,\varepsilon(p)} \quad \leq \quad \tau(p)/\varepsilon(p). \tag{4.29}$$

We recognize the left hand side of this inequality as an average. By the Markov inequality, if we were to choose an index $t \in \{1, \ldots, p-1\}$ uniformly at random, then with probability at least 1/2, $\mathbb{E}[\|\boldsymbol{Q}_t\|_F] \ \leq\ 2\tau(p)/\varepsilon(p)$. In particular, since $[\lfloor(p-1)/2\rfloor, p]$ contains more than half the elements of $\{1, \ldots, p-1\}$, there exists at least one $t_\star \in [\lfloor(p-1)/2\rfloor, p]$ such that

$$\mathbb{E}[\|\boldsymbol{Q}_{t_\star}\|_F] \ \leq\ 2\tau(p)/\varepsilon(p). \tag{4.30}$$

Plugging in the bounds from (4.27) establishes Lemma 4.2. All that remains to do is show that the bounds in (4.27) indeed hold. We establish the bound on $\varepsilon$ in paragraphs (i)-(ii) below, using the convenient split

$$\mathbb{E}\left[\langle \boldsymbol{Q}_{j-1}, \boldsymbol{\lambda}_j\boldsymbol{x}_j^*\rangle \mid \mathcal{F}_{j-1}\right] \ =\ \mathbb{E}\left[\langle \boldsymbol{Q}_{j-1}, \boldsymbol{\lambda}_j^{LS}\boldsymbol{x}_j^*\rangle \mid \mathcal{F}_{j-1}\right] - \mathbb{E}\left[\langle \boldsymbol{Q}_{j-1}, \boldsymbol{\zeta}_j\boldsymbol{x}_j^*\rangle \mid \mathcal{F}_{j-1}\right] \tag{4.31}$$

Finally, in paragraph (iii) below, we establish the bound on $\tau$ claimed in (4.27), completing the proof of Lemma 4.2.

**(i) Upper bounding $\langle \boldsymbol{Q}_{j-1}, \boldsymbol{\lambda}_j^{LS}\boldsymbol{x}_j^*\rangle$.** For $\Omega_j = \{a_1 < a_2 < \cdots < a_k\}$, set $\boldsymbol{U}_{\Omega_j} \doteq [\boldsymbol{e}_{a_1} \mid \boldsymbol{e}_{a_2} \mid \cdots \mid \boldsymbol{e}_{a_k}] \in \mathbb{R}^{n\times k}$, so that $\boldsymbol{P}_{\Omega_j} = \boldsymbol{U}_{\Omega_j}\boldsymbol{U}_{\Omega_j}^*$. Notice that we can write

$$\left\langle \boldsymbol{Q}_{j-1}, \boldsymbol{\lambda}_j^{LS}\boldsymbol{x}_j^*\right\rangle \quad = \quad \left\langle \boldsymbol{Q}_{j-1}\,,\, \boldsymbol{A}_{\Omega_j}(\boldsymbol{A}_{\Omega_j}^*\boldsymbol{A}_{\Omega_j})^{-1}\boldsymbol{U}_{\Omega_j}^*\operatorname{sgn}(\boldsymbol{v}_j)\boldsymbol{v}_j^*\boldsymbol{P}_{\Omega_j}\right\rangle. \tag{4.32}$$

Then, using that $\mathbb{E}[\operatorname{sgn}(\boldsymbol{v}_j)\boldsymbol{v}_j^*] = c_1\sigma\boldsymbol{I}$, we have

$$\begin{aligned}
\mathbb{E}\left[\left\langle \boldsymbol{Q}_{j-1}, \boldsymbol{\lambda}_j^{LS}\boldsymbol{x}_j^*\right\rangle \mid \mathcal{F}_{j-1}\right] \quad &= \quad \mathbb{E}_{\Omega_j}\mathbb{E}_{\boldsymbol{v}_j}\left[\left\langle \boldsymbol{Q}_{j-1}\,,\, \boldsymbol{A}_{\Omega_j}(\boldsymbol{A}_{\Omega_j}^*\boldsymbol{A}_{\Omega_j})^{-1}\boldsymbol{U}_{\Omega_j}^*\operatorname{sgn}(\boldsymbol{v}_j)\boldsymbol{v}_j^*\boldsymbol{P}_{\Omega_j}\right\rangle\right] \\
&= \quad c_1\sigma\mathbb{E}_{\Omega_j}\left[\left\langle \boldsymbol{Q}_{j-1}\,,\, \boldsymbol{A}_{\Omega_j}(\boldsymbol{A}_{\Omega_j}^*\boldsymbol{A}_{\Omega_j})^{-1}\boldsymbol{U}_{\Omega_j}^*\right\rangle\right]. \qquad (4.33)
\end{aligned}$$

Write $(\boldsymbol{A}_{\Omega_j}^*\boldsymbol{A}_{\Omega_j})^{-1} \ =\ \boldsymbol{I} + \boldsymbol{\Delta}(\Omega_j)$. Then, we have

$$\begin{aligned}
\left\langle \boldsymbol{Q}_{j-1}\,,\, \boldsymbol{A}_{\Omega_j}(\boldsymbol{A}_{\Omega_j}^*\boldsymbol{A}_{\Omega_j})^{-1}\boldsymbol{U}_{\Omega_j}^*\right\rangle \quad &= \quad \left\langle \boldsymbol{Q}_{j-1}\boldsymbol{P}_{\Omega_j}\,,\, \boldsymbol{A}_{\Omega_j}(\boldsymbol{A}_{\Omega_j}^*\boldsymbol{A}_{\Omega_j})^{-1}\boldsymbol{U}_{\Omega_j}^*\right\rangle \\
&= \quad \left\langle \boldsymbol{Q}_{j-1}\boldsymbol{P}_{\Omega_j}\,,\, \boldsymbol{A}_{\Omega_j}\boldsymbol{U}_{\Omega_j}^*\right\rangle + \left\langle \boldsymbol{Q}_{j-1}\boldsymbol{P}_{\Omega_j}\,,\, \boldsymbol{A}_{\Omega_j}\boldsymbol{\Delta}(\Omega_j)\boldsymbol{U}_{\Omega_j}^*\right\rangle.
\end{aligned}$$

Since $\boldsymbol{Q}_{j-1} = \Phi[\sum_{l=0}^{j-1}\boldsymbol{\lambda}_l\boldsymbol{x}_l^*] \in \operatorname{range}(\Phi)$, each of the columns of $\boldsymbol{Q}_{j-1} \in \mathbb{R}^{m\times n}$ is orthogonal to the corresponding column of $\boldsymbol{A}$. Since the first inner product in the above equation is simply the inner product of the restriction of $\boldsymbol{A}$ to a subset of its columns and the restriction of $\boldsymbol{Q}_{j-1}$ to a subset of its columns, this term is zero. Applying the Cauchy-Schwarz inequality to the second term of the previous equation gives

$$\left\langle \boldsymbol{Q}_{j-1}\,,\, \boldsymbol{A}_{\Omega_j}(\boldsymbol{A}_{\Omega_j}^*\boldsymbol{A}_{\Omega_j})^{-1}\boldsymbol{U}_{\Omega_j}^*\right\rangle \quad \leq \quad \|\boldsymbol{Q}_{j-1}\boldsymbol{P}_{\Omega_j}\|_F\,\|\boldsymbol{A}_{\Omega_j}\|\|\boldsymbol{\Delta}(\Omega_j)\|_F. \tag{4.34}$$

Standard calculations, given in (C.2) of Appendix C show that $\|\boldsymbol{A}_{\Omega_j}\| \leq (1 + k\mu(\boldsymbol{A}))^{1/2}$ is bounded by a constant, say, $c_2$. A similar calculation in (C.6) shows that $\|\boldsymbol{\Delta}(\Omega_j)\|_F \ \leq\ 2k\mu(\boldsymbol{A})$. Plugging back into (4.33), we have

$$\mathbb{E}\left[\left\langle \boldsymbol{Q}_{j-1}, \boldsymbol{\lambda}_j^{LS}\boldsymbol{x}_j^*\right\rangle \mid \mathcal{F}_{j-1}\right] \ \leq\ 2c_1c_2\sigma k\mu(\boldsymbol{A})\,\mathbb{E}_{\Omega_j}[\|\boldsymbol{Q}_{j-1}\boldsymbol{P}_{\Omega_j}\|_F]. \tag{4.35}$$

For now, we will be content with this expression.

**(ii) Lower bounding $\langle \boldsymbol{Q}_{j-1}, \boldsymbol{\zeta}_j \boldsymbol{x}_j^* \rangle$.** Continuing, we have that

$$\begin{aligned}
\langle \boldsymbol{Q}_{j-1}, \boldsymbol{\zeta}_j \boldsymbol{x}_j^* \rangle &= \langle \boldsymbol{Q}_{j-1}\boldsymbol{x}_j, \boldsymbol{\zeta}_j \rangle && (4.36)\\
&= \tfrac{1}{4} \left\langle \boldsymbol{Q}_{j-1}\boldsymbol{x}_j, \frac{\boldsymbol{\Theta}_j \boldsymbol{Q}_{j-1}\boldsymbol{x}_j}{\|\boldsymbol{\Theta}_j \boldsymbol{Q}_{j-1}\boldsymbol{x}_j\|} \right\rangle && (4.37)\\
&= \tfrac{1}{4} \left\|\boldsymbol{\Theta}_j \boldsymbol{Q}_{j-1}\boldsymbol{x}_j\right\| && (4.38)\\
&\geq \tfrac{1}{4} \left\|\boldsymbol{Q}_{j-1}\boldsymbol{x}_j\right\| - \tfrac{1}{4} \left\|\boldsymbol{P}_{A_{\Omega_j}} \boldsymbol{Q}_{j-1}\boldsymbol{x}_j\right\|, && (4.39)
\end{aligned}$$

where $\boldsymbol{P}_{A_{\Omega_j}} = \boldsymbol{A}_{\Omega_j}(\boldsymbol{A}_{\Omega_j}^*\boldsymbol{A}_{\Omega_j})^{-1}\boldsymbol{A}_{\Omega_j}^* \in \mathbb{R}^{m\times m}$. For the first term of (4.39), applying the Kahane-Khintchine inequality in Corollary B.3, gives

$$\begin{aligned}
\mathbb{E}\left[\left\|\boldsymbol{Q}_{j-1}\boldsymbol{x}_j\right\| \mid \mathcal{F}_{j-1}\right] &= \mathbb{E}_{\Omega_j}\mathbb{E}_{\boldsymbol{v}_j}\left[\|\boldsymbol{Q}_{j-1}\boldsymbol{P}_{\Omega_j}\boldsymbol{v}_j\|\right] && (4.40)\\
&\geq \frac{\sigma}{\sqrt{\pi}} \times \mathbb{E}_{\Omega_j}\left[\|\boldsymbol{Q}_{j-1}\boldsymbol{P}_{\Omega_j}\|_F\right]. && (4.41)
\end{aligned}$$

For the second term of (4.39), writing $\boldsymbol{P}_{\boldsymbol{A}_{\Omega_j}} = \boldsymbol{A}_{\Omega_j}(\boldsymbol{A}_{\Omega_j}^*\boldsymbol{A}_{\Omega_j})^{-1/2} \times (\boldsymbol{A}_{\Omega_j}^*\boldsymbol{A}_{\Omega_j})^{-1/2}\boldsymbol{A}_{\Omega_j}^*$ as a product of matrices with spectral norm one, we have

$$\begin{aligned}
\left\|\boldsymbol{P}_{\boldsymbol{A}_{\Omega_j}} \boldsymbol{Q}_{j-1}\boldsymbol{x}_j\right\| &= \left\|(\boldsymbol{A}_{\Omega_j}^*\boldsymbol{A}_{\Omega_j})^{-1/2}\boldsymbol{A}_{\Omega_j}^*\boldsymbol{Q}_{j-1}\boldsymbol{P}_{\Omega_j}\boldsymbol{v}_j\right\| && (4.42)\\
&\leq \left\|(\boldsymbol{A}_{\Omega_j}^*\boldsymbol{A}_{\Omega_j})^{-1/2}\right\| \left\|\boldsymbol{A}_{\Omega_j}^*\boldsymbol{Q}_{j-1}\boldsymbol{P}_{\Omega_j}\boldsymbol{v}_j\right\|. && (4.43)
\end{aligned}$$

A calculation shows that under the assumption $k\mu(\boldsymbol{A}) < 1/2$, $\|(\boldsymbol{A}_{\Omega_j}^*\boldsymbol{A}_{\Omega_j})^{-1/2}\| \leq \sqrt{2}$, and so

$$\left\|\boldsymbol{P}_{\boldsymbol{A}_{\Omega_j}} \boldsymbol{Q}_{j-1}\boldsymbol{x}_j\right\| \leq \sqrt{2} \times \left\|\boldsymbol{A}_{\Omega_j}^*\boldsymbol{Q}_{j-1}\boldsymbol{P}_{\Omega_j}\boldsymbol{v}_j\right\| = \sqrt{2} \times \left\|\boldsymbol{P}_{\Omega_j}\boldsymbol{A}^*\boldsymbol{Q}_{j-1}\boldsymbol{P}_{\Omega_j}\boldsymbol{v}_j\right\|. \quad (4.44)$$

Applying the Jensen's inequality to bound the expectation of the above expression, we have (via Corollary B.3),

$$\begin{aligned}
\mathbb{E}\left[\left\|\boldsymbol{P}_{\boldsymbol{A}_{\Omega_j}} \boldsymbol{Q}_{j-1}\boldsymbol{x}_j\right\| \mid \mathcal{F}_{j-1}\right] &\leq \sqrt{2} \times \mathbb{E}\left[\left\|\boldsymbol{P}_{\Omega_j}\boldsymbol{A}^*\boldsymbol{Q}_{j-1}\boldsymbol{P}_{\Omega_j}\boldsymbol{v}_j\right\| \mid \mathcal{F}_{j-1}\right] && (4.45)\\
&= \sqrt{2} \times \mathbb{E}_{\Omega_j}\mathbb{E}_{\boldsymbol{v}_j}\left[\left\|\boldsymbol{P}_{\Omega_j}\boldsymbol{A}^*\boldsymbol{Q}_{j-1}\boldsymbol{P}_{\Omega_j}\boldsymbol{v}_j\right\|\right] && (4.46)\\
&\leq \sigma\sqrt{2} \times \mathbb{E}_{\Omega_j}\left[\left\|\boldsymbol{P}_{\Omega_j}\boldsymbol{A}^*\boldsymbol{Q}_{j-1}\boldsymbol{P}_{\Omega_j}\right\|_F\right]. && (4.47)
\end{aligned}$$

Notice that because each column of $\boldsymbol{Q}_{j-1}$ is orthogonal to the corresponding column of $\boldsymbol{A}$, the diagonal elements of $\boldsymbol{A}^*\boldsymbol{Q}_{j-1}$ are zero. Under this condition, we can invoke a decoupling lemma given as Lemma E.1 to remove the first $\boldsymbol{P}_{\Omega_j}$, giving

$$\mathbb{E}_{\Omega_j}\left[\left\|\boldsymbol{P}_{\Omega_j}\boldsymbol{A}^*\boldsymbol{Q}_{j-1}\boldsymbol{P}_{\Omega_j}\right\|_F\right] \leq 16\sqrt{\frac{k}{n}}\,\mathbb{E}_{\Omega_j}\left[\left\|\boldsymbol{A}^*\boldsymbol{Q}_{j-1}\boldsymbol{P}_{\Omega_j}\right\|_F\right] \leq 16\|\boldsymbol{A}\|\sqrt{\frac{k}{n}}\,\mathbb{E}_{\Omega_j}\left[\left\|\boldsymbol{Q}_{j-1}\boldsymbol{P}_{\Omega_j}\right\|_F\right].$$

Via incoherence, we can show that $\|\boldsymbol{A}\| \leq \sqrt{1+n\mu(\boldsymbol{A})}$ (see (C.1)), and so

$$\mathbb{E}_{\Omega_j}\left[\left\|\boldsymbol{P}_{\Omega_j}\boldsymbol{A}^*\boldsymbol{Q}_{j-1}\boldsymbol{P}_{\Omega_j}\right\|_F\right] \leq c_3\sqrt{k/n + k\mu(\boldsymbol{A})}\,\mathbb{E}_{\Omega_j}\left[\left\|\boldsymbol{Q}_{j-1}\boldsymbol{P}_{\Omega_j}\right\|_F\right], \quad (4.48)$$

for appropriate $c_3$. Combining bounds, we have shown that

$$\mathbb{E}\left[\langle \boldsymbol{Q}_{j-1}, \boldsymbol{\lambda}_j\boldsymbol{x}_j^* \rangle \mid \mathcal{F}_{j-1}\right] \leq \sigma\left(2c_1c_2k\mu(\boldsymbol{A}) + \frac{c_3}{4}\sqrt{k/n + k\mu(\boldsymbol{A})} - \frac{1}{4\sqrt{\pi}}\right)\mathbb{E}_{\Omega_j}\left[\left\|\boldsymbol{Q}_{j-1}\boldsymbol{P}_{\Omega_j}\right\|_F\right].$$

Assuming $k/n$ and $k\mu(\boldsymbol{A})$ are bounded below appropriately small constants, we have

$$\begin{aligned}\mathbb{E}\left[\langle \boldsymbol{Q}_{j-1}, \boldsymbol{\lambda}_j \boldsymbol{x}_j^* \rangle \mid \mathcal{F}_{j-1}\right] &\leq -c_4 \sigma \mathbb{E}_{\Omega_j}\left[\left\|\boldsymbol{Q}_{j-1}\boldsymbol{P}_{\Omega_j}\right\|_F\right] \leq -c_4\sigma \left\|\mathbb{E}_{\Omega_j}\left[\boldsymbol{Q}_{j-1}\boldsymbol{P}_{\Omega_j}\right]\right\|_F \\ &\leq -c_4\sigma(k/n)\|\boldsymbol{Q}_{j-1}\|_F = -c_4\sqrt{k/np}\,\|\boldsymbol{Q}_{j-1}\|_F,\end{aligned}$$

where we have used Jensen's inequality and the facts that $\mathbb{E}_{\Omega_j}[\boldsymbol{P}_{\Omega_j}] = (k/n)\boldsymbol{I}$ and $\sigma = \sqrt{n/kp}$. This establishes the first part of (4.27).

**(iii) Bounding $\|\boldsymbol{\lambda}_j \boldsymbol{x}_j^*\|$.** We next bound $\mathbb{E}\left[\left\|\Phi[\boldsymbol{\lambda}_j \boldsymbol{x}_j^*]\right\|_F^2 \mid \mathcal{F}_{j-1}\right]$. We have already shown that under the conditions of the lemma, $\|\boldsymbol{\lambda}_j\|_2 \leq c_5\sqrt{k} + 1/4 \leq c_6\sqrt{k}$. So,

$$\left\|\Phi\left[\boldsymbol{\lambda}_j \boldsymbol{x}_j^*\right]\right\|_F^2 \leq \left\|\boldsymbol{\lambda}_j \boldsymbol{x}_j^*\right\|_F^2 = \|\boldsymbol{\lambda}_j\|^2 \|\boldsymbol{x}_j\|^2 \leq c_6 k \|\boldsymbol{x}_j\|^2. \tag{4.49}$$

Since $\mathbb{E}[\|\boldsymbol{x}_j\|_2^2] = n/p$, we have the simple bound

$$\mathbb{E}\left[\left\|\Phi[\boldsymbol{\lambda}_j \boldsymbol{x}_j^*]\right\|_F^2 \mid \mathcal{F}_{j-1}\right] \leq c_6 kn/p. \tag{4.50}$$

This establishes the second part of (4.27), completing the proof of Lemma 4.2. □

# 5 Balancedness Property

In this section, we show that for any $(\boldsymbol{\Delta}_A, \boldsymbol{\Delta}_X)$ in the tangent space to $\mathcal{M}$ at $(\boldsymbol{A}, \boldsymbol{X})$,

$$\|\mathcal{P}_{\Omega^c}\boldsymbol{\Delta}_X\|_F \geq \alpha\|\boldsymbol{\Delta}_A\|_F \tag{5.1}$$

for appropriate $\alpha > 0$. This property essentially says that if we locally perturb the basis, we are guaranteed to pay some penalty, in terms of the norm of $\mathcal{P}_{\Omega^c}\boldsymbol{\Delta}_X$. Hence, it can be viewed as a step in the direction of Theorem 2.1. By itself, it is not sufficient to establish Theorem 2.1, however, since it does not rule out the possibility that as $\boldsymbol{A}$ changes, $\|\mathcal{P}_{\Omega}\boldsymbol{\Delta}_X\|_1$ might decrease faster than $\|\mathcal{P}_{\Omega^c}\boldsymbol{\Delta}_X\|_1$ increases – for this purpose we need the golfing scheme of the previous section. On a technical level, however, (5.1) makes the golfing scheme possible, by allowing us to open a "hole" around the constraint $\Phi[\boldsymbol{\Lambda}\boldsymbol{X}^*] = \boldsymbol{0}$, and construct dual certificates $\boldsymbol{\Lambda}$ that only satisfy $\Phi[\boldsymbol{\Lambda}\boldsymbol{X}^*] \approx \boldsymbol{0}$.

More precisely, we next show that

**Theorem 5.1.** *There exist numerical constants $C_1, \dots, C_8 > 0$ such that the following occurs. If*

$$k \leq C_1 \times \min\left\{n, \frac{1}{\mu(\boldsymbol{A})}\right\}, \tag{5.2}$$

*then whenever $p \geq C_2 n^2$, with probability at least*

$$1 - C_3 p^{-4} - C_4\, n \exp\left(-\frac{C_5\, p}{n \log p}\right) - C_6 n^2 \exp\left(-\frac{C_7 k^2 p}{n^2}\right), \tag{5.3}$$

*all pairs $(\boldsymbol{\Delta}_A, \boldsymbol{\Delta}_X)$ satisfying (3.1) obey the estimate*

$$\|\mathcal{P}_{\Omega^c}\boldsymbol{\Delta}_X\|_F \geq C_8\|\boldsymbol{\Delta}_A\|_F/\|\boldsymbol{A}\|^2. \tag{5.4}$$

**Organization.** The proof of Theorem 5.1 contains essentially two parts: algebraic manipulations that show that the desired property holds whenever the random matrix $\boldsymbol{X}$ satisfies two particular properties, and then probabilistic reasoning to show that these properties hold with the stated probability. The first property, stated in Lemma 5.2, simply involves a bound on the extreme eigenvalues of $\boldsymbol{X}\boldsymbol{X}^*$. This lemma is proved in Appendix F. The second probabilistic property involves controlling the difference between a certain operator and its large sample limit. The quantities involved will arise naturally in the proof of Theorem 5.1, and the claim will be formally stated in Lemma 5.3 below. The proof of Lemma 5.3 is a bit technical, requiring us to apply the matrix Chernoff bound conditional on $\Omega$. This proof is given in Section 6.

## 5.1 Proof of Theorem 5.1.

Before commencing the proof of Theorem 5.1, we introduce one additional definition. Fix $0 < t < 1/2$, and let $\mathcal{E}_{eig}(t)$ denote the event:

$$\mathcal{E}_{eig}(t) \;\doteq\; \{\, \omega \mid \|\boldsymbol{X}\boldsymbol{X}^* - \boldsymbol{I}\| < t \} \tag{5.5}$$

In particular, on $\mathcal{E}_{eig}$, $\|\boldsymbol{X}\boldsymbol{X}^*\| < 1 + t < 2$, $\|(\boldsymbol{X}\boldsymbol{X}^*)^{-1}\| = (\lambda_{min}(\boldsymbol{X}\boldsymbol{X}^*))^{-1} < 1/(1-t) < 2$. It should not be particularly surprising that this event is highly likely. The matrix $\boldsymbol{X}$ has iid columns, and it is easy to see that $\mathbb{E}[\boldsymbol{X}\boldsymbol{X}^*] = \boldsymbol{I}$. The following lemma shows $\boldsymbol{X}\boldsymbol{X}^*$ is also close to $\boldsymbol{I}$ in the operator norm, with high probability:

**Lemma 5.2.** *Fix any $0 < t < 1/2$, and let $\mathcal{E}_{eig}(t)$ denote the event that the following bound holds:*

$$\|\boldsymbol{X}\boldsymbol{X}^* - \boldsymbol{I}\| < t. \tag{5.6}$$

*Then there exist numerical constants $C_1, C_2, C_3$ all strictly positive such that for all $p \geq C_1(n/t)^{1/4}$,*

$$\mathbb{P}\left[\mathcal{E}_{eig}(t)\right] \;\geq\; 1 - C_2\, n \,\exp\left(-\frac{C_3\, t^2\, p}{n \log p}\right) - p^{-7}. \tag{5.7}$$

Lemma 5.2 is essentially a consequence of the matrix Chernoff bound of [Tro10]. Its proof is a bit technical, and so is delayed to Section F of the appendix. For now, we take this result as given and commence the proof of Theorem 5.1.

*Proof of Theorem 5.1.* On the event $\mathcal{E}_{eig}$, $\boldsymbol{X}\boldsymbol{X}^*$ is invertible, and any pair $(\boldsymbol{\Delta}_A, \boldsymbol{\Delta}_X)$ satisfying (3.1) also satisfies

$$\boldsymbol{\Delta}_A = -\boldsymbol{A}\boldsymbol{\Delta}_X\boldsymbol{X}^*(\boldsymbol{X}\boldsymbol{X}^*)^{-1}. \tag{5.8}$$

Hence, using that[5]

$$\|\boldsymbol{X}^*(\boldsymbol{X}\boldsymbol{X}^*)^{-1}\| = 1/\sqrt{\lambda_{min}(\boldsymbol{X}\boldsymbol{X}^*)}$$

and that for any matrices $\boldsymbol{P}, \boldsymbol{Q}, \boldsymbol{R}$,

$$\|\boldsymbol{P}\boldsymbol{Q}\boldsymbol{R}\|_F \leq \|\boldsymbol{P}\|\|\boldsymbol{R}\|\|\boldsymbol{Q}\|_F,$$

on $\mathcal{E}_{eig}(1/2)$ we have

$$\|\boldsymbol{\Delta}_A\|_F \;\leq\; \frac{\|\boldsymbol{A}\|}{\sqrt{\lambda_{min}(\boldsymbol{X}\boldsymbol{X}^*)}}\|\boldsymbol{\Delta}_X\|_F \;\leq\; \sqrt{2}\,\|\boldsymbol{A}\|\,\|\boldsymbol{\Delta}_X\|_F. \tag{5.9}$$

We next show that for any pair $(\boldsymbol{\Delta}_A, \boldsymbol{\Delta}_X)$ satisfying (3.1), $\boldsymbol{\Delta}_X$ cannot be too concentrated on $\Omega$. More precisely, we will show $\exists \alpha' < \infty$ such that for any such $\boldsymbol{\Delta}_X$,

$$\|\mathcal{P}_\Omega[\boldsymbol{\Delta}_X]\|_F \;\leq\; \alpha'\|\mathcal{P}_{\Omega^c}[\boldsymbol{\Delta}_X]\|_F. \tag{5.10}$$

[5] This can be shown via the singular value decomposition of $\boldsymbol{X}$.

On $\mathcal{E}_{eig}$, the inverse in (5.8) is justified, and (5.8) holds. We can plug this relationship into the tangent space constraint $\boldsymbol{\Delta}_A \boldsymbol{X} + \boldsymbol{A}\boldsymbol{\Delta}_X = \boldsymbol{0}$, giving

$$\boldsymbol{0} \;=\; \boldsymbol{\Delta}_A \boldsymbol{X} + \boldsymbol{A}\boldsymbol{\Delta}_X \;=\; -\boldsymbol{A}\boldsymbol{\Delta}_X \boldsymbol{X}^*(\boldsymbol{X}\boldsymbol{X}^*)^{-1}\boldsymbol{X} + \boldsymbol{A}\boldsymbol{\Delta}_X \;=\; \boldsymbol{A}\boldsymbol{\Delta}_X \left(\boldsymbol{I} - \boldsymbol{X}^*(\boldsymbol{X}\boldsymbol{X}^*)^{-1}\boldsymbol{X}\right).$$

Above, $\boldsymbol{P}_X \doteq \boldsymbol{X}^*(\boldsymbol{X}\boldsymbol{X}^*)^{-1}\boldsymbol{X}$ is the projection matrix onto the range of $\boldsymbol{X}^*$. We have one further constraint $\boldsymbol{A}_i^* \boldsymbol{\Delta}_A \boldsymbol{e}_i = 0 \;\forall\, i$. We introduce a more concise notation for this constraint, by letting $\mathcal{C}_A : \mathbb{R}^n \to \mathbb{R}^{m \times n}$ via

$$\mathcal{C}_A[\boldsymbol{z}] = \boldsymbol{A}\,\mathrm{diag}(\boldsymbol{z}). \tag{5.11}$$

For $\boldsymbol{U} = [\boldsymbol{u}_1 \mid \boldsymbol{u}_2 \mid \cdots \mid \boldsymbol{u}_n] \in \mathbb{R}^{m \times n}$, the action of the adjoint of $\mathcal{C}_A$ is given by

$$\mathcal{C}_A^*\left[\boldsymbol{U}\right] = [\langle \boldsymbol{A}_1, \boldsymbol{u}_1\rangle, \dots, \langle \boldsymbol{A}_n, \boldsymbol{u}_n\rangle]^* \in \mathbb{R}^n. \tag{5.12}$$

Hence, our second constraint can be expressed concisely via $\mathcal{C}_A^*[\boldsymbol{\Delta}_A] = \boldsymbol{0} \in \mathbb{R}^n$.

On $\mathcal{E}_{eig}$, any $\boldsymbol{\Delta}_X$ participating in a pair $(\boldsymbol{\Delta}_A, \boldsymbol{\Delta}_X) \in T_{x_\star}\boldsymbol{M}$ must satisfy

$$\boldsymbol{A}\boldsymbol{\Delta}_X(\boldsymbol{I} - \boldsymbol{P}_X) = \boldsymbol{0} \quad \text{and} \quad \mathcal{C}_A^*[\boldsymbol{A}\boldsymbol{\Delta}_X \boldsymbol{X}^*(\boldsymbol{X}\boldsymbol{X}^*)^{-1}] = \boldsymbol{0}. \tag{5.13}$$

It is convenient to temporarily express the constraint (5.13) in vector form, as a constraint on $\boldsymbol{\delta}_x \doteq \mathrm{vec}[\boldsymbol{\Delta}_X] \in \mathbb{R}^{np}$. In vector notation, (5.13) is equivalent to $\boldsymbol{M}\boldsymbol{\delta}_x = \boldsymbol{0}$, with

$$\boldsymbol{M} \doteq \begin{bmatrix} (\boldsymbol{I} - \boldsymbol{P}_X) \otimes \boldsymbol{A} \\ \boldsymbol{C}_A^*((\boldsymbol{X}\boldsymbol{X}^*)^{-1}\boldsymbol{X} \otimes \boldsymbol{A}) \end{bmatrix} \in \mathbb{R}^{(mp+n) \times np}. \tag{5.14}$$

In forming $\boldsymbol{M}$, we have used the familiar identity $\mathrm{vec}[\boldsymbol{Q}\boldsymbol{R}\boldsymbol{S}] = (\boldsymbol{S}^* \otimes \boldsymbol{Q})\,\mathrm{vec}[\boldsymbol{R}]$, for matrices $\boldsymbol{Q}$, $\boldsymbol{R}$, and $\boldsymbol{S}$ of compatible size. We have used $\boldsymbol{C}_A$ to denote the matrix version of the operator $\mathcal{C}_A$, uniquely defined via[6]

$$\mathrm{vec}[\mathcal{C}_A[\boldsymbol{z}]] \;=\; \boldsymbol{C}_A \boldsymbol{z} \quad \forall\, \boldsymbol{z} \in \mathbb{R}^{m \times n}. \tag{5.15}$$

It will be easier to work with a symmetric variant of the equation $\boldsymbol{M}\boldsymbol{\delta}_x = \boldsymbol{0}$. Set

$$\boldsymbol{T} \;\doteq\; \boldsymbol{M}^*\boldsymbol{M} \;=\; (\boldsymbol{I} - \boldsymbol{P}_X) \otimes \boldsymbol{A}^*\boldsymbol{A} + \left(\boldsymbol{X}^*(\boldsymbol{X}\boldsymbol{X}^*)^{-1} \otimes \boldsymbol{A}^*\right) \boldsymbol{C}_A \boldsymbol{C}_A^* \left((\boldsymbol{X}\boldsymbol{X}^*)^{-1}\boldsymbol{X} \otimes \boldsymbol{A}\right), \tag{5.16}$$

then $\boldsymbol{M}\boldsymbol{\delta}_x = \boldsymbol{0}$ if and only if

$$\boldsymbol{T}\boldsymbol{\delta}_x = \boldsymbol{0}. \tag{5.17}$$

Splitting $\boldsymbol{\delta}_x$ as $\boldsymbol{\delta}_x = \boldsymbol{P}_\Omega \boldsymbol{\delta}_x + \boldsymbol{P}_{\Omega^c}\boldsymbol{\delta}_x$, and multiplying (5.17) on the left by $\boldsymbol{P}_\Omega$ gives

$$\boldsymbol{P}_\Omega \boldsymbol{T} \boldsymbol{P}_\Omega \boldsymbol{\delta}_x = -\boldsymbol{P}_\Omega \boldsymbol{T} \boldsymbol{P}_{\Omega^c} \boldsymbol{\delta}_x, \tag{5.18}$$

or

$$[\boldsymbol{P}_\Omega \boldsymbol{T} \boldsymbol{P}_\Omega]\,(\boldsymbol{P}_\Omega \boldsymbol{\delta}_x) = -\,[\boldsymbol{P}_\Omega \boldsymbol{T} \boldsymbol{P}_{\Omega^c}]\,(\boldsymbol{P}_{\Omega^c}\boldsymbol{\delta}_x). \tag{5.19}$$

Now, although the matrix $\boldsymbol{P}_\Omega \boldsymbol{T} \boldsymbol{P}_\Omega$ is rank-deficient, as we will see, its nullspace does not contain any vectors $\boldsymbol{z}$ supported on $\Omega$. More quantitatively, let $S_\Omega \subset \mathbb{R}^{np}$ denote the subspace of vectors whose support is contained in $\Omega$ (i.e., the solution space of $\boldsymbol{P}_\Omega \boldsymbol{z} = \boldsymbol{z}$), and define

$$\xi \;\doteq\; \inf_{\boldsymbol{z} \in S_\Omega \setminus \{\boldsymbol{0}\}} \frac{\|\boldsymbol{P}_\Omega \boldsymbol{T} \boldsymbol{P}_\Omega \boldsymbol{z}\|}{\|\boldsymbol{z}\|}, \tag{5.20}$$

Then if $\xi > 0$, by (5.19) we have

$$\|\boldsymbol{P}_\Omega \boldsymbol{\delta}_x\| \;\le\; \xi^{-1}\,\|\boldsymbol{P}_\Omega \boldsymbol{T} \boldsymbol{P}_\Omega \boldsymbol{\delta}_x\| \;=\; \xi^{-1}\,\|[\boldsymbol{P}_\Omega \boldsymbol{T} \boldsymbol{P}_{\Omega^c}]\,\boldsymbol{P}_{\Omega^c}\boldsymbol{\delta}_x\| \;\le\; \xi^{-1}\,\|\boldsymbol{P}_\Omega \boldsymbol{T} \boldsymbol{P}_{\Omega^c}\|\,\|\boldsymbol{P}_{\Omega^c}\boldsymbol{\delta}_x\|.$$

[6]In particular, it is not difficult to see that $\boldsymbol{C}_A \in \mathbb{R}^{mn \times n}$ is a block diagonal matrix whose blocks are the columns of $\boldsymbol{A}$.

A calculation[7] shows that $\|\boldsymbol{P}_\Omega \boldsymbol{T} \boldsymbol{P}_{\Omega^c}\| \;\le\; C\|\boldsymbol{A}\|$, and hence, thus far we have shown

$$\begin{aligned} \|\boldsymbol{\Delta}_A\|_F &\le \sqrt{2}\|\boldsymbol{A}\|\|\boldsymbol{\Delta}_X\|_F \;\le\; \sqrt{2}\|\boldsymbol{A}\|\left(\|\boldsymbol{P}_\Omega \boldsymbol{\Delta}_X\|_F + \|\mathcal{P}_{\Omega^c}\boldsymbol{\Delta}_X\|_F\right) \\ &\le \sqrt{2}\|\boldsymbol{A}\|\left(1 + C\xi^{-1}\|\boldsymbol{A}\|\right)\|\mathcal{P}_{\Omega^c}\boldsymbol{\Delta}_X\|_F. \end{aligned} \tag{5.21}$$

Our only remaining tasks are to lower bound $\xi$ to complete the bound on $\alpha$, and then verify that the failure probability is indeed small. We carry out these tasks below, with some technical details associated with bounding $\xi$ delayed to Section 6.

The expression for $\boldsymbol{T}$ in (5.16) is quite complicated. Notice, however, that as $p \to \infty$, $\boldsymbol{X}\boldsymbol{X}^* \to \boldsymbol{I}$ almost surely. If we can replace $(\boldsymbol{X}\boldsymbol{X}^*)^{-1}$ with $\boldsymbol{I}$ in (5.16), the expression will simplify significantly. We introduce $\hat{\boldsymbol{T}}$, this simplified approximation, given by

$$\begin{aligned} \hat{\boldsymbol{T}} &\doteq (\boldsymbol{I} - \boldsymbol{X}^*\boldsymbol{X}) \otimes \boldsymbol{A}^*\boldsymbol{A} + (\boldsymbol{X}^* \otimes \boldsymbol{A}^*)\,\boldsymbol{C}_A\boldsymbol{C}_A^*\,(\boldsymbol{X} \otimes \boldsymbol{A}) \\ &= \boldsymbol{I} \otimes \boldsymbol{A}^*\boldsymbol{A} - (\boldsymbol{X}^* \otimes \boldsymbol{A}^*)(\boldsymbol{I} - \boldsymbol{C}_A\boldsymbol{C}_A^*)(\boldsymbol{X} \otimes \boldsymbol{A}). \end{aligned} \tag{5.22}$$

The matrix $\boldsymbol{I} \otimes \boldsymbol{A}^*\boldsymbol{A}$ is quite simple: for a matrix $\boldsymbol{Z}$, $(\boldsymbol{I} \otimes \boldsymbol{A}^*\boldsymbol{A})\mathrm{vec}[\boldsymbol{Z}] = \mathrm{vec}[(\boldsymbol{A}^*\boldsymbol{A})\boldsymbol{Z}]$, and so $(\boldsymbol{A}^*\boldsymbol{A})$ simply acts columnwise on $\boldsymbol{Z}$. We will see that if the columns of $\boldsymbol{Z}$ are appropriately *sparse*, then because $\boldsymbol{A}$ is incoherent, $\boldsymbol{A}^*\boldsymbol{A} \approx \boldsymbol{I}$ will approximately preserve their norms. Hence, the restricted singular value $\xi$ associated with the matrix $\boldsymbol{I} \otimes \boldsymbol{A}^*\boldsymbol{A}$ is well-behaved.

We therefore let $\boldsymbol{R}$ denote the nuisance term in (5.22)

$$\boldsymbol{R} \;\doteq\; (\boldsymbol{X}^* \otimes \boldsymbol{A}^*)(\boldsymbol{I} - \boldsymbol{C}_A\boldsymbol{C}_A^*)(\boldsymbol{X} \otimes \boldsymbol{A}). \tag{5.23}$$

so that we have $\hat{\boldsymbol{T}} = \boldsymbol{I} \otimes \boldsymbol{A}^*\boldsymbol{A} - \boldsymbol{R}$, and

$$\boldsymbol{T} = \boldsymbol{I} \otimes \boldsymbol{A}^*\boldsymbol{A} - \boldsymbol{R} + (\boldsymbol{T} - \hat{\boldsymbol{T}}). \tag{5.24}$$

In terms of these variables,

$$\begin{aligned} \xi &= \inf_{\boldsymbol{z} \in S_\Omega \setminus \{\boldsymbol{0}\}} \left\{ \frac{\|\boldsymbol{P}_\Omega(\boldsymbol{I} \otimes \boldsymbol{A}^*\boldsymbol{A} - \boldsymbol{R} + \boldsymbol{T} - \hat{\boldsymbol{T}})\boldsymbol{P}_\Omega \boldsymbol{z}\|}{\|\boldsymbol{z}\|} \right\} \\ &\ge \inf_{\boldsymbol{z} \in S_\Omega \setminus \{\boldsymbol{0}\}} \left\{ \frac{\|\boldsymbol{P}_\Omega(\boldsymbol{I} \otimes \boldsymbol{A}^*\boldsymbol{A})\boldsymbol{P}_\Omega \boldsymbol{z}\|}{\|\boldsymbol{z}\|} \right\} - \sup_{\boldsymbol{z} \neq \boldsymbol{0}} \left\{ \frac{\|\boldsymbol{P}_\Omega \boldsymbol{R} \boldsymbol{P}_\Omega \boldsymbol{z}\|}{\|\boldsymbol{z}\|} \right\} - \sup_{\boldsymbol{z} \neq \boldsymbol{0}} \left\{ \frac{\|\boldsymbol{P}_\Omega(\boldsymbol{T} - \hat{\boldsymbol{T}})\boldsymbol{P}_\Omega \boldsymbol{z}\|}{\|\boldsymbol{z}\|} \right\} \\ &= \inf_{\boldsymbol{z} \in S_\Omega \setminus \{\boldsymbol{0}\}} \left\{ \frac{\|\boldsymbol{P}_\Omega(\boldsymbol{I} \otimes \boldsymbol{A}^*\boldsymbol{A})\boldsymbol{P}_\Omega \boldsymbol{z}\|}{\|\boldsymbol{z}\|} \right\} - \|\boldsymbol{P}_\Omega \boldsymbol{R} \boldsymbol{P}_\Omega\| - \|\boldsymbol{P}_\Omega(\boldsymbol{T} - \hat{\boldsymbol{T}})\boldsymbol{P}_\Omega\|. \end{aligned} \tag{5.25}$$

The first and third terms above require relatively little manipulation to control. In particular, in paragraph (i) below, we will show that

$$\inf_{\boldsymbol{z} \in S_\Omega \setminus \{\boldsymbol{0}\}} \left\{ \frac{\|\boldsymbol{P}_\Omega(\boldsymbol{I} \otimes \boldsymbol{A}^*\boldsymbol{A})\boldsymbol{P}_\Omega \boldsymbol{z}\|}{\|\boldsymbol{z}\|} \right\} \;\ge\; 1 - k\mu(\boldsymbol{A}). \tag{5.26}$$

In paragraph (ii) below, we will show that there is a constant $t_\star > 0$ such that on $\mathcal{E}_{eig}(t_\star)$,

$$\|\boldsymbol{P}_\Omega(\boldsymbol{T} - \hat{\boldsymbol{T}})\boldsymbol{P}_\Omega\| \;\le\; 1/8. \tag{5.27}$$

[7] Notice that $\|\boldsymbol{P}_\Omega \boldsymbol{T} \boldsymbol{P}_{\Omega^c}\| \le \|\boldsymbol{P}_\Omega \boldsymbol{T}\|\|\boldsymbol{P}_{\Omega^c}\| = \|\boldsymbol{P}_\Omega \boldsymbol{T}\|$. Using (5.16), write

$$\begin{aligned} \|\boldsymbol{P}_\Omega \boldsymbol{T}\| &\le \|\boldsymbol{P}_\Omega(\boldsymbol{I} \otimes \boldsymbol{A}^*)\|\left\|(\boldsymbol{I} - \boldsymbol{P}_X) \otimes \boldsymbol{A} + (\boldsymbol{X}^*(\boldsymbol{X}\boldsymbol{X}^*)^{-1} \otimes \boldsymbol{I})\boldsymbol{C}_A\boldsymbol{C}_A^*((\boldsymbol{X}\boldsymbol{X}^*)^{-1}\boldsymbol{X} \otimes \boldsymbol{A})\right\| \\ &\le \|\boldsymbol{P}_\Omega(\boldsymbol{I} \otimes \boldsymbol{A}^*)\|\|(\boldsymbol{I} - \boldsymbol{P}_X) \otimes \boldsymbol{I} + (\boldsymbol{X}^*(\boldsymbol{X}\boldsymbol{X}^*)^{-1} \otimes \boldsymbol{I})\boldsymbol{C}_A\boldsymbol{C}_A^*((\boldsymbol{X}\boldsymbol{X}^*)^{-1}\boldsymbol{X} \otimes \boldsymbol{I})\|\|\boldsymbol{A}\|, \\ &\le \|\boldsymbol{P}_\Omega(\boldsymbol{I} \otimes \boldsymbol{A}^*)\|(1 + 1/\lambda_{min}(\boldsymbol{X}\boldsymbol{X}^*))\|\boldsymbol{A}\|. \end{aligned}$$

Now, $\boldsymbol{P}_\Omega(\boldsymbol{I} \otimes \boldsymbol{A}^*)$ is a block-diagonal matrix, with blocks given by $\boldsymbol{A}^*_{\Omega_1}, \dots, \boldsymbol{A}^*_{\Omega_p}$. By (C.2), the operator norm of each of these blocks is bounded by a constant, say, $c_1$. Hence, $\|\boldsymbol{P}_\Omega(\boldsymbol{I} \otimes \boldsymbol{A}^*)\|$ is bounded by $c_1$ as well. Similarly, on $\mathcal{E}_{eig}$, $\lambda_{min}^{-1}(\boldsymbol{X}\boldsymbol{X}^*)$ is also bounded by a constant, giving the desired expression.

The analysis of $\boldsymbol{P}_\Omega \boldsymbol{R} \boldsymbol{P}_\Omega$ is a bit trickier, requiring both additional algebraic manipulations and additional probability estimates. For now, we will *define* an event $\mathcal{E}_R$, on which the norm of this term is small:

$$\mathcal{E}_R \doteq \{\omega \mid \|\boldsymbol{P}_\Omega \boldsymbol{R} \boldsymbol{P}_\Omega\| \leq 1/8\}. \tag{5.28}$$

In Section 6, we prove the following lemma, which shows that $\mathcal{E}_R$ is indeed likely:

**Lemma 5.3.** *Let $\mathcal{E}_R$ be the event that $\|\boldsymbol{P}_\Omega \boldsymbol{R} \boldsymbol{P}_\Omega\| \leq 1/8$. Then there exist positive numerical constants $C_1, \dots, C_6$ such that whenever*

$$k \leq \min\left\{C_1 n, \frac{C_2}{\mu(\boldsymbol{A})}\right\} \tag{5.29}$$

*and $p > C_3 n^2$ we have*

$$\mathbb{P}[\mathcal{E}_R] \geq 1 - C_4 p^{-4} - C_5 n^2 \exp\left(-C_6 k^2 p/n^2\right). \tag{5.30}$$

Plugging (5.26), (5.27) and (5.28) into (5.25), we obtain that on $\mathcal{E}_{eig}(t_\star) \cap \mathcal{E}_R$

$$\xi \geq \frac{3}{4} - k\mu(\boldsymbol{A}). \tag{5.31}$$

So, assuming $C_1$ in the statement of Theorem 5.1 is such that $k\mu(\boldsymbol{A}) < 1/2$, we have $\xi > 1/4$. Plugging this value for $\xi$ into (5.21), we finally obtain that on $\mathcal{E}_{eig}(t_\star) \cap \mathcal{E}_R$, for any pair $(\boldsymbol{\Delta}_A, \boldsymbol{\Delta}_X)$ in the tangent space

$$\|\boldsymbol{\Delta}_A\|_F \leq \sqrt{2}\|\boldsymbol{A}\|(1 + C'\|\boldsymbol{A}\|)\|\mathcal{P}_{\Omega^c}[\boldsymbol{\Delta}_X]\|_F \leq C'' \|\boldsymbol{A}\|^2 \|\mathcal{P}_{\Omega^c}[\boldsymbol{\Delta}_X]\|_F. \tag{5.32}$$

Hence, the bound claimed in Theorem 5.1 holds with probability at least $1 - \mathbb{P}[\mathcal{E}_{eig}(t_\star)^c] - \mathbb{P}[\mathcal{E}_R^c]$. When the constants $C_1$ and $C_2$ in Theorem 5.1 are chosen appropriately, the conditions of Lemmas 5.2 of Section F and 5.3 of Section 6 are verified. From Lemma 5.2,

$$\mathbb{P}[\mathcal{E}_{eig}(t_\star)^c] < c_1 n \exp(-c_2 p/n \log(p)) + p^{-7}.$$

In Lemma 5.3 we show that

$$\mathbb{P}[\mathcal{E}_R^c] < c_3 p^{-4} + c_4 n^2 \exp(-c_5 k^2 p/n^2).$$

Combining the probabilities and consolidating polynomial terms establishes the desired result. It remains only to show that the two bounds in (5.26) and (5.27) indeed hold.

**(i) Establishing** (5.26)**.** For this term, it is more convenient to work with matrices and the Frobenius norm, rather than vectors and the $\ell^2$ norm. Fix any $\boldsymbol{Z} \in \mathbb{R}^{n \times p}$ with $\boldsymbol{z} \doteq \mathrm{vec}[\boldsymbol{Z}] \in S_\Omega$ (i.e., $\boldsymbol{Z}$ has support contained in $\Omega$). Then

$$\begin{aligned} \|\boldsymbol{P}_\Omega(\boldsymbol{I} \otimes \boldsymbol{A}^*\boldsymbol{A})\boldsymbol{P}_\Omega \boldsymbol{z}\|^2 &= \|\mathcal{P}_\Omega[\, \boldsymbol{A}^*\boldsymbol{A}\, \mathcal{P}_\Omega[\boldsymbol{Z}]\,]\|_F^2 = \sum_{j=1}^p \left\|\boldsymbol{A}_{\Omega_j}^* \boldsymbol{A}_{\Omega_j} \boldsymbol{Z}(\Omega_j, j)\right\|_2^2 \\ &\geq \sigma_{min}^2(\boldsymbol{A}_{\Omega_j}^* \boldsymbol{A}_{\Omega_j}) \sum_j \|\boldsymbol{Z}(\Omega_j, j)\|_2^2 \geq \|\boldsymbol{Z}\|_F^2 (1 - k\mu(\boldsymbol{A}))^2, \end{aligned} \tag{5.33}$$

where in the final step we have used that $\boldsymbol{Z}$ is supported on $\Omega$, where we have used the bound $\sigma_{min}(\boldsymbol{A}_{\Omega_j}^* \boldsymbol{A}_{\Omega_j}) > 1 - k\mu(\boldsymbol{A})$, shown in (C.3) of Appendix C. The bound (5.33) holds for all $\boldsymbol{z}$ supported on $\Omega$, and so (5.26) holds.

**(ii) Establishing** (5.27). For the term $\boldsymbol{P}_\Omega(\boldsymbol{T}-\hat{\boldsymbol{T}})\boldsymbol{P}_\Omega$, write $\boldsymbol{\Xi} \doteq (\boldsymbol{X}\boldsymbol{X}^*)^{-1} - \boldsymbol{I}$, and notice that $\boldsymbol{T}-\hat{\boldsymbol{T}}$ can be written as

$$\begin{aligned}
\boldsymbol{T}-\hat{\boldsymbol{T}} &= \boldsymbol{X}^*\boldsymbol{X}\otimes\boldsymbol{A}^*\boldsymbol{A} - \big(\boldsymbol{X}^*(\boldsymbol{X}\boldsymbol{X}^*)^{-1}\boldsymbol{X}\big)\otimes\boldsymbol{A}^*\boldsymbol{A} \\
&\quad + \big(\boldsymbol{X}^*(\boldsymbol{X}\boldsymbol{X}^*)^{-1}\otimes\boldsymbol{A}^*\big)\boldsymbol{C}_A\boldsymbol{C}_A^*\big((\boldsymbol{X}\boldsymbol{X}^*)^{-1}\boldsymbol{X}\otimes\boldsymbol{A}\big) \\
&\quad - (\boldsymbol{X}^*\otimes\boldsymbol{A}^*)\,\boldsymbol{C}_A\boldsymbol{C}_A^*\,(\boldsymbol{X}\otimes\boldsymbol{A}) \\
&= (\boldsymbol{X}^*\otimes\boldsymbol{A}^*)(-\boldsymbol{\Xi}\otimes\boldsymbol{I})(\boldsymbol{X}\otimes\boldsymbol{A}) \\
&\quad + \big(\boldsymbol{X}^*(\boldsymbol{X}\boldsymbol{X}^*)^{-1}\otimes\boldsymbol{A}^*\big)\boldsymbol{C}_A\boldsymbol{C}_A^*\big((\boldsymbol{X}\boldsymbol{X}^*)^{-1}\boldsymbol{X}\otimes\boldsymbol{A}\big) \\
&\quad - (\boldsymbol{X}^*\otimes\boldsymbol{A}^*)\,\boldsymbol{C}_A\boldsymbol{C}_A^*\big((\boldsymbol{X}\boldsymbol{X}^*)^{-1}\boldsymbol{X}\otimes\boldsymbol{A}\big) \\
&\quad + (\boldsymbol{X}^*\otimes\boldsymbol{A}^*)\,\boldsymbol{C}_A\boldsymbol{C}_A^*\big((\boldsymbol{X}\boldsymbol{X}^*)^{-1}\boldsymbol{X}\otimes\boldsymbol{A}\big) \\
&\quad - (\boldsymbol{X}^*\otimes\boldsymbol{A}^*)\,\boldsymbol{C}_A\boldsymbol{C}_A^*\,(\boldsymbol{X}\otimes\boldsymbol{A}) \\
&= (\boldsymbol{X}^*\otimes\boldsymbol{A}^*)(-\boldsymbol{\Xi}\otimes\boldsymbol{I})(\boldsymbol{X}\otimes\boldsymbol{A}) \\
&\quad + (\boldsymbol{X}^*\otimes\boldsymbol{A}^*)(\boldsymbol{\Xi}\otimes\boldsymbol{I})\boldsymbol{C}_A\boldsymbol{C}_A^*\big((\boldsymbol{X}\boldsymbol{X}^*)^{-1}\boldsymbol{X}\otimes\boldsymbol{A}\big) \\
&\quad + (\boldsymbol{X}^*\otimes\boldsymbol{A}^*)\boldsymbol{C}_A\boldsymbol{C}_A^*(\boldsymbol{\Xi}\otimes\boldsymbol{I})(\boldsymbol{X}\otimes\boldsymbol{A}) \\
&= (\boldsymbol{X}^*\otimes\boldsymbol{A}^*)\Big((\boldsymbol{C}_A\boldsymbol{C}_A^*-\boldsymbol{I})\,\boldsymbol{\Xi}\otimes\boldsymbol{I} + (\boldsymbol{\Xi}\otimes\boldsymbol{I})\boldsymbol{C}_A\boldsymbol{C}_A^*(\boldsymbol{X}\boldsymbol{X}^*)^{-1}\Big)(\boldsymbol{X}\otimes\boldsymbol{A}).
\end{aligned}$$

Therefore, using $\|\boldsymbol{C}_A\boldsymbol{C}_A^*-\boldsymbol{I}\| = 1$ and $\|\boldsymbol{C}_A\| = 1$, we have the estimate

$$\begin{aligned}
\|\boldsymbol{P}_\Omega(\boldsymbol{T}-\hat{\boldsymbol{T}})\boldsymbol{P}_\Omega\| &\le \|\boldsymbol{P}_\Omega(\boldsymbol{X}^*\otimes\boldsymbol{A}^*)\|^2\times\big(\|\boldsymbol{\Xi}\| + \|\boldsymbol{X}\boldsymbol{X}^*\|^{-1}\|\boldsymbol{\Xi}\|\big) \\
&\le \|\boldsymbol{P}_\Omega(\boldsymbol{I}\otimes\boldsymbol{A}^*)\|^2\|\boldsymbol{X}\|^2\big(1+\|(\boldsymbol{X}\boldsymbol{X}^*)^{-1}\|\big)\|\boldsymbol{\Xi}\| \\
&\le 6\times\|\boldsymbol{P}_\Omega(\boldsymbol{I}\otimes\boldsymbol{A}^*)\|^2\|\boldsymbol{\Xi}\|, \qquad (5.34)
\end{aligned}$$

where the last bound holds on $\mathcal{E}_{eig}(t)$ for small enough $t$ (e.g., $t < 1/2$ is sufficient). From the incoherence of $\boldsymbol{A}$ (i.e., (C.2)),

$$\|\boldsymbol{P}_\Omega(\boldsymbol{I}\otimes\boldsymbol{A}^*)\|^2 \;=\; \max_j \|\boldsymbol{A}_{\Omega_j}\|^2 \;\le\; 1 + k\mu(\boldsymbol{A}) < 2. \qquad (5.35)$$

Hence, on the event $\mathcal{E}_{eig}$, $\|\boldsymbol{P}_\Omega(\boldsymbol{T}-\hat{\boldsymbol{T}})\boldsymbol{P}_\Omega\| \le 12\,\|\boldsymbol{\Xi}\|$. Finally, on $\mathcal{E}_{eig}(t)$, $\|\boldsymbol{\Xi}\| \le t/(1-t)$; choosing $t$ small enough guarantees that on $\mathcal{E}_{eig}(t)$, $\|\boldsymbol{P}_\Omega(\boldsymbol{T}-\hat{\boldsymbol{T}})\boldsymbol{P}_\Omega\| \le 1/8$ as desired (in particular, $t < 1/97$ suffices).

Thus, (5.26) and (5.27) hold, and Theorem 5.1 is established. □

# 6 Controlling the residual $\boldsymbol{P}_\Omega\boldsymbol{R}\boldsymbol{P}_\Omega$

In this section, we estimate the norm of the residual $\boldsymbol{P}_\Omega\boldsymbol{R}\boldsymbol{P}_\Omega$, where $\boldsymbol{R}$ was defined in (5.23), and show that with high probability it is bounded by a small constant. To establish this result, in Section 6.1 below we first develop a more convenient expression for $\boldsymbol{P}_\Omega\boldsymbol{R}\boldsymbol{P}_\Omega$ as a sum of random semidefinite matrices that are independent conditioned on $\Omega$.

## 6.1 Proof of Lemma 5.3

*Proof.* We begin by introducing an additional bit of notation. For $\boldsymbol{X}\in\mathbb{R}^{n\times p}$, we write

$$\boldsymbol{x}^i = \boldsymbol{e}_i^*\boldsymbol{X}\in\mathbb{R}^{1\times p} \qquad (6.1)$$

for the $i$-th row of $\boldsymbol{X}$, and

$$\boldsymbol{x}_j = \boldsymbol{X}\boldsymbol{e}_j\in\mathbb{R}^n \qquad (6.2)$$

for the $j$-th column of $\boldsymbol{X}$. Similarly, we let

$$\Omega^i = \{j \mid (i,j) \in \Omega\} \subseteq [p], \tag{6.3}$$

and

$$\Omega_j \doteq \{i \mid (i,j) \in \Omega\} \subset [n]. \tag{6.4}$$

Recalling the definition (5.23) and using the familiar identity, $(\boldsymbol{P} \otimes \boldsymbol{Q}) = (\boldsymbol{P} \otimes \boldsymbol{I})(\boldsymbol{I} \otimes \boldsymbol{Q})$, we have

$$\boldsymbol{R} \;=\; (\boldsymbol{X}^* \otimes \boldsymbol{I})(\boldsymbol{I} \otimes \boldsymbol{A}^*)(\boldsymbol{I} - \boldsymbol{C}_A \boldsymbol{C}_A^*)(\boldsymbol{I} \otimes \boldsymbol{A})(\boldsymbol{X} \otimes \boldsymbol{I}). \tag{6.5}$$

The product of the middle three terms is a block diagonal matrix

$$(\boldsymbol{I} \otimes \boldsymbol{A}^*)(\boldsymbol{I} - \boldsymbol{C}_A \boldsymbol{C}_A^*)(\boldsymbol{I} \otimes \boldsymbol{A}) \;=\; \begin{bmatrix} \boldsymbol{A}^*(\boldsymbol{I} - \boldsymbol{A}_1 \boldsymbol{A}_1^*)\boldsymbol{A} & & \\ & \ddots & \\ & & \boldsymbol{A}^*(\boldsymbol{I} - \boldsymbol{A}_n \boldsymbol{A}_n^*)\boldsymbol{A} \end{bmatrix}. \tag{6.6}$$

For compactness, let $\boldsymbol{P}_i = \boldsymbol{I} - \boldsymbol{A}_i \boldsymbol{A}_i^*$; notice that this is the projection matrix onto the orthogonal complement of $\boldsymbol{A}_i$. Then, expanding the product in (6.5) more explicitly, we have

$$\boldsymbol{R} \;=\; \begin{bmatrix} \sum_{b=1}^n X_{b,1} X_{b,1} \boldsymbol{A}^* \boldsymbol{P}_b \boldsymbol{A} & \dots & \sum_{b=1}^n X_{b,1} X_{b,p} \boldsymbol{A}^* \boldsymbol{P}_b \boldsymbol{A} \\ \vdots & \ddots & \vdots \\ \sum_{b=1}^n X_{b,p} X_{b,1} \boldsymbol{A}^* \boldsymbol{P}_b \boldsymbol{A} & \dots & \sum_{b=1}^n X_{b,p} X_{b,p} \boldsymbol{A}^* \boldsymbol{P}_b \boldsymbol{A} \end{bmatrix}. \tag{6.7}$$

Breaking this sum up into $n$ terms (indexed by common $b$), we have

$$\boldsymbol{P}_\Omega \boldsymbol{R} \boldsymbol{P}_\Omega = \sum_{b=1}^n \boldsymbol{P}_\Omega \left( \boldsymbol{x}^{b^*} \boldsymbol{x}^b \otimes \boldsymbol{A}^* \boldsymbol{P}_b \boldsymbol{A} \right) \boldsymbol{P}_\Omega. \tag{6.8}$$

If we set

$$\begin{aligned} \boldsymbol{\Psi}_i \;&\doteq\; \boldsymbol{P}_\Omega \left( \boldsymbol{x}^{i^*} \boldsymbol{x}^i \otimes \boldsymbol{A}^* \boldsymbol{P}_i \boldsymbol{A} \right) \boldsymbol{P}_\Omega \\ &=\; \boldsymbol{P}_\Omega \left( \boldsymbol{P}_{\Omega^i} \boldsymbol{v}^{i^*} \boldsymbol{v}^i \boldsymbol{P}_{\Omega^i} \otimes \boldsymbol{A}^* \boldsymbol{P}_i \boldsymbol{A} \right) \boldsymbol{P}_\Omega \end{aligned} \tag{6.9}$$

then we have

$$\boldsymbol{P}_\Omega \boldsymbol{R} \boldsymbol{P}_\Omega \;=\; \sum_{i=1}^n \boldsymbol{\Psi}_i. \tag{6.10}$$

This is a sum of random positive semidefinite matrices. Moreover, from (6.9), we observe that conditioned on $\Omega$, the $\boldsymbol{\Psi}_i$ are fixed functions of independent random vectors $\boldsymbol{v}^i$, and hence the $\boldsymbol{\Psi}_i$ are conditionally independent. We would like to apply a matrix tail bound to this sum, conditioned on $\Omega$. To do this, we first need to understand how the support $\Omega$ affects the expected size of $\boldsymbol{\Psi}_i$.

With high probability, the support $\Omega$ is quite regular. If we fix any $i \in [n]$, the expected size of $\Omega^i$ is simply $pk/n$. In fact, because the $\Omega_j$ are independent (and hence the events $j \in \Omega^i$ are independent), $|\Omega^i|$ concentrates near this value. Moreover, if $i$ and $i'$ are distinct, $|\Omega^i \cap \Omega^{i'}|$ concentrates about *its* expectation, which is bounded by $k^2 p/n^2$. We define a set of "desirable" supports, for which these quantities do not greatly exceed their expectations:

$$\mathcal{O} \doteq \left\{ \Omega \subset [n] \times [p] \;\middle|\; \begin{array}{l} \max_{i=1,\dots,n} |\Omega^i| \le 3pk/2n \\ \max_{i \ne i'} |\Omega^i \cap \Omega^{i'}| \le 3pk^2/2n^2 \end{array} \right\}. \tag{6.11}$$

It is not difficult to show that the event $\Omega \in \mathcal{O}$ is overwhelmingly likely:

**Lemma 6.1.** *With overwhelming probability,* $\Omega \in \mathcal{O}$*:*

$$\mathbb{P}\left[\Omega \in \mathcal{O}\right] \;\geq\; 1 - n^2 \exp\left(-\frac{pk^2}{10n^2}\right). \tag{6.12}$$

We prove Lemma 6.1 in Section 6.2 below.

Now, when $\Omega \in \mathcal{O}$, the norms of the rows of $\boldsymbol{X}$ also concentrate about their conditional expectations. We define $n$ events, on which the rows $\boldsymbol{x}^i$ are not too large in norm, and also do not concentrate too strongly on the intersection $\Omega^{i'} \cap \Omega^i$ for any $i' \neq i$:

$$\mathcal{E}_i \doteq \left\{ \omega \;\middle|\; \max_{a \neq i} \|\boldsymbol{x}^i \boldsymbol{P}_{\Omega^a}\| \;\leq\; 2\sqrt{k/n}, \;\; \text{and} \;\; \|\boldsymbol{x}^i\| \;\leq\; 2 \right\}. \tag{6.13}$$

We further set

$$\mathcal{E}_X \doteq \cap_{i=1}^n \mathcal{E}_i. \tag{6.14}$$

It is not hard to show that $\mathcal{E}_X$ is overwhelmingly likely:

**Lemma 6.2.** *For any* $\Omega \in \mathcal{O}$*,*

$$\mathbb{P}[\mathcal{E}_X \mid \Omega] \;\geq\; 1 - n^2 \exp\left(-\frac{k^2 p}{4n^2}\right). \tag{6.15}$$

We prove Lemma 6.2 in Section 6.3 below. This lemma is useful because whenever $\mathcal{E}_i$ occurs, $\boldsymbol{\Psi}_i$ is indeed small in norm:

**Lemma 6.3.** *Let* $\mathcal{E}_i$ *be the event defined in* (6.13)*, and let* $\boldsymbol{\Psi}_i$ *denote the* $i$*-th residual term:*

$$\boldsymbol{\Psi}_i \;=\; \boldsymbol{P}_\Omega \left( {\boldsymbol{x}^i}^* \boldsymbol{x}^i \otimes \boldsymbol{A}^* \boldsymbol{P}_i \boldsymbol{A} \right) \boldsymbol{P}_\Omega \tag{6.16}$$

*Then on event* $\mathcal{E}_i$*, we have*

$$\|\boldsymbol{\Psi}_i\| \;\leq\; 4k/n + 24\, k\mu(\boldsymbol{A}). \tag{6.17}$$

We prove Lemma 6.3 in Section 6.4 below. For now, however, we show how the previous three lemmas, together with a matrix Chernoff bound, imply the desired result. For convenience, let $\boldsymbol{\Psi} \doteq \boldsymbol{P}_\Omega \boldsymbol{R} \boldsymbol{P}_\Omega = \sum_i \boldsymbol{\Psi}_i$. Set

$$\bar{\boldsymbol{\Psi}}_i = \boldsymbol{\Psi}_i \times \mathbb{1}_{\mathcal{E}_i}, \tag{6.18}$$

where $\mathbb{1}_{\mathcal{E}_i}$ denotes the indicator random variable for the event $\mathcal{E}_i$. By Lemma 6.3, $\bar{\boldsymbol{\Psi}}_i$ always satisfies

$$\|\bar{\boldsymbol{\Psi}}_i\| \;\leq\; 4k/n + 24\, k\mu(\boldsymbol{A}) \;\doteq\; B. \tag{6.19}$$

Conditioned on $\Omega$, each $\bar{\boldsymbol{\Psi}}_i$ is a function of $\boldsymbol{v}^i$ only, and hence the $\bar{\boldsymbol{\Psi}}_i$ are independent conditioned on $\Omega$.

We apply a sequence of manipulations to reduce the problem of bounding the probability that $\|\boldsymbol{\Psi}\|$ exceeds $1/8$ to the problem of bounding the probability that $\|\bar{\boldsymbol{\Psi}}\|$ exceeds $1/8$:

$$\begin{aligned}
\mathbb{P}\left[\|\boldsymbol{\Psi}\| \geq 1/8\right] &= \mathbb{P}\left[\|\boldsymbol{\Psi}\| \geq 1/8 \mid \Omega \in \mathcal{O}\right] \mathbb{P}\left[\Omega \in \mathcal{O}\right] + \mathbb{P}\left[\|\boldsymbol{\Psi}\| \geq 1/8 \mid \Omega \in \mathcal{O}^c\right] \mathbb{P}\left[\Omega \in \mathcal{O}^c\right] \\
&\leq \mathbb{P}\left[\|\boldsymbol{\Psi}\| \geq 1/8 \mid \Omega \in \mathcal{O}\right] \;+\; \mathbb{P}\left[\Omega \in \mathcal{O}^c\right] \\
&\leq \max_{\Omega_0 \in \mathcal{O}} \mathbb{P}\left[\|\boldsymbol{\Psi}\| \geq 1/8 \mid \Omega_0\right] \;+\; \mathbb{P}\left[\Omega \in \mathcal{O}^c\right] \\
&\leq \max_{\Omega_0 \in \mathcal{O}} \left\{ \mathbb{P}\left[\|\bar{\boldsymbol{\Psi}}\| \geq 1/8 \mid \Omega_0\right] + \mathbb{P}\left[\boldsymbol{\Psi} \neq \bar{\boldsymbol{\Psi}} \mid \Omega_0\right] \right\} \;+\; \mathbb{P}\left[\Omega \in \mathcal{O}^c\right] \\
&\leq \max_{\Omega_0 \in \mathcal{O}} \left\{ \mathbb{P}\left[\|\bar{\boldsymbol{\Psi}}\| \geq 1/8 \mid \Omega_0\right] + \mathbb{P}\left[\cup_i \mathcal{E}_i^c \mid \Omega_0\right] \right\} \;+\; \mathbb{P}\left[\Omega \in \mathcal{O}^c\right] \qquad (6.20) \\
&= \max_{\Omega_0 \in \mathcal{O}} \left\{ \mathbb{P}\left[\|\bar{\boldsymbol{\Psi}}\| \geq 1/8 \mid \Omega_0\right] + \mathbb{P}\left[\mathcal{E}_X^c \mid \Omega_0\right] \right\} \;+\; \mathbb{P}\left[\Omega \in \mathcal{O}^c\right]. \qquad (6.21)
\end{aligned}$$

In (6.20), we have used that by definition on $\mathcal{E}_i$, $\bar{\boldsymbol{\Psi}}_i = \boldsymbol{\Psi}_i \mathbb{1}_{\mathcal{E}_i}$ is equal to $\boldsymbol{\Psi}_i$, while in the following line we have used the definition of $\mathcal{E}_X = \cap_i \mathcal{E}_i$. Now, Lemma 6.2 shows that conditioned on any $\Omega_0 \in \mathcal{O}$, $\mathcal{E}_X^c$ is overwhelmingly unlikely, while Lemma 6.1 shows that the event $\Omega \in \mathcal{O}^c$ is overwhelmingly unlikely. Plugging in the bounds from those two lemmas, we have that

$$\begin{aligned}
\mathbb{P}\left[\|\boldsymbol{\Psi}\| \geq 1/8\right] \;&\leq\; \max_{\Omega_0 \in \mathcal{O}} \mathbb{P}\left[\|\bar{\boldsymbol{\Psi}}\| \geq 1/8 \mid \Omega_0\right] \;+\; n^2 \exp\left(-\frac{k^2 p}{4n^2}\right) \;+\; n^2 \exp\left(-\frac{k^2 p}{10 n^2}\right) \\
&\leq\; \max_{\Omega_0 \in \mathcal{O}} \mathbb{P}\left[\|\bar{\boldsymbol{\Psi}}\| \geq 1/8 \mid \Omega_0\right] \;+\; 2n^2 \exp\left(-\frac{k^2 p}{10 n^2}\right). \qquad (6.22)
\end{aligned}$$

We complete the proof by applying the matrix Chernoff bound (B.4) to the first term. Fix any $\Omega_0 \in \mathcal{O}$. We need to estimate $\mu_{max} = \|\mathbb{E}[\bar{\boldsymbol{\Psi}} \mid \Omega_0]\|$. Since $\mathbf{0} \preceq \bar{\boldsymbol{\Psi}} \preceq \boldsymbol{\Psi}$ always,

$$\mu_{max} \;=\; \|\mathbb{E}[\bar{\boldsymbol{\Psi}} \mid \Omega_0]\| \;\leq\; \|\mathbb{E}[\boldsymbol{\Psi} \mid \Omega_0]\|. \qquad (6.23)$$

This conditional expectation can be easily evaluated using the expression for $\boldsymbol{\Psi}$ in (6.9) and the fact that $\mathbb{E}[\boldsymbol{v}^{i*}\boldsymbol{v}^i] = (n/kp)\,\boldsymbol{I}$:

$$\mathbb{E}[\boldsymbol{\Psi} \mid \Omega] \;=\; \mathbb{E}_V\left[\sum_{i=1}^n \boldsymbol{P}_\Omega \left(\boldsymbol{P}_{\Omega^i} \boldsymbol{v}^{i*} \boldsymbol{v}^i \boldsymbol{P}_{\Omega^i} \otimes \boldsymbol{A}^* \boldsymbol{P}_i \boldsymbol{A}\right) \boldsymbol{P}_\Omega\right] \;=\; \frac{n}{kp} \sum_{i=1}^n \boldsymbol{P}_\Omega \left(\boldsymbol{P}_{\Omega^i} \otimes \boldsymbol{A}^* \boldsymbol{P}_i \boldsymbol{A}\right) \boldsymbol{P}_\Omega.$$

Notice that for each $i$, $\boldsymbol{P}_{\Omega^i} \otimes \boldsymbol{A}^* \boldsymbol{P}_i \boldsymbol{A} \preceq \boldsymbol{P}_{\Omega^i} \otimes \boldsymbol{A}^* \boldsymbol{A}$, and so

$$\mathbb{E}[\,\boldsymbol{\Psi} \mid \Omega\,] \;\preceq\; \frac{n}{kp} \sum_{i=1}^n \boldsymbol{P}_\Omega \left(\boldsymbol{P}_{\Omega^i} \otimes \boldsymbol{A}^* \boldsymbol{A}\right) P_\Omega \;=\; \frac{n}{kp} \boldsymbol{P}_\Omega \left(\sum_{i=1}^n \boldsymbol{P}_{\Omega^i} \otimes \boldsymbol{A}^* \boldsymbol{A}\right) \boldsymbol{P}_\Omega \qquad (6.24)$$

The matrix $\sum_i \boldsymbol{P}_{\Omega^i}$ is diagonal; its $(j,j)$ element simply counts the number of nonzero entries $i$ in the $j$-th column of $\Omega$. This number is a constant $k$, so $\sum_i \boldsymbol{P}_{\Omega^i} = k\boldsymbol{I}$, and

$$\mathbb{E}[\,\boldsymbol{\Psi} \mid \Omega\,] \;\preceq\; \frac{n}{p} \boldsymbol{P}_\Omega \left(\boldsymbol{I} \otimes \boldsymbol{A}^* \boldsymbol{A}\right) \boldsymbol{P}_\Omega. \qquad (6.25)$$

The matrix $\boldsymbol{P}_\Omega(\boldsymbol{I} \otimes \boldsymbol{A}^*\boldsymbol{A})\boldsymbol{P}_\Omega$ is block-diagonal, with $j$-th block $\boldsymbol{P}_{\Omega_j} \boldsymbol{A}^* \boldsymbol{A} \boldsymbol{P}_{\Omega_j}$. This block has norm bounded by $\|\boldsymbol{A}_{\Omega_j}\|^2$. Using a calculation given in (C.2), this is in turn bounded by $3/2$, provided $k\mu(\boldsymbol{A}) < 1/2$. Hence, we have

$$\mu_{max} \;\leq\; 3n/2p. \qquad (6.26)$$

We apply the matrix Chernoff bound (B.3) with $t\mu_{max} = 1/8$, and hence $t \geq p/12n$ gives

$$\mathbb{P}\left[\left\|\bar{\boldsymbol{\Psi}}\right\| \geq 1/8 \mid \Omega\right] \;\leq\; np \left(\frac{12en}{p}\right)^{1/8B}, \qquad (6.27)$$

where we recall $B$ is the bound on the norm of the summands $\bar{\boldsymbol{\Psi}}_i$. By choosing the constant $C_1 > 0$ in the statement of Lemma 5.3, we can make the exponent $\nu = 1/8B$ as large as desired; the probability that $\|\bar{\boldsymbol{\Psi}}\|$ exceeds $1/8$ is bounded as

$$\mathbb{P}\left[\left\|\bar{\boldsymbol{\Psi}}\right\| \geq 1/8 \mid \Omega\right] \;\leq\; C(\nu)\, n^{1+\nu} p^{1-\nu}. \qquad (6.28)$$

Assuming $p \geq Cn^2$, and by appropriate choice of $\nu$, we can make the right hand side smaller than $C'p^{-4}$ (here, the exponent 4 is clearly arbitrary). Plugging into (6.22) completes the proof.

□

### 6.2 Proof of Lemma 6.1

*Proof.* Notice that $|\Omega^i| = \sum_{j=1}^p \mathbb{1}_{(i,j)\in\Omega}$ is a sum of $p$ independent Bernoulli$(k/n)$ random variables. Write

$$Z_j \;=\; \mathbb{1}_{(i,j)\in\Omega} - \mathbb{E}[\mathbb{1}_{(i,j)\in\Omega}] \;=\; \mathbb{1}_{(i,j)\in\Omega} - k/n.$$

Then $|\Omega^i| = pk/n + \sum_{j=1}^p Z_j$. The $Z_j$ are independent, zero mean, with magnitude bounded by 1 and variance

$$\mathbb{E}[Z_j^2] \;=\; \mathrm{Var}[\mathbb{1}_{(i,j)\in\Omega}] \;\le\; \mathbb{E}\left[\left(\mathbb{1}_{(i,j)\in\Omega}\right)^2\right] \;=\; \frac{k}{n}. \tag{6.29}$$

By Bernstein's inequality, for any $\epsilon > 0$,

$$\mathbb{P}\Big[\sum_j Z_j > \epsilon p\Big] \;\le\; \exp\left(-\frac{p\epsilon^2}{2\mathbb{E}[Z_j^2] + 2\epsilon/3}\right) \;\le\; \exp\left(-\frac{p\epsilon^2}{2k/n + 2\epsilon/3}\right) \tag{6.30}$$

Setting $\epsilon = k/2n$ and then taking a union bound over $i \in [n]$ gives

$$\mathbb{P}\left[\max_i |\Omega^i| \;\ge\; \frac{3}{2}\frac{kp}{n}\right] \;\le\; n \exp\left(-\frac{p\,k}{10\,n}\right). \tag{6.31}$$

Above, we have used $8 + 4/3 < 10$ to simplify the constant. Similarly, notice that

$$|\Omega^i \cap \Omega^{i'}| \;=\; \sum_{j=1}^p \mathbb{1}_{(i,j)\in\Omega}\mathbb{1}_{(i',j)\in\Omega} \;\doteq\; \sum_j H_j$$

is a sum of independent Bernoulli random variables $H_j$ which take on value one with probability

$$\mathbb{E}[H_j] \;=\; \mathbb{P}[H_j = 1] \;=\; \binom{n-2}{k-2} / \binom{n}{k} \;\le\; \frac{k^2}{n^2}. \tag{6.32}$$

Set $Z_j = H_j - \mathbb{E}[H_j]$. Then, the bound $|Z_j| \le 1$ always holds, and furthermore

$$\mathbb{E}[Z_j^2] \;=\; \mathrm{Var}[H_j] \;\le\; \mathbb{E}[H_j^2] \;\le\; \frac{k^2}{n^2}. \tag{6.33}$$

With these definitions,

$$|\Omega^i \cap \Omega^{i'}| \;\le\; pk^2/n^2 + \sum_j Z_j.$$

Again applying Bernstein's inequality to $\sum_j Z_j$, setting $\epsilon = k^2/2n^2$ and taking a union bound over the $\binom{n}{2}$ choices of distinct $(i, i')$, we have

$$\mathbb{P}\left[\max_{i \ne i'} |\Omega^i \cap \Omega^{i'}| \;\ge\; \frac{3}{2}\frac{k^2 p}{n^2}\right] \;\le\; \binom{n}{2} \exp\left(-\frac{pk^2}{10n^2}\right). \tag{6.34}$$

Summing the failure probabilities in (6.31) and (6.34) (using that $\binom{n}{2} + n < n^2$ and that $\exp(-pk/10n) < \exp(-pk^2/10n^2)$), completes the proof. □

### 6.3 Proof of Lemma 6.2

*Proof.* This proof is an exercise in Gaussian measure concentration. Equation (2.35) of [Led01] implies that if $\boldsymbol{v}$ is an iid sequence of $\mathcal{N}(0, \sigma^2)$ random variables, and $f$ is a positively homogeneous, 1-Lipschitz function, then

$$\mathbb{P}\left[f(\boldsymbol{v}) \ge \mathbb{E}[f(\boldsymbol{v})] + t\right] \le \exp\left(-\frac{t^2}{2\sigma^2}\right). \tag{6.35}$$

Now, with $\Omega$ fixed, $\|\boldsymbol{x}^i\| = \|\boldsymbol{v}^i \boldsymbol{P}_{\Omega^i}\| \doteq f(\boldsymbol{v}^i)$ is a 1-Lipschitz function of the iid $\mathcal{N}(0, n/kp)$ vector $\boldsymbol{v}^i$. Since for any $\Omega \in \mathcal{O}$, $|\Omega^i| \le 3pk/2n$,

$$\mathbb{E}\left[\|\boldsymbol{x}^i\| \mid \Omega\right] \quad \le \quad \sqrt{\mathbb{E}\left[\|\boldsymbol{x}^i\|^2 \mid \Omega\right]} \quad = \quad \sqrt{|\Omega^i| n/kp} \quad \le \quad \sqrt{3/2}. \tag{6.36}$$

Hence, for $\Omega \in \mathcal{O}$,

$$\mathbb{P}[\|\boldsymbol{x}^i\| \ge 2 \mid \Omega] \quad \le \quad \mathbb{P}\left[f(\boldsymbol{v}^i) \ge \mathbb{E}[f(\boldsymbol{v}^i) \mid \Omega] + (2 - \sqrt{3/2}) \mid \Omega\right] \;\le\; \exp\left(-\frac{kp}{4n}\right), \tag{6.37}$$

where we have used that $(2 - \sqrt{3/2})^2/2 \approx 0.3005 > 1/4$ to simplify the constant.

Now, fix $i' \ne i$. We apply exactly the same reasoning to

$$\|\boldsymbol{x}^i \boldsymbol{P}_{\Omega^{i'}}\| = \|\boldsymbol{v}^i \boldsymbol{P}_{\Omega^i} \boldsymbol{P}_{\Omega^{i'}}\| = \|\boldsymbol{v}^i \boldsymbol{P}_{\Omega^i \cap \Omega^{i'}}\| \doteq g(\boldsymbol{v}^i).$$

Again, $g(\cdot)$ is a 1-Lipschitz function of $\boldsymbol{v}^i$. Furthermore, for $\Omega \in \mathcal{O}$, $|\Omega^i \cap \Omega^{i'}| \le 3pk^2/2n^2$, and

$$\mathbb{E}[g(\boldsymbol{v}^i) \mid \Omega] \;\le\; \sqrt{3k/2n}. \tag{6.38}$$

Again,

$$\mathbb{P}\left[g(\boldsymbol{v}^i) \ge 2\sqrt{k/n} \mid \Omega\right] \;\le\; \mathbb{P}\left[g(\boldsymbol{v}^i) \ge \mathbb{E}[g(\boldsymbol{v}^i) \mid \Omega] + (2 - \sqrt{3/2})\sqrt{k/n} \mid \Omega\right] \;\le\; \exp\left(-\frac{k^2 p}{4n^2}\right).$$

A union bound over all $n$ choices of $i$ in (6.37) and all $n(n-1)$ ordered pairs $(i, i')$ completes the proof. □

## 6.4 Proof of Lemma 6.3

*Proof.* We will show the calculations for $i = 1$. An identical argument works for $i = 2, \dots, n$ as well. Since $\boldsymbol{A}$ is incoherent, $\boldsymbol{A}^* \boldsymbol{P}_1 \boldsymbol{A} \;=\; \boldsymbol{A}^*\boldsymbol{A} - \boldsymbol{A}^*\boldsymbol{A}_1\boldsymbol{A}_1^*\boldsymbol{A} \;\approx\; \boldsymbol{I} - \boldsymbol{e}_1\boldsymbol{e}_1^*$, and so we set

$$\boldsymbol{\Delta} \;\doteq\; \boldsymbol{A}^*\boldsymbol{P}_1\boldsymbol{A} - (\boldsymbol{I} - \boldsymbol{e}_1\boldsymbol{e}_1^*) \in \mathbb{R}^{n\times n}. \tag{6.39}$$

We notice that since $\mu(\boldsymbol{A}) \le 1$,

$$\|\boldsymbol{\Delta}\|_\infty \;\le\; \|\boldsymbol{A}^*\boldsymbol{A} - \boldsymbol{I}\|_\infty + \|\boldsymbol{A}^*\boldsymbol{A}_1\boldsymbol{A}_1^*\boldsymbol{A} - \boldsymbol{e}_1\boldsymbol{e}_1^*\|_\infty \;\le\; 2\mu(\boldsymbol{A}). \tag{6.40}$$

Write

$$\begin{aligned} \|\boldsymbol{\Psi}_1\| \quad &= \quad \left\|\boldsymbol{P}_\Omega\left({\boldsymbol{x}^1}^*\boldsymbol{x}^1 \otimes (\boldsymbol{I} - \boldsymbol{e}_1\boldsymbol{e}_1^* + \boldsymbol{\Delta})\right)\boldsymbol{P}_\Omega\right\| \\ &\le \quad \left\|\boldsymbol{P}_\Omega\left({\boldsymbol{x}^1}^*\boldsymbol{x}^1 \otimes (\boldsymbol{I} - \boldsymbol{e}_1\boldsymbol{e}_1^*)\right)\boldsymbol{P}_\Omega\right\| + \left\|\boldsymbol{P}_\Omega({\boldsymbol{x}^1}^*\boldsymbol{x}^1 \otimes \boldsymbol{\Delta})\boldsymbol{P}_\Omega\right\| \end{aligned} \tag{6.41}$$

We handle the two terms individually. For the first term,

$$\boldsymbol{L} \;\doteq\; \boldsymbol{P}_\Omega\left({\boldsymbol{x}^1}^*\boldsymbol{x}^1 \otimes (\boldsymbol{I} - \boldsymbol{e}_1\boldsymbol{e}_1^*)\right)\boldsymbol{P}_\Omega \;\in\; \mathbb{R}^{np\times np}, \tag{6.42}$$

we let $\mathcal{L} : \mathbb{R}^{n\times p} \to \mathbb{R}^{n\times p}$ be the equivalent linear operator such that for all $\boldsymbol{Q} \in \mathbb{R}^{n\times p}$,

$$\mathrm{vec}\left[\,\mathcal{L}[\boldsymbol{Q}]\,\right] \;=\; \boldsymbol{L}\,\mathrm{vec}[\boldsymbol{Q}]. \tag{6.43}$$

The norm of $\boldsymbol{L}$ as a linear operator from $\ell^2$ to $\ell^2$ is the same as the induced norm on $\mathcal{L}$:

$$\|\boldsymbol{L}\| \;=\; \|\mathcal{L}\| \;\doteq\; \sup_{\boldsymbol{Q}\ne\boldsymbol{0}} \frac{\|\mathcal{L}[\boldsymbol{Q}]\|_F}{\|\boldsymbol{Q}\|_F}.$$

From (6.42) and the relationship $\mathrm{vec}[\boldsymbol{PQR}] = (\boldsymbol{R}^* \otimes \boldsymbol{P})\,\mathrm{vec}[\boldsymbol{Q}]$, the operator $\mathcal{L}[\boldsymbol{Q}]$ is given by

$$\mathcal{L}[\boldsymbol{Q}] \;=\; \mathcal{P}_\Omega \left[ (\boldsymbol{I} - \boldsymbol{e}_1\boldsymbol{e}_1^*) \; \mathcal{P}_\Omega[\boldsymbol{Q}] \; {\boldsymbol{x}^1}^* \boldsymbol{x}^1 \right]. \tag{6.44}$$

For any $\boldsymbol{H} \in \mathbb{R}^{n\times p}$, we can express the projection $\mathcal{P}_\Omega$ of $\boldsymbol{H}$ onto $\Omega$ via its action on the rows of $\boldsymbol{H}$:

$$\mathcal{P}_\Omega[\boldsymbol{H}] \;=\; \sum_{a=1}^n \boldsymbol{e}_a \boldsymbol{e}_a^* \boldsymbol{H} \boldsymbol{P}_{\Omega^a}. \tag{6.45}$$

Applying this expression twice, (6.44) becomes

$$\begin{aligned} \mathcal{L}[\boldsymbol{Q}] \quad &= \quad \sum_{a=1}^n \boldsymbol{e}_a \boldsymbol{e}_a^* \left[ (\boldsymbol{I} - \boldsymbol{e}_1\boldsymbol{e}_1^*) \, \mathcal{P}_\Omega[\boldsymbol{Q}] \; {\boldsymbol{x}^1}^* \boldsymbol{x}^1 \right] \boldsymbol{P}_{\Omega^a} \quad &&= \quad \sum_{a=2}^n \boldsymbol{e}_a \boldsymbol{e}_a^* \; \mathcal{P}_\Omega[\boldsymbol{Q}] \; {\boldsymbol{x}^1}^* \boldsymbol{x}^1 \boldsymbol{P}_{\Omega^a} \\ &= \quad \sum_{a=2}^n \boldsymbol{e}_a \boldsymbol{e}_a^* \left( \sum_{b=1}^n \boldsymbol{e}_b \boldsymbol{e}_b^* \; \boldsymbol{Q} \; \boldsymbol{P}_{\Omega^b} \right) {\boldsymbol{x}^1}^* \boldsymbol{x}^1 \boldsymbol{P}_{\Omega^a} \quad &&= \quad \sum_{a=2}^n \boldsymbol{e}_a \boldsymbol{e}_a^* \; \boldsymbol{Q} \; \boldsymbol{P}_{\Omega^a} {\boldsymbol{x}^1}^* \boldsymbol{x}^1 \boldsymbol{P}_{\Omega^a}. \end{aligned} \tag{6.46}$$

Since the $a$-th summand occupies only the $a$-th row, we can write $\|\mathcal{L}[\boldsymbol{Q}]\|_F^2$ as the sum of the squared $\ell^2$ norms of the terms in the above expression:

$$\|\mathcal{L}[\boldsymbol{Q}]\|_F^2 \;=\; \sum_{a=2}^n \|\boldsymbol{e}_a^* \; \boldsymbol{Q} \; \boldsymbol{P}_{\Omega^a} {\boldsymbol{x}^1}^* \boldsymbol{x}^1 \boldsymbol{P}_{\Omega^a}\|^2 \;\le\; \sum_{a=2}^n \|\boldsymbol{e}_a^* \boldsymbol{Q}\|^2 \|\boldsymbol{x}^1 \boldsymbol{P}_{\Omega^a}\|^4 \;\le\; 16 \frac{k^2}{n^2} \|\boldsymbol{Q}\|_F^2, \tag{6.47}$$

where above we have used that on $\mathcal{E}_1$, $\|\boldsymbol{x}^1 \boldsymbol{P}_{\Omega^a}\| \le 2\sqrt{k/n}$ for every $a \neq 1$. We conclude that

$$\|\boldsymbol{L}\| \;\le\; 4k/n. \tag{6.48}$$

We next address the second term in (6.41). Define

$$\boldsymbol{W} \;\doteq\; \boldsymbol{P}_\Omega \left( {\boldsymbol{x}^1}^* \boldsymbol{x}^1 \otimes \boldsymbol{\Delta} \right) \boldsymbol{P}_\Omega \;\in\; \mathbb{R}^{np\times np}. \tag{6.49}$$

We need to bound the operator norm of $\boldsymbol{W}$. As above, we associate a linear map $\mathcal{W} : \mathbb{R}^{n\times p} \to \mathbb{R}^{n\times p}$, given by

$$\mathcal{W}[\boldsymbol{Q}] \quad = \quad \mathcal{P}_\Omega \left[ \boldsymbol{\Delta} \; \mathcal{P}_\Omega[\boldsymbol{Q}] \; {\boldsymbol{x}^1}^* \boldsymbol{x}^1 \right] \quad = \quad \sum_{a,b=1}^n \boldsymbol{e}_a \boldsymbol{e}_a^* \; \boldsymbol{\Delta} \; \boldsymbol{e}_b \boldsymbol{e}_b^* \; \boldsymbol{Q} \; \boldsymbol{P}_{\Omega^b} \; {\boldsymbol{x}^1}^* \boldsymbol{x}^1 \boldsymbol{P}_{\Omega^a}. \tag{6.50}$$

In the above expression, terms for which $a, b \neq 1$ will be easily handled, since on $\mathcal{E}_1$, $\|\boldsymbol{x}^1 \boldsymbol{P}_{\Omega^a}\| \le 2\sqrt{k/n}$ for every $a \neq 1$. We therefore break the above summation into four terms:

$$\boldsymbol{T}_1 \;\doteq\; \boldsymbol{e}_1 \boldsymbol{e}_1^* \; \boldsymbol{\Delta} \; \boldsymbol{e}_1 \boldsymbol{e}_1^* \; \boldsymbol{Q} \; \boldsymbol{P}_{\Omega^1} {\boldsymbol{x}^1}^* \boldsymbol{x}^1 \boldsymbol{P}_{\Omega^1}, \tag{6.51}$$

$$\boldsymbol{T}_2 \;\doteq\; \sum_{b=2}^n \boldsymbol{e}_1 \boldsymbol{e}_1^* \; \boldsymbol{\Delta} \; \boldsymbol{e}_b \boldsymbol{e}_b^* \; \boldsymbol{Q} \; \boldsymbol{P}_{\Omega^b} {\boldsymbol{x}^1}^* \boldsymbol{x}^1 \boldsymbol{P}_{\Omega^1}, \tag{6.52}$$

$$\boldsymbol{T}_3 \;\doteq\; \sum_{a=2}^n \boldsymbol{e}_a \boldsymbol{e}_a^* \; \boldsymbol{\Delta} \; \boldsymbol{e}_1 \boldsymbol{e}_1^* \; \boldsymbol{Q} \; \boldsymbol{P}_{\Omega^1} {\boldsymbol{x}^1}^* \boldsymbol{x}^1 \boldsymbol{P}_{\Omega^a}, \tag{6.53}$$

$$\text{and} \qquad \boldsymbol{T}_4 \;\doteq\; \sum_{a,b=2}^n \boldsymbol{e}_a \boldsymbol{e}_a^* \; \boldsymbol{\Delta} \; \boldsymbol{e}_b \boldsymbol{e}_b^* \; \boldsymbol{Q} \; \boldsymbol{P}_{\Omega^b} {\boldsymbol{x}^1}^* \boldsymbol{x}^1 \boldsymbol{P}_{\Omega^a}. \tag{6.54}$$

In terms of these four quantities,

$$\mathcal{W}[\boldsymbol{Q}] \;=\; \boldsymbol{T}_1 + \boldsymbol{T}_2 + \boldsymbol{T}_3 + \boldsymbol{T}_4. \tag{6.55}$$

Now, $\boldsymbol{e}_1^*\boldsymbol{\Delta}\boldsymbol{e}_1 = \boldsymbol{A}_1^*\boldsymbol{A}_1 - (\boldsymbol{A}_1^*\boldsymbol{A}_1)^2 = 0$, since $\boldsymbol{A}_1^*\boldsymbol{A}_1 = 1$. Plugging $\boldsymbol{e}_1^*\boldsymbol{\Delta}\boldsymbol{e}_1 = 0$ into (6.51), we have $\boldsymbol{T}_1 = \boldsymbol{0}$. Below, we show that on $\mathcal{E}_1$, the following bounds on the terms $\boldsymbol{T}_2, \boldsymbol{T}_3, \boldsymbol{T}_4$ hold:

$$\begin{aligned} \|\boldsymbol{T}_2\|_F &\leq 8\mu(\boldsymbol{A})\sqrt{k}\|\boldsymbol{Q}\|_F, &(6.56)\\ \|\boldsymbol{T}_3\|_F &\leq 8\mu(\boldsymbol{A})\sqrt{k}\|\boldsymbol{Q}\|_F, &(6.57)\\ \|\boldsymbol{T}_4\|_F &\leq 8\mu(\boldsymbol{A})k\|\boldsymbol{Q}\|_F. &(6.58)\end{aligned}$$

Hence, on $\mathcal{E}_1$,

$$\|\mathcal{W}[\boldsymbol{Q}]\|_F \leq \left(16\mu(\boldsymbol{A})\sqrt{k} + 8\mu(\boldsymbol{A})k\right)\|\boldsymbol{Q}\|_F, \tag{6.59}$$

and so $\|\boldsymbol{W}\| \leq 16\mu(\boldsymbol{A})\sqrt{k} + 8\mu(\boldsymbol{A})k \leq 24k\mu(\boldsymbol{A})$. Combining this observation with (6.48) gives the desired result, (6.17). We just have to establish the three inequalities (6.56)-(6.58). Paragraphs (i)-(iii) below do this.

**(i) Establishing** (6.56)**.** For the term $\boldsymbol{T}_2$ defined in (6.52), notice

$$\begin{aligned} \|\boldsymbol{T}_2\|_F &= \left\|\boldsymbol{e}_1^*\boldsymbol{\Delta}\Big(\sum_b \boldsymbol{e}_b\boldsymbol{e}_b^*\boldsymbol{Q}\,\boldsymbol{P}_{\Omega^b}\boldsymbol{x}^{1*}\Big)\boldsymbol{x}^1\boldsymbol{P}_{\Omega^1}\right\|_2 \\ &\leq \left\|\boldsymbol{e}_1^*\boldsymbol{\Delta}\right\|_2\left\|\sum_b \boldsymbol{e}_b\boldsymbol{e}_b^*\boldsymbol{Q}\,\boldsymbol{P}_{\Omega^b}\boldsymbol{x}^{1*}\right\|_2\left\|\boldsymbol{x}^1\boldsymbol{P}_{\Omega^1}\right\|_2 &(6.60)\\ &\leq \sqrt{n}\|\boldsymbol{e}_1^*\boldsymbol{\Delta}\|_\infty \times \left\|\sum_b \boldsymbol{e}_b\boldsymbol{e}_b^*\boldsymbol{Q}\,\boldsymbol{P}_{\Omega^b}\boldsymbol{x}^{1*}\right\|_2 \times 2 &(6.61)\\ &\leq 4\sqrt{n}\,\mu(\boldsymbol{A})\left\|\sum_b \boldsymbol{e}_b\boldsymbol{e}_b^*\boldsymbol{Q}\,\boldsymbol{P}_{\Omega^b}\boldsymbol{x}^{1*}\right\|_2. &(6.62)\end{aligned}$$

Above, we have used: the Cauchy-Schwarz inequality in (6.60), the bound $\|\boldsymbol{X}^1\| \leq 2$ on $\mathcal{E}_1$ in (6.61), and the bound $\|\boldsymbol{\Delta}\|_\infty \leq 2\mu(\boldsymbol{A})$ in (6.62). Now,

$$\begin{aligned} \Big\|\sum_b \boldsymbol{e}_b\boldsymbol{e}_b^*\boldsymbol{Q}\,\boldsymbol{P}_{\Omega^b}\boldsymbol{x}^{1*}\Big\|_2^2 &= \sum_b (\boldsymbol{e}_b^*\boldsymbol{Q}\,\boldsymbol{P}_{\Omega^b}\boldsymbol{x}^{1*})^2 \\ &\leq \sum_b \|\boldsymbol{e}_b^*\boldsymbol{Q}\|_2^2\|\boldsymbol{x}^1\boldsymbol{P}_{\Omega^b}\|_2^2 \\ &\leq 4k\,\|\boldsymbol{Q}\|_F^2\,/n. &(6.63)\end{aligned}$$

Combining (6.62) and (6.63) establishes (6.56).

**(ii) Establishing** (6.57)**.** For the term $\boldsymbol{T}_3$ defined in (6.53), we have

$$\begin{aligned} \|\boldsymbol{T}_3\|_F^2 &= \sum_{a=2}^n (\boldsymbol{e}_a^*\boldsymbol{\Delta}\boldsymbol{e}_1)^2\left\|\boldsymbol{e}_1^*\boldsymbol{Q}\;\;\boldsymbol{P}_{\Omega^1}\boldsymbol{x}^{1*}\;\;\boldsymbol{x}^1\boldsymbol{P}_{\Omega^a}\right\|^2 \\ &\leq 4\mu^2(\boldsymbol{A})\sum_{a=2}^n \|\boldsymbol{e}_1^*\boldsymbol{Q}\|^2\|\boldsymbol{x}^1\boldsymbol{P}_{\Omega^1}\|^2\|\boldsymbol{x}^1\boldsymbol{P}_{\Omega^a}\|^2, &(6.64)\\ &\leq 4\mu^2(\boldsymbol{A}) \times 4 \times 4(k/n) \times (n-1)\|\boldsymbol{e}_1^*\boldsymbol{Q}\|^2. &(6.65)\end{aligned}$$

In (6.64) we have used the bound $\|\boldsymbol{\Delta}\|_\infty \leq 2\mu(\boldsymbol{A})$ and the Cauchy-Schwarz inequality; in (6.64) we have plugged in the bounds $\|\boldsymbol{X}^1\| \leq 2$ and $\|\boldsymbol{X}^1\boldsymbol{P}_{\Omega^a}\| \leq 2\sqrt{k/n}$. Finally, conservatively bounding $\|\boldsymbol{e}_1^*\boldsymbol{Q}\|$ by $\|\boldsymbol{Q}\|_F$ and taking the square root of both sides gives the desired result, (6.57).

**(iii) Establishing** (6.58)**.** The final term, $\boldsymbol{T}_4$ requires a bit more manipulation. Expressing $\|\boldsymbol{T}_4\|_F^2$ as a sum of squared $\ell^2$ norms of the rows of $\boldsymbol{T}_4$ gives

$$\begin{aligned}
\|\boldsymbol{T}_4\|_F^2 &= \sum_{a=2}^{n}\Big\|\sum_{b=2}^{n}\boldsymbol{e}_a^*\boldsymbol{\Delta}\boldsymbol{e}_b\,\boldsymbol{e}_b^*\boldsymbol{Q}\boldsymbol{P}_{\Omega^b}\boldsymbol{x}^{1^*}\times\boldsymbol{x}^1\boldsymbol{P}_{\Omega^a}\Big\|^2 \\
&= \sum_{a=2}^{n}\|\boldsymbol{x}^1\boldsymbol{P}_{\Omega^a}\|^2\Big(\sum_{b=2}^{n}\boldsymbol{e}_a^*\boldsymbol{\Delta}\boldsymbol{e}_b\times\boldsymbol{e}_b^*\boldsymbol{Q}\boldsymbol{P}_{\Omega^b}\boldsymbol{x}^{1^*}\Big)^2 && (6.66)\\
&\le 4(k/n)\times\sum_{a=2}^{n}\Big(\boldsymbol{e}_a^*\boldsymbol{\Delta}\sum_{b=2}^{n}\boldsymbol{e}_b\boldsymbol{e}_b^*\boldsymbol{Q}\boldsymbol{P}_{\Omega^b}\boldsymbol{x}^{1^*}\Big)^2 && (6.67)\\
&\le 4(k/n)\times\sum_{a=2}^{n}\|\boldsymbol{e}_a^*\boldsymbol{\Delta}\|_2^2\Big\|\sum_{b=2}^{n}\boldsymbol{e}_b\boldsymbol{e}_b^*\boldsymbol{Q}\boldsymbol{P}_{\Omega^b}\boldsymbol{x}^{1^*}\Big\|_2^2 && (6.68)\\
&\le 4(k/n)\times\|\boldsymbol{\Delta}\|_F^2\times\Big\|\sum_{b=2}^{n}\boldsymbol{e}_b\boldsymbol{e}_b^*\boldsymbol{Q}\boldsymbol{P}_{\Omega^b}\boldsymbol{x}^{1^*}\Big\|_2^2 && (6.69)\\
&\le 4(k/n)\times n^2\|\boldsymbol{\Delta}\|_\infty^2\times\sum_{b=2}^{n}\Big(\boldsymbol{e}_b^*\boldsymbol{Q}\boldsymbol{P}_{\Omega^b}\boldsymbol{x}^{1^*}\Big)^2 && (6.70)\\
&\le 16\,k\,n\,\mu^2(\boldsymbol{A})\times\sum_{b=2}^{n}\|\boldsymbol{e}_b^*\boldsymbol{Q}\|_2^2\|\boldsymbol{P}_{\Omega^b}\boldsymbol{x}^{1^*}\|_2^2 && (6.71)\\
&\le 64\,k^2\,\mu^2(\boldsymbol{A})\times\sum_{b=2}^{n}\|\boldsymbol{e}_b^*\boldsymbol{Q}\|^2. && (6.72)
\end{aligned}$$

Bounding the summation by $\|\boldsymbol{Q}\|_F^2$ and taking square roots gives (6.58). This completes the proof of Lemma 6.3. □

# A Conditioning of the Linearized Subproblem

In Section 2, we sketched an argument for why the restricted isometry property would not be useful for analyzing the linearized subproblem in dictionary learning. We demonstrated a perturbation pair $(\boldsymbol{\Delta}_A, \boldsymbol{\Delta}_X)$ lying in nullspace of the linear constraints, such that $\boldsymbol{\Delta}_X$ has the same sparsity as $\boldsymbol{X}$. The reader might wonder whether this analysis can be made rigorous, and justifiably so, since the classical RIP analysis pertains to an $\ell^1$-minimization problem in which all of the variables are weighted equally, whereas in the linearized subproblem, $\boldsymbol{\Delta}_A$ is not penalized. In this section, we show a more precise sense in which the RIP is violated.

To see this, notice that whenever $\boldsymbol{X}$ has full row rank $n$, we can write

$$\boldsymbol{\Delta}_A = -\boldsymbol{A}\boldsymbol{\Delta}_X \boldsymbol{X}^*(\boldsymbol{X}\boldsymbol{X}^*)^{-1}. \tag{A.1}$$

Eliminating $\boldsymbol{\Delta}_A$ yields the equivalent problem

$$\text{minimize } \|\boldsymbol{X}+\boldsymbol{\Delta}_X\|_1 \quad \text{subject to} \quad \boldsymbol{A}\boldsymbol{\Delta}_X(\boldsymbol{I}-\boldsymbol{P}_X) = \boldsymbol{0},\ \langle \boldsymbol{A}_i, \boldsymbol{A}\boldsymbol{\Delta}_X\boldsymbol{X}^*(\boldsymbol{X}\boldsymbol{X}^*)^{-1}\boldsymbol{e}_i\rangle = 0\ \forall\, i, \tag{A.2}$$

where $\boldsymbol{P}_X = \boldsymbol{X}^*(\boldsymbol{X}\boldsymbol{X}^*)^{-1}\boldsymbol{X}$. If we make the substitution $\boldsymbol{Z} = \boldsymbol{X} + \boldsymbol{\Delta}_X$, we obtain an equivalent problem,

$$\begin{array}{lll} \text{minimize } \|\boldsymbol{Z}\|_1 & \text{subject to} & \boldsymbol{A}\boldsymbol{Z}(\boldsymbol{I}-\boldsymbol{P}_X) = \boldsymbol{A}\boldsymbol{X}(\boldsymbol{I}-\boldsymbol{P}_X), \\ & & \langle \boldsymbol{A}_i, \boldsymbol{A}\boldsymbol{Z}\boldsymbol{X}^*(\boldsymbol{X}\boldsymbol{X}^*)^{-1}\boldsymbol{e}_i\rangle = \langle \boldsymbol{A}_i, \boldsymbol{A}\boldsymbol{X}\boldsymbol{X}^*(\boldsymbol{X}\boldsymbol{X}^*)^{-1}\boldsymbol{e}_i\rangle\ \forall\, i. \end{array} \tag{A.3}$$

This is an equality constrained $\ell^1$ norm minimization problem. We wish to know whether $\boldsymbol{Z} = \boldsymbol{X}$ is the unique optimal solution (corresponding to $\boldsymbol{\Delta}_X = \boldsymbol{0}$ being uniquely optimal for the original problem). Let $\pi$ be any permutation of $[n]$ with no fixed point, and let $\boldsymbol{\Pi} \in \mathbb{R}^{n\times n}$ be the corresponding permutation matrix. Then it is easy to verify that if we set $\boldsymbol{H} = \boldsymbol{\Pi}\boldsymbol{X} \in \mathbb{R}^n$, $\boldsymbol{A}\boldsymbol{H}(\boldsymbol{I}-\boldsymbol{P}_X) = \boldsymbol{0}$, and

$$|\langle \boldsymbol{A}_i, \boldsymbol{A}\boldsymbol{H}\boldsymbol{X}^*(\boldsymbol{X}\boldsymbol{X}^*)^{-1}\boldsymbol{e}_i\rangle| \le \mu(\boldsymbol{A}).$$

It is also obvious that $\boldsymbol{H}$ has the same number of nonzeros, and in fact the same number of nonzeros in each column as the desired solution $\boldsymbol{X}$. Hence, if $\mu(\boldsymbol{A}) = 0$ (i.e., $\boldsymbol{A}$ is an orthonormal basis), $\boldsymbol{H}$ lies in the nullspace of the linear constraints in (A.3), and so the RIP cannot hold for this problem. Hence, in what is arguably the best possible case for recovering $\boldsymbol{A}$ and $\boldsymbol{X}$, the linearized subproblem cannot have the RIP. Moreover, applying linear operations to the constraints in (A.3) cannot help, since $\boldsymbol{H}$ lies strictly in the nullspace of these constraints. When $\mu(\boldsymbol{A})$ is nonzero but small, as in our above problem, $\boldsymbol{H}$ still lies very near the nullspace, and the RIP does not hold with any useful constant for (A.3). Of course, strictly speaking, when $\mu(\boldsymbol{A}) > 0$ this argument does not preclude the possiblity that there is some linear transformation of the equality constraints in (A.3) that does have the RIP.

# B Technical Tools

In this section, we quote two results used in our arguments. The first, which plays a key role, is the matrix Chernoff bound of Tropp [Tro10]. This convenient and powerful result builds on ideas introduced by Ahlswede and Winter [AW02].

**Theorem B.1** (Matrix Chernoff Bound, [Tro10] Theorem 2.5)**.** *Let $\boldsymbol{M}_1, \dots, \boldsymbol{M}_n$ be a finite sequence of independent random positive-semidefinite matrices of dimension $d$. Suppose that for each $\boldsymbol{M}_i$, $\lambda_{max}(\boldsymbol{M}_i) \le B$ almost surely. Set $\mu_{min} = \lambda_{min}(\sum_i \mathbb{E}[\boldsymbol{M}_i])$ and $\mu_{max} = \lambda_{max}(\sum_i \mathbb{E}[\boldsymbol{M}_i])$. Then the following two bounds hold:*

$$\mathbb{P}\Big[\lambda_{min}\Big(\sum_i \boldsymbol{M}_i\Big) \le t\mu_{min}\Big] \quad \le \quad d\,\exp\left(-(1-t)^2\mu_{min}/2B\right), \quad \forall\, t \in [0,1), \tag{B.1}$$

$$\mathbb{P}\Big[\lambda_{max}\Big(\sum_i \boldsymbol{M}_i\Big) \ge (1+t)\mu_{max}\Big] \quad \le \quad d\left(\frac{e^t}{(1+t)^{1+t}}\right)^{\mu_{max}/B}, \quad \forall\, t \ge 0. \tag{B.2}$$

Two simplifications of the upper tail are useful:

$$\mathbb{P}\Big[\ \Big\|\sum_i \boldsymbol{M}_i\Big\|\ \geq\ (1+t)\,\mu_{max}\ \Big]\quad\leq\quad d\exp\left(-t^2\mu_{max}/4B\right),\quad \forall\, t\in[0,1], \tag{B.3}$$

and

$$\mathbb{P}\Big[\ \Big\|\sum_i \boldsymbol{M}_i\Big\|\ \geq\ t\,\mu_{max}\ \Big]\quad\leq\quad d\,\left(\frac{e}{t}\right)^{t\mu_{max}/B},\quad \forall\, t>e. \tag{B.4}$$

The second is given in [Tro10], while the first follows from (B.2) and the inequality $t-(1+t)\log(1+t)\leq -t^2/4$, which by convexity (or calculus) can be shown to hold on $[0,1]$.

The second result we quote here is the classical Kahane-Khintchine inequality, here with constant $1/\sqrt{2}$ found by Latala and Oleszkiewicz [LO94]:

**Theorem B.2** (Kahane-Khintchine Inequality [Kah64], [LO94] Theorem 1)**.** *Let $\sigma_1,\dots,\sigma_n$ be an iid sequence of Rademacher random variables (i.e., variables that take on $\pm 1$ with equal probability), and let $\boldsymbol{x}_1,\dots,\boldsymbol{x}_n$ be a fixed sequence of vectors in a normed space $V$. Then*

$$\frac{1}{\sqrt{2}}\left(\mathbb{E}\Big[\Big\|\sum_i \sigma_i\boldsymbol{x}_i\Big\|_V^2\Big]\right)^{1/2}\ \leq\ \mathbb{E}\Big[\Big\|\sum_i \sigma_i\boldsymbol{x}_i\Big\|_V\Big]\ \leq\ \left(\mathbb{E}\Big[\Big\|\sum_i \sigma_i\boldsymbol{x}_i\Big\|_V^2\Big]\right)^{1/2} \tag{B.5}$$

This result has the following useful consequence:

**Corollary B.3.** *Let $\boldsymbol{M}\in\mathbb{R}^{m\times n}$ be any fixed matrix, and $\boldsymbol{v}\in\mathbb{R}^n$ be an iid $\mathcal{N}(0,\sigma^2)$ vector. Then*

$$\frac{\sigma}{\sqrt{\pi}}\|\boldsymbol{M}\|_F\ \leq\ \mathbb{E}\left[\|\boldsymbol{M}\boldsymbol{v}\|_2\right]\ \leq\ \sigma\|\boldsymbol{M}\|_F. \tag{B.6}$$

# C Consequences of Incoherence

In this section, we assemble several useful consequences of the assumption that $\boldsymbol{A}$ has low mutual coherence. All of the following bounds are well known [Fuc04]; we record their statements and (very simple) proofs here for completeness. As above, let $\boldsymbol{A}\in\mathbb{R}^{m\times n}$ be a matrix with unit norm columns and mutual coherence $\mu(\boldsymbol{A})$. A bound on the mutual coherence immediately implies about on the norm of $\boldsymbol{A}$. Set $\boldsymbol{\Delta}=\boldsymbol{A}^*\boldsymbol{A}-\boldsymbol{I}$. Then

$$\|\boldsymbol{A}\|^2=\|\boldsymbol{A}^*\boldsymbol{A}\|=\|\boldsymbol{I}+\boldsymbol{\Delta}\|\leq 1+\|\boldsymbol{\Delta}\|\leq 1+\|\boldsymbol{\Delta}\|_F\leq 1+n\|\boldsymbol{\Delta}\|_\infty=1+n\mu(\boldsymbol{A}). \tag{C.1}$$

For submatrices of $\boldsymbol{A}$, tighter bounds can be obtained in a similar manner: if $L\in\binom{[n]}{k}$, the same argument shows

$$\|\boldsymbol{A}_L\|^2=\|\boldsymbol{A}_L^*\boldsymbol{A}_L\|\leq 1+k\mu(\boldsymbol{A}). \tag{C.2}$$

Similarly, via eigenvalue perturbation bounds,

$$\lambda_{min}(\boldsymbol{A}_L^*\boldsymbol{A}_L)\geq 1-k\mu(\boldsymbol{A}). \tag{C.3}$$

In particular, if we assume that $k\mu(\boldsymbol{A})<1/2$, we have

$$\|(\boldsymbol{A}_L^*\boldsymbol{A}_L)^{-1}\|\leq 2. \tag{C.4}$$

We can obtain a tighter result by using the Neumann series representation of the inverse. Suppose that $k\mu(\boldsymbol{A})<1/2$, and write $\boldsymbol{A}_L^*\boldsymbol{A}_L=\boldsymbol{I}+\boldsymbol{H}$, and note that $\|\boldsymbol{H}\|_F<k\mu(\boldsymbol{A})$. Then using

$$(\boldsymbol{A}_L^*\boldsymbol{A}_L)^{-1}=\sum_{t=0}^{\infty}(-1)^t\boldsymbol{H}^t, \tag{C.5}$$

we have

$$\|(\boldsymbol{A}_L^*\boldsymbol{A}_L)^{-1}-\boldsymbol{I}\|_F\ =\ \|\sum_{t=1}^{\infty}(-\boldsymbol{H})^t\|_F\ \leq\ \sum_{t=1}^{\infty}\|-\boldsymbol{H}\|_F^t\ \leq\ k\mu(\boldsymbol{A})/(1-k\mu(\boldsymbol{A}))\ <\ 2k\mu(\boldsymbol{A}). \tag{C.6}$$

# D Local Properties

In this section, we prove Lemma 3.1, which reduces the question of whether $\boldsymbol{x}$ is a local optimum of $f$ over $\mathcal{M}$ to one of whether $\boldsymbol{\delta} = \mathbf{0}$ minimized $f(\boldsymbol{x}+\boldsymbol{\delta})$ over $T_{\boldsymbol{x}}\mathcal{M}$. The reasoning behind this lemma is simple. The function $f(\boldsymbol{A},\boldsymbol{X}) = \|\boldsymbol{X}\|_1$ is Lipschitz (its Lipschitz constant $L$ is at most $\sqrt{np}$). At the same time, $f$ is a polyhedral semi-norm. This means that the (unbounded) unit ball of $f$, $B \doteq \{\boldsymbol{x} \mid f(\boldsymbol{x}) \le 1\}$ is a polyhedral set. Let $r = \min\{f(\boldsymbol{x}+\boldsymbol{\delta}) \mid \boldsymbol{\delta} \in T_{\boldsymbol{x}}\mathcal{M}\}$. The set of optimal $\boldsymbol{\delta}$ precisely correspond to the the points $-\boldsymbol{x} + rB \cap T_{\boldsymbol{x}}\mathcal{M}$. In particular, if the optimizer is unique, then $rB$ intersects $\boldsymbol{x} + T_{\boldsymbol{x}}\mathcal{M}$ at a single point. Since $B$ is polyhedral, this intersection is "sharp": $f$ increases linearly as we move away from the optimal point. This property, sometimes referred to in the optimization literature as *strong uniqueness* implies that higher order terms due to the curvature of $\mathcal{M}$ are negligible; if $\boldsymbol{\delta} = \mathbf{0}$ is a unique optimum over the tangent space, $\boldsymbol{x}$ is locally optimal over $\mathcal{M}$. We now formally prove Lemma 3.1.

*Proof.* For any $\boldsymbol{\delta} \in T_{\boldsymbol{x}}\mathcal{M}$, let $\boldsymbol{x}_{\boldsymbol{\delta}} : (-\varepsilon, \varepsilon) \to \mathcal{M}$ be the unique geodesic satisfying $\boldsymbol{x}_{\boldsymbol{\delta}}(0) = \boldsymbol{x}$ and $\dot{\boldsymbol{x}}_{\boldsymbol{\delta}}(0) = \boldsymbol{\delta}$. Then

$$\boldsymbol{x}_{\boldsymbol{\delta}}(t) = \boldsymbol{x} + t\boldsymbol{\delta} + O(t^2). \tag{D.1}$$

We first prove that optimality over the tangent space is necessary. Indeed, suppose there exists $\boldsymbol{\delta} \in T_{\boldsymbol{x}}\mathcal{M}$ with $f(\boldsymbol{x}+\boldsymbol{\delta}) < f(\boldsymbol{x})$. Set $\tau = f(\boldsymbol{x}) - f(\boldsymbol{x}+\boldsymbol{\delta}) > 0$. By convexity, for $\eta \in [0,1]$,

$$f(\boldsymbol{x}+\eta\boldsymbol{\delta}) \le f(\boldsymbol{x}) - \eta\tau. \tag{D.2}$$

But,

$$\begin{aligned} f(\boldsymbol{x}_{\boldsymbol{\delta}}(t)) &= f(\boldsymbol{x}+\eta\boldsymbol{\delta} + (\boldsymbol{x}_{\boldsymbol{\delta}}(t) - (\boldsymbol{x}+\eta\boldsymbol{\delta}))) \le f(\boldsymbol{x}+\eta\boldsymbol{\delta}) + L\|\boldsymbol{x}_{\boldsymbol{\delta}}(t) - \boldsymbol{x} - \eta\boldsymbol{\delta}\|_2 \\ &\le f(\boldsymbol{x}) - \eta\tau + L\|(t-\eta)\boldsymbol{\delta}\|_2 + O(Lt^2). \end{aligned} \tag{D.3}$$

When t is sufficiently small and let $\eta = t$, this value is strictly smaller than $f(\boldsymbol{x})$.

Conversely, suppose that $\boldsymbol{\delta} = \mathbf{0}$ is the unique minimizer of $f(\boldsymbol{x}+\boldsymbol{\delta})$ over $\boldsymbol{\delta} \in T_{\boldsymbol{x}}\mathcal{M}$. We will show that this minimizer is *strongly unique* (see e.g., [JO80]), i.e., $\exists\, \beta > 0$ such that

$$f(\boldsymbol{x}+\boldsymbol{\delta}) \ge f(\boldsymbol{x}) + \beta\|\boldsymbol{\delta}\| \quad \forall\, \boldsymbol{\delta} \in T_{\boldsymbol{x}}\mathcal{M}. \tag{D.4}$$

To see this, notice that if we write $\boldsymbol{x} = (\boldsymbol{A},\boldsymbol{X})$ and $\boldsymbol{\delta} = (\boldsymbol{\Delta}_A, \boldsymbol{\Delta}_X)$, then $f(\boldsymbol{x}) = \|\boldsymbol{X}\|_1$. Hence, if we set $r_0 = \min\{|X_{ij}| \mid X_{ij} \ne 0\} > 0$, whenever $\|\boldsymbol{\Delta}_X\|_\infty < r_0$ and $t < 1$, we have

$$\begin{aligned} \|\boldsymbol{X} + t\boldsymbol{\Delta}_X\|_1 &= \|\boldsymbol{X}\|_1 + t\langle \boldsymbol{\Sigma}, \boldsymbol{\Delta}_X\rangle + t\|\mathcal{P}_{\Omega^c}\boldsymbol{\Delta}_X\|_1 \\ &= \|\boldsymbol{X}\|_1 + t\langle \boldsymbol{\Sigma} + \operatorname{sign}(\mathcal{P}_{\Omega^c}\boldsymbol{\Delta}_X), \boldsymbol{\Delta}_X\rangle. \end{aligned}$$

Set $\beta(\boldsymbol{\delta}) \doteq \langle \boldsymbol{\Sigma} + \operatorname{sign}(\mathcal{P}_{\Omega^c}\boldsymbol{\Delta}_X), \boldsymbol{\Delta}_X\rangle$, and notice that $\beta$ is a continuous function of $\boldsymbol{\delta}$. Let

$$\beta^\star = \inf_{\boldsymbol{\delta} \in T_{\boldsymbol{x}}\mathcal{M}, \|\boldsymbol{\delta}\| = r_0/2} \beta(\boldsymbol{\delta}) \ge 0. \tag{D.5}$$

Then for all $\boldsymbol{\delta} \in T_{\boldsymbol{x}}\mathcal{M}$ with $\|\boldsymbol{\delta}\| \le r_0/2$ we have

$$f(\boldsymbol{x}+\boldsymbol{\delta}) \ge f(\boldsymbol{x}) + (2\beta^\star/r_0)\|\boldsymbol{\delta}\|. \tag{D.6}$$

Moreover, by convexity of $f$, the same bound holds for all $\boldsymbol{\delta} \in T_{\boldsymbol{x}}\mathcal{M}$ (regardless of $\|\boldsymbol{\delta}\|$). It remains to show that $\beta^\star$ is strictly larger than zero. On the contrary, suppose $\beta^\star = 0$. Since the infimum in (D.5) is taken over a compact set, it is achieved by some $\boldsymbol{\delta}^\star \in T_{\boldsymbol{x}}\mathcal{M}$. Hence, if $\beta^\star = 0$, $f(\boldsymbol{x}+\boldsymbol{\delta}^\star) = f(\boldsymbol{x})$, contradicting the uniqueness of the minimizer $\boldsymbol{x}$. This establishes (D.4).

Hence, continuing forward, we have

$$f(\boldsymbol{x}_{\boldsymbol{\delta}}(\eta)) \ge f(\boldsymbol{x}) + \eta\beta\|\boldsymbol{\delta}\| - O(\beta\eta^2). \tag{D.7}$$

For $\eta$ sufficiently small the right hand side is strictly greater than $f(\boldsymbol{x})$. □

## E A Decoupling Lemma

The following technical lemma concerns the expected norm of the restriction $\boldsymbol{P}_\Omega \boldsymbol{M} \boldsymbol{P}_\Omega$ of a matrix $\boldsymbol{M}$ with no diagonal elements to a random diagonal block. Its proof is an application of a well-known decoupling technique [LT91]. In particular, several steps are quite similar to manipulations in the proof of Proposition 2.1 of [Tro08], and are repeated here for completeness.

**Lemma E.1.** *Fix $\boldsymbol{M} \in \mathbb{R}^{n\times n}$ with all diagonal elements equal to zero. Let $\Omega \sim \mathrm{uni}\binom{[n]}{k}$ be a uniform random subset of size $k$. Then the following estimate holds:*

$$\mathbb{E}\left[\|\boldsymbol{P}_\Omega \boldsymbol{M} \boldsymbol{P}_\Omega\|_F\right] \;\le\; 16\sqrt{\frac{k}{n}}\,\mathbb{E}\left[\|\boldsymbol{M}\boldsymbol{P}_\Omega\|_F\right]. \tag{E.1}$$

*Proof.* Let $\boldsymbol{\Lambda}$ denote a diagonal matrix whose entries are iid Bernoulli random variables taking on value 1 with probability $k/n$. Let $k' \doteq \mathrm{tr}[\boldsymbol{\Lambda}]$ denote the number of nonzeros in $\boldsymbol{\Lambda}$; $k'$ is a binomial random variable. Then

$$\begin{aligned}
\mathbb{E}\left[\|\boldsymbol{\Lambda}\boldsymbol{M}\boldsymbol{\Lambda}\|_F\right] &= \sum_{s=0}^{n} \mathbb{P}[k'=s]\,\mathbb{E}\left[\|\boldsymbol{\Lambda}\boldsymbol{M}\boldsymbol{\Lambda}\|_F \mid k'=s\right], && \text{(E.2)}\\
&\ge \sum_{s=k}^{n} \mathbb{P}[k'=s]\,\mathbb{E}\left[\|\boldsymbol{\Lambda}\boldsymbol{M}\boldsymbol{\Lambda}\|_F \mid k'=s\right]. && \text{(E.3)}
\end{aligned}$$

Now, conditioned on $k'=s$, the nonzeros on the diagonal of $\boldsymbol{\Lambda}$ are distributed according to a uniform distribution on $\binom{[n]}{s}$. Furthermore, whenever $\Omega \subset \mathrm{support}(\boldsymbol{\Lambda})$, $\|\boldsymbol{P}_\Omega \boldsymbol{M}\boldsymbol{P}_\Omega\|_F \le \|\boldsymbol{\Lambda}\boldsymbol{M}\boldsymbol{\Lambda}\|_F$, since $\Omega$ restricts to a smaller submatrix. Hence,

$$\forall\, s \ge k, \quad \mathbb{E}\left[\|\boldsymbol{\Lambda}\boldsymbol{M}\boldsymbol{\Lambda}\|_F \mid k'=s\right] \;\ge\; \mathbb{E}\left[\|\boldsymbol{P}_\Omega \boldsymbol{M}\boldsymbol{P}_\Omega\|_F\right]. \tag{E.4}$$

Plugging into (E.3), and using that $k$ is a median of the binomial random variable $k'$, we have

$$\begin{aligned}
\mathbb{E}\left[\|\boldsymbol{\Lambda}\boldsymbol{M}\boldsymbol{\Lambda}\|_F\right] &\ge \sum_{s=k}^{n} \mathbb{P}[k'=s]\,\mathbb{E}\left[\|\boldsymbol{P}_\Omega \boldsymbol{M}\boldsymbol{P}_\Omega\|_F\right], && \text{(E.5)}\\
&= \mathbb{P}[k'\ge k]\,\mathbb{E}\left[\|\boldsymbol{P}_\Omega \boldsymbol{M}\boldsymbol{P}_\Omega\|_F\right], && \text{(E.6)}\\
&\ge \frac{1}{2}\,\mathbb{E}\left[\|\boldsymbol{P}_\Omega \boldsymbol{M}\boldsymbol{P}_\Omega\|_F\right]. && \text{(E.7)}
\end{aligned}$$

Hence, we have

$$\mathbb{E}\left[\|\boldsymbol{P}_\Omega \boldsymbol{M}\boldsymbol{P}_\Omega\|_F\right] \;\le\; 2\,\mathbb{E}\left[\|\boldsymbol{\Lambda}\boldsymbol{M}\boldsymbol{\Lambda}\|_F\right]. \tag{E.8}$$

Similar to [Tro08], for each $i,j$, let $\boldsymbol{M}_{ij} \in \mathbb{R}^{n\times n}$ be a matrix whose $(i,j)$ entry is equal to the $(i,j)$ entry of $\boldsymbol{M}$, and whose other entries are equal to zero. Write

$$\mathbb{E}\left[\|\boldsymbol{\Lambda}\boldsymbol{M}\boldsymbol{\Lambda}\|_F\right] \;=\; \mathbb{E}\Big[\Big\|\sum_{i>j}\lambda_i\lambda_j(\boldsymbol{M}_{ij}+\boldsymbol{M}_{ji})\Big\|_F\Big] \tag{E.9}$$

Introduce an independent sequence of Bernoulli random variables $\eta_1,\dots,\eta_n$, each taking on value 1 with probability 1/2, and write

$$\begin{aligned}
\mathbb{E}\Big[\Big\|\sum_{i>j}\lambda_i\lambda_j(\boldsymbol{M}_{ij}+\boldsymbol{M}_{ji})\Big\|_F\Big] &= 2\,\mathbb{E}_{\boldsymbol{\Lambda}}\Big[\Big\|\mathbb{E}_\eta\Big[\sum_{i>j}\Big(\eta_i(1-\eta_j)+\eta_j(1-\eta_i)\Big)\lambda_i\lambda_j(\boldsymbol{M}_{ij}+\boldsymbol{M}_{ji})\Big]\Big\|_F\Big]\\
&\le 2\,\mathbb{E}_{\boldsymbol{\Lambda}}\mathbb{E}_\eta\Big[\Big\|\sum_{i>j}\Big(\eta_i(1-\eta_j)+\eta_j(1-\eta_i)\Big)\lambda_i\lambda_j(\boldsymbol{M}_{ij}+\boldsymbol{M}_{ji})\Big\|_F\Big], && \text{(E.10)}\\
&= 2\,\mathbb{E}_\eta\mathbb{E}_{\boldsymbol{\Lambda}}\Big[\Big\|\sum_{i>j}\Big(\eta_i(1-\eta_j)+\eta_j(1-\eta_i)\Big)\lambda_i\lambda_j(\boldsymbol{M}_{ij}+\boldsymbol{M}_{ji})\Big\|_F\Big]. && \text{(E.11)}
\end{aligned}$$

In (E.10), we used Jensen's inequality to pull the expectation out of the norm. Now, there must be at least one sequence $\eta^\star$ such that the right hand side of (E.11) is no smaller than its expectation over $\eta$. Let $T \subset [n]$ be the indices of the nonzeros in $\eta^\star$, and let $T^c$ be its complement. Then combining (E.8) and (E.11), we have

$$\begin{aligned}
\mathbb{E}\Big[\|\boldsymbol{P}_\Omega \boldsymbol{M} \boldsymbol{P}_\Omega\|_F\Big] &\le 4\,\mathbb{E}_{\boldsymbol{\Lambda}}\Big[\Big\|\sum_{i>j}\Big(\eta_i^\star(1-\eta_j^\star)+\eta_j^\star(1-\eta_i^\star)\Big)\lambda_i\lambda_j(\boldsymbol{M}_{ij}+\boldsymbol{M}_{ji})\Big\|_F\Big] && \text{(E.12)}\\
&= 4\,\mathbb{E}_{\Lambda}\Big[\Big\|\sum_{i\in T, j\in T^c}\lambda_i\lambda_j(\boldsymbol{M}_{ij}+\boldsymbol{M}_{ji})\Big\|_F\Big] && \text{(E.13)}\\
&\le 4\,\mathbb{E}_{\boldsymbol{\Lambda}}\Big[\Big\|\sum_{i\in T, j\in T^c}\lambda_i\lambda_j\boldsymbol{M}_{ij}\Big\|_F\Big] + 4\,\mathbb{E}_{\boldsymbol{\Lambda}}\Big[\Big\|\sum_{i\in T, j\in T^c}\lambda_i\lambda_j\boldsymbol{M}_{ji}\Big\|_F\Big] && \text{(E.14)}
\end{aligned}$$

Now, let $\boldsymbol{\Lambda}'$ be an independent copy of $\boldsymbol{\Lambda}$. Then the above is equal to

$$\begin{aligned}
&4\,\mathbb{E}_{\boldsymbol{\Lambda},\boldsymbol{\Lambda}'}\Big[\Big\|\sum_{i\in T, j\in T^c}\lambda_i'\lambda_j\boldsymbol{M}_{ij}\Big\|_F\Big] + 4\,\mathbb{E}_{\boldsymbol{\Lambda},\boldsymbol{\Lambda}'}\Big[\Big\|\sum_{i\in T, j\in T^c}\lambda_i\lambda_j'\boldsymbol{M}_{ji}\Big\|_F\Big]\\
&\le 8\,\mathbb{E}_{\boldsymbol{\Lambda},\boldsymbol{\Lambda}'}\Big[\Big\|\sum_{i,j=1}^{n}\lambda_i'\lambda_j\boldsymbol{M}_{ij}\Big\|_F\Big] \;=\; 8\,\mathbb{E}_{\boldsymbol{\Lambda},\boldsymbol{\Lambda}'}\Big[\|\boldsymbol{\Lambda}'\boldsymbol{M}\boldsymbol{\Lambda}\|_F\Big] && \text{(E.15)}\\
&\le 8\,\mathbb{E}_{\boldsymbol{\Lambda}}\Big(\mathbb{E}_{\boldsymbol{\Lambda}'}\Big[\|\boldsymbol{\Lambda}'\boldsymbol{M}\boldsymbol{\Lambda}\|_F^2\Big]\Big)^{1/2} && \text{(E.16)}\\
&= 8\sqrt{k/n}\,\mathbb{E}_{\boldsymbol{\Lambda}}\Big[\|\boldsymbol{M}\boldsymbol{\Lambda}\|_F\Big]. && \text{(E.17)}
\end{aligned}$$

Above, we have used the fact that the Frobenius norm does not increase when a matrix is restricted to a subset of its elements to move from a summation over $(i,j) \in T \times T^c$ to a summation over all pairs $(i,j)$.

We now just have to move from the Bernoulli model back to the uniform model. For a fixed value $s$ of $k'$ (i.e., a fixed number of nonzeros in $\boldsymbol{\Lambda}$), we can divide support($\boldsymbol{\Lambda}$) into $a = \lceil k'/k \rceil$ random subsets $S_1, \dots, S_a$ of size at most $k$. Conditioned on $k' = s$, the marginal distribution of each $S_i$ is uniform on $\binom{[n]}{|S_i|}$, and hence

$$\mathbb{E}_{\boldsymbol{\Lambda}}\Big[\|\boldsymbol{M}\boldsymbol{P}_{S_i}\|_F \mid k' = s\Big] \;\le\; \begin{cases} \mathbb{E}_\Omega\left[\|\boldsymbol{M}\boldsymbol{P}_\Omega\|_F\right] & (i-1)k < s \\ 0 & \text{else} \end{cases}. \qquad \text{(E.18)}$$

The condition on $i$ in the first line above simply implies that $S_i$ is nonempty. So,

$$\begin{aligned}
\mathbb{E}_{\boldsymbol{\Lambda}}\Big[\,\|\boldsymbol{M}\boldsymbol{\Lambda}\|_F\Big] &\le \mathbb{E}_{\boldsymbol{\Lambda}}\Big[\sum_i \|\boldsymbol{M}\boldsymbol{P}_{S_i}\|_F\Big] && \text{(E.19)}\\
&= \sum_{s=0}^{n}\sum_i \mathbb{E}_\Lambda\left[\|\boldsymbol{M}\boldsymbol{P}_{S_i}\|_F \mid k'=s\right]\,\mathbb{P}[k'=s]. && \text{(E.20)}\\
&\le \sum_{s=0}^{n}\left\lfloor \frac{s}{k}+1\right\rfloor\,\mathbb{E}_\Omega\left[\|\boldsymbol{M}\boldsymbol{P}_\Omega\|_F\right]\,\mathbb{P}[k'=s] && \text{(E.21)}\\
&= \mathbb{E}_\Omega\left[\,\|\boldsymbol{M}\boldsymbol{P}_\Omega\|_F\,\right] \times \mathbb{E}\left[k'/k+1\right] \;=\; 2\,\mathbb{E}_\Omega\left[\,\|\boldsymbol{M}\boldsymbol{P}_\Omega\|_F\,\right]. && \text{(E.22)}
\end{aligned}$$

Combining (E.17) and (E.22) completes the proof. □

# F Proof of Lemma 5.2

We will establish Lemma 5.2 by applying the matrix Chernoff bound to the extreme eigenvalues of the sum of independent positive semidefinite matrices

$$\boldsymbol{X}\boldsymbol{X}^* = \sum_{j=1}^{p} \boldsymbol{x}_j \boldsymbol{x}_j^*. \tag{F.1}$$

A bit of care needs to be taken because the summands $\boldsymbol{x}_j \boldsymbol{x}_j^*$ have unbounded norm; we handle this by replacing them with a sequence of truncated terms $\bar{\boldsymbol{x}}_j \bar{\boldsymbol{x}}_j^*$ that are equivalent to $\boldsymbol{x}_j \boldsymbol{x}_j^*$ with very high probability.

*Proof.* Set

$$\bar{\boldsymbol{x}}_j = \begin{cases} \boldsymbol{x}_j & \|\boldsymbol{x}_j\| \;\le\; (1+\beta)\sqrt{n\log p/p} \\ \boldsymbol{0} & \text{else} \end{cases} \tag{F.2}$$

were $\beta > 0$ is a constant to be chosen later. It is not difficult to show[8] that for each $j$

$$\mathbb{P}\left[ \|\boldsymbol{x}_j\| > (1+\beta)\sqrt{\frac{n\log p}{p}} \right] \;<\; p^{-\beta^2/2}. \tag{F.4}$$

So, $\max_j \|\boldsymbol{x}_j\|$ is bounded by $(1+\beta)\sqrt{n\log p/p}$ with probability at least $1 - p^{1-\beta^2/2}$. Hence, with at least this probability $\bar{\boldsymbol{x}}_j = \boldsymbol{x}_j \;\forall\, j$, and $\sum_j \boldsymbol{x}_j \boldsymbol{x}_j^* = \sum_j \bar{\boldsymbol{x}}_j \bar{\boldsymbol{x}}_j^*$. Thanks to truncation, the following bound always holds:

$$\|\bar{\boldsymbol{x}}_j \bar{\boldsymbol{x}}_j^*\| \;\le\; B \;\doteq\; (1+\beta)^2 \frac{n\log p}{p}. \tag{F.5}$$

Since $\boldsymbol{x}_j \boldsymbol{x}_j^* \succeq \bar{\boldsymbol{x}}_j \bar{\boldsymbol{x}}_j^*$ always, $\mathbb{E}[\boldsymbol{x}_j \boldsymbol{x}_j^*] \succeq \mathbb{E}[\bar{\boldsymbol{x}}_j \bar{\boldsymbol{x}}_j^*]$, and so

$$\mu_{max} \;\doteq\; \left\| \mathbb{E}\Big[ \sum_j \bar{\boldsymbol{x}}_j \bar{\boldsymbol{x}}_j^* \Big] \right\| \;\le\; \left\| \mathbb{E}\Big[ \sum_j \boldsymbol{x}_j \boldsymbol{x}_j^* \Big] \right\| \;=\; \|\boldsymbol{I}\| \;=\; 1. \tag{F.6}$$

Plugging in to the bound (B.3), we have

$$\mathbb{P}\Big[\; \lambda_{max}\Big( \sum_j \bar{\boldsymbol{x}}_j \bar{\boldsymbol{x}}_j^* \Big) \;\ge\; 1 + t \;\Big] \quad\le\quad n \exp\left( - \frac{t^2\, \mu_{max}\, p}{4\,(1+\beta)^2\, n\, \log p} \right). \tag{F.7}$$

Notice that there is still a dependence on $\mu_{max} \le 1$ in the exponent. This will be resolved by developing a lower bound on $\mu_{min} \le \mu_{max}$.

---

[8] Conditioned on $\Omega_j$, $\|\boldsymbol{x}_j\| = \|\boldsymbol{P}_{\Omega_j} \boldsymbol{v}_j\| \doteq f(\boldsymbol{v}_j)$ is a 1-Lipschitz function of $\boldsymbol{v}_j$. From Jensen's inequality, $\mathbb{E}[f(\boldsymbol{v}_j)] \le \sqrt{\mathbb{E}[f^2(\boldsymbol{v}_j)]} \le \sqrt{n/p}$. Hence, from Gaussian measure concentration,

$$\mathbb{P}\left[\, f(\boldsymbol{v}_j) \ge \mathbb{E}\left[ f(\boldsymbol{v}_j) \mid \Omega_j \right] + t \mid \Omega_j \,\right] \;\le\; \exp\left( - \frac{t^2 k p}{2n} \right). \tag{F.3}$$

We set $t = \beta\sqrt{n\log p/p}$ and then use the fact that the bound holds for all $\Omega_j$ to remove the conditioning, giving (F.4).

The smallest eigenvalue requires a bit more work, since it decreases under truncation:

$$\begin{aligned}
\mu_{min} &\doteq \lambda_{min}\Big(\mathbb{E}\Big[\sum_j \bar{\boldsymbol{x}}_j \bar{\boldsymbol{x}}_j^*\Big]\Big) && \text{(F.8)}\\
&\geq \lambda_{min}\Big(\mathbb{E}\Big[\sum_j \boldsymbol{x}_j \boldsymbol{x}_j^*\Big]\Big) - \Big\|\mathbb{E}\Big[\sum_j \bar{\boldsymbol{x}}_j \bar{\boldsymbol{x}}_j^* - \boldsymbol{x}_j \boldsymbol{x}_j^*\Big]\Big\| && \text{(F.9)}\\
&\geq \lambda_{min}\Big(\mathbb{E}\Big[\sum_j \boldsymbol{x}_j \boldsymbol{x}_j^*\Big]\Big) - \sum_j \mathbb{E}\Big[\Big\|\bar{\boldsymbol{x}}_j \bar{\boldsymbol{x}}_j^* - \boldsymbol{x}_j \boldsymbol{x}_j^*\Big\|\Big] && \text{(F.10)}\\
&= \lambda_{min}(\boldsymbol{I}) - \sum_j \mathbb{E}\left[\left\|\boldsymbol{x}_j \boldsymbol{x}_j^*\right\| \mathbf{1}_{\|\boldsymbol{x}_j\| > \sqrt{B}}\right] && \text{(F.11)}\\
&= 1 - \sum_j \mathbb{E}\left[\|\boldsymbol{x}_j\|_2^2 \mathbf{1}_{\|\boldsymbol{x}_j\| > \sqrt{B}}\right] && \text{(F.12)}\\
&\geq 1 - \sum_j \sqrt{\mathbb{E}[\|\boldsymbol{x}_j\|_2^4]}\sqrt{\mathbb{E}[(\mathbf{1}_{\|\boldsymbol{x}_j\| > \sqrt{B}})^2]} && \text{(F.13)}\\
&= 1 - p\sqrt{\mathbb{E}[\|\boldsymbol{x}_1\|_2^4]}\sqrt{\mathbb{P}[\|\boldsymbol{x}_1\| > \sqrt{B}]} && \text{(F.14)}\\
&\geq 1 - p \times \sqrt{3}\, n/p \times p^{-\beta^2/4} \quad = \quad 1 - \sqrt{3}\, n p^{-\beta^2/4}. && \text{(F.15)}
\end{aligned}$$

Above, in (F.10) we have used Jensen's inequality, while in (F.11) we have used the definition of $\bar{\boldsymbol{x}}_j$. In (F.13) we have used the Cauchy-Schwarz inequality. The manipulations are completed using the fact that for $Z \sim \mathcal{N}(0, \sigma^2)$, $\mathbb{E}[Z^4] = 3\sigma^4$ to give the following bound on $\mathbb{E}\|\boldsymbol{x}_1\|^4$:

$$\begin{aligned}
\mathbb{E}\left[\|\boldsymbol{x}_1\|_2^4\right] &= \mathbb{E}\left[\big(\sum_{a \in \Omega_1} V_{a1}^2\big)^2\right] = \sum_{a,b \in \Omega_1} \mathbb{E}\left[V_{a1}^2 V_{b1}^2\right] \;=\; k(k-1)\sigma^4 + k\mathbb{E}[V_{11}^4]\\
&= (k^2 + 2k)\sigma^4 \;\leq\; 3k^2\sigma^4 \;=\; 3\,n^2/p^2.
\end{aligned}$$

From the above, write $\mu_{min} \geq 1 - g(p)$. Then Tropp's bound gives

$$\mathbb{P}\Big[\lambda_{min}\Big(\sum_j \bar{\boldsymbol{x}}_j \bar{\boldsymbol{x}}_j^*\Big) < 1 - t\Big] \;\leq\; n \exp\left(-\frac{(t - g(p))^2}{2(\beta+1)^2} \frac{p}{n \log p}\right). \tag{F.16}$$

For concreteness, we choose $\beta = 4$. Then provided $p > (Cn/t)^{1/4}$, $g(p) < t/2 < 1/2$, completing the proof. □